\documentclass{article}

\usepackage{arxiv}
\usepackage{natbib}

\usepackage[utf8]{inputenc} 
\usepackage[T1]{fontenc}    
\usepackage{hyperref}       
\usepackage{url}            
\usepackage{booktabs}       
\usepackage{amsfonts}       
\usepackage{nicefrac}       
\usepackage{microtype}      
\usepackage{lipsum}

\usepackage{soul}

\usepackage{makecell}
\usepackage{bm}
\usepackage{bbm}
\usepackage{multirow}
\usepackage{lineno,subcaption}
\usepackage{graphicx}
\usepackage{amsmath}

\title{Dynamic and interpretable hazard-based models of traffic incident durations}

\author{
  Kieran~Kalair \\
  Centre for Complexity Science\\
  University of Warwick\\
  Gibbet Hill Road, Coventry CV4 7AL \\ 
  United Kingdom. \\
  \texttt{k.kalair@warwick.ac.uk} \\
   \And
 Colm~Connaughton\thanks{London Mathematical Laboratory, London, United Kingdom} \\
  Mathematics Institute\\
  University of Warwick\\
  Gibbet Hill Road, Coventry CV4 7AL \\ 
  United Kingdom. \\
  \texttt{c.p.connaughton@warwick.ac.uk} \\
}

\begin{document}
\maketitle

\begin{abstract}
Understanding and predicting the duration or ``return-to-normal" time of traffic incidents is important for system-level management and optimisation of road transportation networks. Increasing real-time availability of multiple data sources characterising the state of urban traffic networks, together with advances in machine learning offer the opportunity for new and improved approaches to this problem that go beyond static statistical analyses of incident duration. In this paper we consider two such improvements: dynamic update of incident duration predictions as new information about incidents becomes available and automated interpretation of the factors responsible for these predictions. For our use case, we take one year of incident data and traffic state time-series data from the M25 motorway in London. We use it to train models that predict the probability distribution of incident durations, utilising both time-invariant and time-varying features of the data. The latter allow predictions to be updated as an incident progresses, and more information becomes available. For dynamic predictions, time-series features are fed into the Match-Net algorithm, a temporal convolutional hitting-time network, recently developed for dynamical survival analysis in clinical applications. The predictions are benchmarked against static regression models for survival analysis and against an established dynamic technique known as landmarking and found to perform favourably by several standard comparison measures. To provide interpretability, we utilise the concept of Shapley values recently developed in the domain of interpretable artificial intelligence to rank the features most relevant to the model predictions at different time horizons.  For example, the time of day is always a significantly influential time-invariant feature, whereas the time-series features strongly influence predictions at 5 and 60-minute horizons. Although we focus here on traffic incidents, the methodology we describe can be applied to many survival analysis problems where time-series data is to be combined with time-invariant features.
\end{abstract}


\section{Introduction}\label{sec:Introduction}

Managing and reducing congestion on roads is a fundamental challenge faced across the world.
In many countries, one being the UK, there is limited scope to build additional infrastructure to cope with the demand for road traffic.
Instead, the focus is on an approach referred to as `Intelligent Mobility', broadly described as combining real-time data and modelling to improve the management of existing physical infrastructure.
An appropriate test-bed for such approaches is the UK Strategic Road Network (SRN), which constitutes around 4,400 miles of motorways and major trunk roads across England \citep[see][]{strategic_road_network_stats}.
Further, the SRN carries 30\% of all traffic in the UK, with 4 million vehicles using it everyday and 1 billion tonnes of freight being transported across it each year \citep[see][]{highways_england_srn_report}.
There is little chance that in the short term, the SRN will see significant infrastructure changes, however data describing the traffic state across the network is already being collected and made available for analysis.
Whilst a significant component of the UK transport infrastructure, congestion remains a major problem on the network, with the cited report suggesting 75\% of businesses consider tackling congestion on the SRN is important or critical to their business.

Traffic congestion can be broadly separated into two types: recurrent and non-recurrent.
Recurrent congestion is simply the result of the demand regularly exceeding capacity on busy sections of road during `rush hour' periods.
Non-recurrent congestion on the other hand is mainly caused by traffic incidents and rare incidents \citep{overview_of_traffic_incident_duration_analysis_and_prediction}.
To better manage traffic during these incidents, traffic operators require reliable estimates of how long a particular incident will last.
Whilst there is significant existing work on modelling incident duration, many fundamental challenges remain that are both of practical interest to traffic management centres and remain active areas of research in an academic sense. 
A review of existing work on this problem is found in \citet{overview_of_traffic_incident_duration_analysis_and_prediction}, where six future challenges for incident duration prediction are listed.
These are: combining multiple data-sources, time sequential prediction models, outlier prediction, improvement of prediction methods through machine learning or alternative frameworks, combining recovery times and accounting for unobserved factors.
Our work aims to address five of these challenges, combining time-series from sensor networks with incident reports to issue dynamically updated duration predictions. 
Inspired by approaches adopted in medical applications, we consider classical survival approaches and non-linear, machine learning methods for prediction, understanding where gains in performance are attained.
Finally, as we are able to observe traffic behaviour over long periods of time through the sensor network, we are able to judge not just when a traffic incident has been cleared, but when the traffic state has returned to normal operating conditions, thereby combining recovery times into our modelling approach.

The rest of this paper is structured as follows.
Firstly, we overview existing work on incident duration prediction, specifying how the duration is defined, existing methods and offer more detail on challenges highlighted in the literature.
Secondly, we detail our dataset, its collection and processing and an initial exploratory analysis of it.
We then describe the considered modelling approaches, both for static and dynamic predictions, and results for our dataset.
Finally, we consider what variables are important for the models and end by summarising our main findings.

Note that throughout this paper, an `event' is defined as the traffic state on a section of road returning to a baseline behaviour.
As such, when an event has `occurred' we really mean the traffic state has recovered. 
This is just a note of terminology commonly used in the survival analysis literature.

\section{Background}\label{sec:LitReview}

In this section, we summarise existing work on traffic incident duration analysis that is relevant to our own, along with relevant research from other disciplines that has influenced our approach.

Before any methodologies are considered, it is first important to define exactly what is being modelled.
A traffic incident is considered to have four different time-phases: the time taken to detect and report an incident, the time to dispatch an operator to the scene, the travel time of the operator to the scene, and finally the time to clear an incident.
Such a framework is described in \cite{overview_of_traffic_incident_duration_analysis_and_prediction}.
We consider an incident to have `started' when the incident is reported by the human operator, who would use a series of cameras to observe the state of the road  in the case of the SRN.
Our focus is to model the time from this point until both the incident has been physically cleared, and traffic behaviour on the road has returned to some sense of `normal' behaviour.
We describe exactly how we determine normal behaviour in section \ref{sec:DataDrivenBaseline}. 
In brief, we use a seasonal model of the speed time-series to estimate when the traffic speed on a section of road has recovered to a level close to what would be normally be expected for that section at that time of day.
The idea to use such a speed profile is also considered in \citet{modelling_total_duration_of_traffic_incidents_including_incident_detection_and_recovery_time}, where they define the `total incident duration' to be the time from incident start until the speed has recovered to the profile.

Whatever explicit definition of duration is used, there is an enormous amount of work on predicting incident durations.
An initial step of many of these is to determine an appropriate distributional form that the durations take.
These are typically heavy tailed and empirically show significant variation.
Examples of this include modelling the distribution of incident durations as log-normal in \cite{an_analysis_of_the_severity_and_incident_duration_of_truck_involved_freeway_accidents} and \cite{analytical_method_to_estimate_accident_duration_using_archived_speed_profile_and_its_statistical_analysis}, log-logistic in \cite{analysis_of_cascading_incident_event_duraitons_on_urban_freeways} and \cite{estimating_freeway_incident_duraiton_using_accelerated_failure_time_modelling}, Weibull in \cite{modelling_total_duration_of_traffic_incidents_including_incident_detection_and_recovery_time} and \cite{response_time_of_highway_traffic_accidents_in_abu_Dhabi_investigation_with_hazard_based_duration_models}, and generalised F in \cite{examination_of_factors_affecting_freeway_incident_clearance_times_a_comparision_of_the_generalized_F_model_and_several_alternative_nested_models}.
In the latter, it is noted that the generalised F distribution can be equivalent to many other distributional forms for particular parameter choices, including the exponential, Weibull, log-normal, log-logistic and gamma distributions. 
Hence, it offers more freedom than choosing any single one of these forms. 
Indeed, the authors state that the increased flexibility it offers allows it to fit the data better.
Even further flexibility in the distributional choice is given in \citet{application_of_finite_mixture_models_for_analysing_freeway_incident_clearance_time}, where it was shown that modelling the distribution as a mixture, that is the sum of multiple components, may improve model performance. 
Specifically, they consider a 2-component log-logistic mixture model, where the final distribution is the weighted average of two log-logistic distributions.
Finally, other authors \cite{forecasting_the_clearance_time_of_freeway_accidents} have had difficulty finding statistically defensible distributional fits to their data, although it should be noted that different definitions of incident durations will likely impact this.

Using common probability distribution is appealing as it limits a model's freedom and can be easier to fit to data.
However, it is clear that authors are exploring more complex distributional forms to better model the data and seeing better results when they do so, as in \citet{examination_of_factors_affecting_freeway_incident_clearance_times_a_comparision_of_the_generalized_F_model_and_several_alternative_nested_models} and \citet{application_of_finite_mixture_models_for_analysing_freeway_incident_clearance_time}.
Mixture distributions are a naturally appealing form, as we assume the data is generated by multiple sub-populations, and can have different effects of covariates for different populations.
We incorporate ideas from this section of the literature by considering models that assume log-normal and Weibull distributions on incident durations, as-well as mixture distributions where one supposes the data is generated by one of many sub-populations, each of which has a parametric form.
Recent applications of survival analysis in healthcare \cite{deephit_a_deep_learning_approach_to_survival_analysis_with_competing_risks} have removed distributional assumptions entirely, and instead formulate models that output distributions with no closed form.
This is done by treating the output space as discrete, and treating the model output as a probability mass function (PMF) defined over it, allowing for construction of a fully non-parametric estimate.
Such an approach offers even more freedom, and as we see more complex distributions used in the traffic literature to provide more freedom, one could ask if removing the distribution assumption entirely can improve model performance.

After a distribution is determined, many works focus on methods from survival analysis, with a common choice being the Accelerated Failure Time (AFT) model.
Example applications of this are given in \cite{an_exploratory_hazard_based_analysis_of_highway_incident_duration}, \cite{modelling_accident_duration_and_its_mitigation_strategies_on_south_korean_freeway_systems} and \cite{simultaneous_equation_modelling_of_freeway_accident_duration_and_lanes_blocked}.
Such a model assumes that each covariate either accelerates or decelerates the life-time of a particular individual.
These models are widely used, and offer an interpretable means of investigating what factors strongly or weakly influence incident duration.
However, it can be difficult to incorporate time-series features into them.
Whilst it is possible to model time-varying effects of covariates, for example in \cite{efficient_estimation_for_the_accelerated_faliure_time_model}, it is more complex to derive optimal features from a time-series that are also interpretable.
Clearly, AFT models are useful in that they produce interpretable outputs and relationships between variables, and can model the non-Gaussian distributions empirically observed in incident duration data.
The alternative and well known classical survival model one could apply is a Cox regression model \cite{regression_models_and_life_tables}. 
Such a model assumes a baseline hazard function for the population, describing the instantaneous rate of incidents, from which survival probabilities can be calculated, see section \ref{sec:SurvivalAnalysisMethods} for more details.
Covariate vectors for individuals shift this baseline hazard allowing for individualised predictions.
Applications of this to transportation problems are given in \cite{determination_of_the_risk_factors_that_influence_occurence_time_of_traffic_accidents_with_survival_analysis}, \cite{the_effect_of_earlier_or_automatic_collision_notification_on_traffic_mortality_by_survival_analysis} and \cite{competing_risks_analysis_on_traffic_accident_duration_time}.

Whilst these two methods are widely used, a number of alternatives exist.
One such method is a sequential regression approach, an early example being \citet{a_simple_time_sequential_procedure_for_predicting_freeway_incident_duration}.
Importantly, the authors identify that more information describing an incident will become available over time, and hence consider a series of models to make sequential predictions. 
However, their sequential information was more descriptive from an operational stand-point, for example identifying damage to the road and response time of rescue vehicles.
We do not have this, and instead focus on the minute-to-minute updates provided through traffic time-series recorded by sensors along the road, and how to engineer features from these.
Truncated regression approaches were discussed in \citet{analysis_of_cascading_incident_event_duraitons_on_urban_freeways}, where there was a specific effort to model `cascading' incidents (referred to as primary and secondary incidents in some literature).
The thought here was that incidents that occur nearby in time and space would lead to a significantly longer clearance time for the road segment, and hence this should be accounted for in modelling.
We performed an extensive analysis of primary and secondary incidents in our dataset in \cite{a_non_parametric_hawkes_process_model_of_primary_and_secondary_accidents_on_a_uk_smart_motorway}. Building on this and the previous cited work, we include a cascade variable in the models, allowing this to influence duration predictions.

Further regression approaches are explored in \cite{estimating_magnitude_and_duration_of_incident_delays} and \cite{cluster_based_lognormal_distribution_model_for_accident_duration}, and switching regression models are used in \cite{exploring_the_influential_factors_in_incident_clearance_time_disentangling_causation_from_self_selection_bias}.
Note that in \cite{cluster_based_lognormal_distribution_model_for_accident_duration}, the authors first cluster the incident data, then use this clustering as additional features for a model, further suggesting that there is some element of sub-population structure in the data.
A final relevant regression based work is \cite{modelling_traffic_incident_duration_using_quantile_regression}, where quantile regression is used to model incident durations.
This is a natural choice, as there is a clear skew in the empirically observed duration distributions, and if one does not want to assume a particular distributional form, they can instead model properties of the distribution, in this case quantiles of the data.

Further methodologies to note are those based on tree models or ensembles of them, particularly because we apply such a method later in this work.
Tree based models are discussed in \cite{forecasting_the_clearance_time_of_freeway_accidents}, where the authors compare models that assume particular incident duration distributions, a k-nearest-neighbour approach and a classification tree method based on predicting `short', `medium' and `long' incidents.
They concluded that no model provided accurate enough results on their dataset to warrant industrial implementation, but found the classification tree was the preferred model of those considered.
Further, \cite{prediction_of_lane_clearance_time_of_freeway_incidents_using_the_M5p_tree_algorithm} considered a regression tree approach, where the terminal nodes of each tree were themselves multivariate linear models.
Such an approach avoids binning of incidents into pre-defined categories, and achieved 42.70\% mean absolute percentage error, better than the compared reference models.
From an interdisciplinary setting, alternative tree methods have been considered, namely one known as `random survival forests' \cite{random_survival_forests} as an extension to random forests to a survival analysis setting.
In such a framework, the terminal nodes of each tree specify cumulative hazard functions for all data-points that fall into that node, and these hazards are combined across many trees to determine an ensemble hazard.
There is no defined distributional assumption in such a model, again leaning towards the side of freedom in allowing the data to construct its own hazard function estimate rather than parametrizing an estimated form.

Neural networks are a rapidly developing methodology in machine learning, and have been used extensively in incident duration prediction and form the basis for some of our considered approaches.
Examples of this include \cite{vehicle_breakdown_duration_modelling}, \cite{applying_data_fusion_techniques_to_traveler_information_services_in_highway_network} and \cite{a_computerized_feature_selection_method_using_genetic_algorithms}.
Each of these applies feed forward neural networks to determine estimates of incident durations, and particularly in \cite{a_computerized_feature_selection_method_using_genetic_algorithms} sequential prediction was considered, using two models.
The first took standard inputs, and the second took these along with detector information near the incident.
These were input into feed-forward neural networks, and used to generate point predictions.
Additional neural network applications are given in \cite{a_comparative_study_of_models_for_the_incident_duration_prediction}, where their performance is compared to that of linear regressions, decision trees, support vector machines and k-nearest-neighbour methods.
The authors find that different models have optimal performance at different incident durations, suggesting there is still much to improve on feed forward networks.
A final point to note is that neural networks have been applied to survival analysis problems in healthcare a number of times.
Examples of this include \cite{deep_surv_personalized_treatment_recommender_system_using_a_cox} which develops a Cox model, replacing linear regression with a neural network output, \cite{deephit_a_deep_learning_approach_to_survival_analysis_with_competing_risks} which removes any distributional assumptions, and \cite{dynamic_prediction_in_clinical_survival_analysis_using_temporal_convolutional_networks} which uses a sliding window mechanism and temporal convolutions for dynamic predictions.
We consider if the later two are useful in the application to traffic incidents later in this work.
Specifically, using \cite{dynamic_prediction_in_clinical_survival_analysis_using_temporal_convolutional_networks} offers an automated way to engineer features from our sensor network data, whilst being able to model a parametric or non-parametric output.

Whilst we have discussed a number of different methodological approaches, the actual features used to make these predictions, regardless of approach, appear quite consistent across different works.
In \cite{estimating_magnitude_and_duration_of_incident_delays}, the authors state that using number of lanes affected, number of vehicles involved, truck involvement, time of day, police response time and weather condition, one can explain 81\% of variation in incident duration.
An overview of various feature types is given in \cite{overview_of_traffic_incident_duration_analysis_and_prediction}, identifying incident characteristics, environmental conditions, temporal characteristics, road characteristics, traffic flow measurements, operator reactions and vehicle characteristics as important factors when modelling incident durations.

We note that we are not the first to use sensor data in incident duration analysis. 
Speed data collected from roads was used in \cite{examination_of_factors_affecting_freeway_incident_clearance_times_a_comparision_of_the_generalized_F_model_and_several_alternative_nested_models}, where the authors included two features based on the speed series: if the difference between the 15th and 85th percentiles of the speed data was greater than 7mph and if the 85th percentile for speed was less than 70mph.
Further, in \cite{sequential_forecast_of_incident_duraiton_using_artifical_neural_network_models} and \cite{a_computerized_feature_selection_method_using_genetic_algorithms} the authors train two feed forward neural networks, with input features that include the speed and flow for detectors near the incident.
The first model provided a forecast just before the incident occurred, where as new data was fed into the second whenever available, updating predictions as time progressed.
In the second paper, the focus was on reducing the dimensionality of the problem through feature selection via a genetic algorithm.
We fundamentally differ from these for multiple reasons. 
With regard to \cite{examination_of_factors_affecting_freeway_incident_clearance_times_a_comparision_of_the_generalized_F_model_and_several_alternative_nested_models}, this was an analysis of which factors impact incident duration the most, and where as they used hazard based models for analysis, we use the sensor data to engineer dynamic features, either manually or through temporal convolutions.
Additionally, we differ from \cite{sequential_forecast_of_incident_duraiton_using_artifical_neural_network_models} and \cite{a_computerized_feature_selection_method_using_genetic_algorithms} through how we determine features and network structure, and through predicting an output distribution, not just a point estimate.

With dynamic prediction being highlighted as an area to address in the literature, some recent papers have looked at this problem through different methods to those already discussed.
One example of this is \cite{competing_risk_mixture_model_and_text_analysis_for_sequential_incident_duration_prediction}, where a topic model was used to interpret written report messages and predictions were made as new textual information arrived.
Further, multiple regression models were built in \cite{dynamic_prediction_of_the_incident_duration_using_adaptive_feature_set}, and as different features became available, data-points were assigned clusters, and a prediction was generated using a regression model tailored to each cluster.
Lastly \cite{sequential_prediction_for_large_scale_traffic_incident_duraiton_application_and_comparision_of_survival_models} considered a five stage approach, where a prediction was made at each stage and different features were available defining these stages. 
These included vehicles involved, agency response time and number of agencies responding.
While this structured approach addresses some aspects of dynamic prediction, the purely data-driven approach which we present in this paper provides much more flexibility. 

\section{Data Collection \& Processing}\label{sec:Data}

As discussed, various traffic data for the SRN is provided by a system called the National Traffic Information Service (NTIS)\footnote{Technical details of the NTIS data feeds are available at \url{http://www.trafficengland.com/services-info}}.
This is both historic and real-time data.
Whilst the SRN includes all motorways and major A-roads in the UK, we focus our analysis on one the busiest UK motorways, the M25 London Orbital, pictured in \textbf{Figure 1}.
\begin{figure}[ht!]
	\centering
	\includegraphics[width=0.48\textwidth]{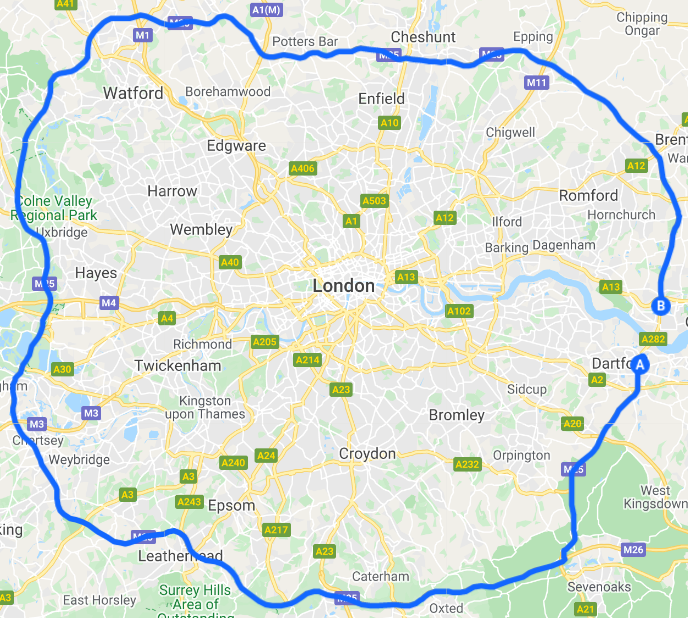}
	\caption{The M25 London Orbital, roughly 180 kilometres in length. The Dartford crossing, located in the east, is a short segment of road that is not included in the data set.}\label{fig:M25Plot}
\end{figure}
Inside NTIS, roads are represented by a directed graph, with edges (referred to as links from now on) being segments of road that have a constant number of lanes and no slip roads joining or leaving. 
We extract all links that lie on the M25 in the clockwise direction, yielding a subset of the road network to collect data for.
Placed along links are physical sensors called `loops', which record passing vehicles and report averaged quantities each minute.

The most relevant components of NTIS to our work are incident flags that are manually entered by traffic operators.
These flags specify an incident type, for example accident or obstruction, the start and end time, date and the link the incident occurred on.
Accompanying this information are further details on the incident, for example what vehicles it involved and how many lanes were blocked if any.
We extract all incidents that occurred on the chosen links between September 1st 2017 and September 30th 2018.
Along with this, NTIS provides time-series data for each link, recording the average travel time, speed and flow for vehicles and publishing it at 1 minute intervals.
These values are determined by a combination of floating vehicle data and loop sensors.
As well as the incident flags, we extract the time-series of these quantities in the specified time period.
In total, our dataset has 4415 incidents that we train on, and 2011 incidents that we use for out of sample validation of the models.
Note that of these 4415 training incidents, we use 1324 as hold-out data to judge when to stop our training machine learning models.

\subsection{Establishing A Data-Driven Baseline}\label{sec:DataDrivenBaseline}

As discussed, incident duration consists of 4 distinct phases and we want to model the time it takes for a link to recover to some baseline state.
How to determine this baseline, and ensure it is robust to outliers whilst retaining important features of the data is an open problem.
However a natural way to approach it is to develop some seasonal model of behaviour on a link and use this seasonality as a baseline behaviour.
We define such a baseline by first taking the speed time-series and pre-filtering it by removing the periods impacted by incidents.
After, we account for any potential missing incident flags by further removing any remaining periods with a severity higher than 0.3.
Severity is defined as in \cite{anomaly_detection_and_classification_in_traffic_flow_data_from_fluctuations_in_the_flow_density_relationship}, which in short, considers the joint behaviour of the speed-flow time-series and questions what points correspond to large fluctuations from a region of typical behaviour in this relationship.
We then take this filtered series and extract the seasonal components, in our case daily and weekly components, to capture natural variability on the link.
Note that inspection of the data shows no trend.

To construct a seasonal estimate, we consider simple phase averaging, taking the median of data collected at a given time-point in a week, and STL decomposition \cite{STL_a_seasonal_trend_decompostion_procedure_based_on_loess}.
We see little difference between the two methods, so choose to use the phase average baseline for simplicity.
We define one speed baseline for each link, establishing a robust profile describing the speed behaviour on a `typical week', and replicate this over the entire data period.
It is robust in the sense that we have pre-filtered the extreme outliers. It also captures the clear seasonality in the problem  and can be applied to new test data without any difficulty.
An example of this profile for a single link, along with the residuals from it are given in \textbf{Figure 2A}, along with an example incident in \textbf{Figure 2B}.
We specifically mark where the NTIS flag was raised, where it was closed, and where the speed behaviour returned to the baseline.
\begin{figure}[ht!]
	\centering
	\begin{minipage}[t]{.47\textwidth}
		\includegraphics[width=\textwidth]{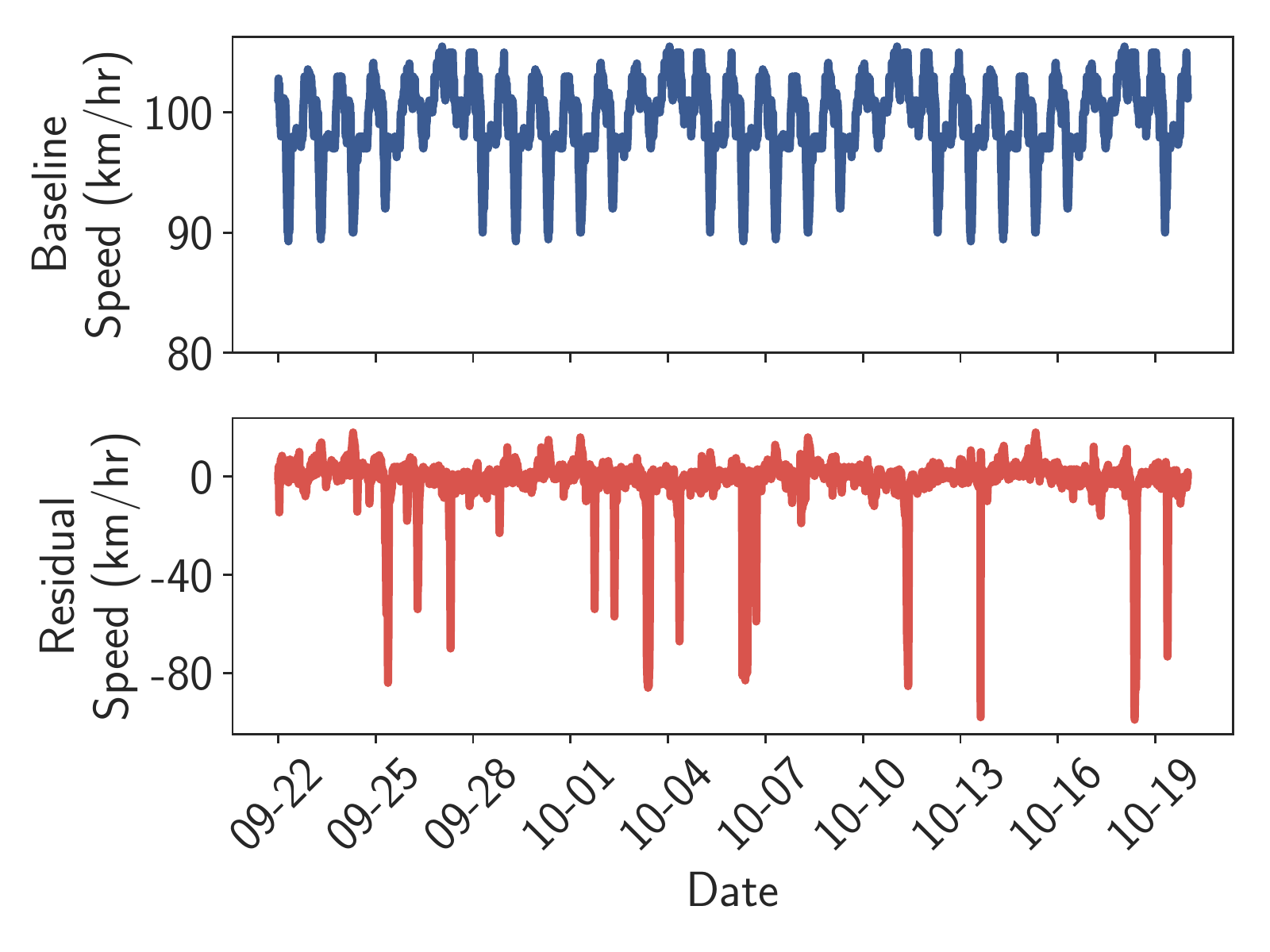}
		\centering \textbf{A}
	\end{minipage}
	~
	\begin{minipage}[t]{.47\textwidth}
		\includegraphics[width=\textwidth]{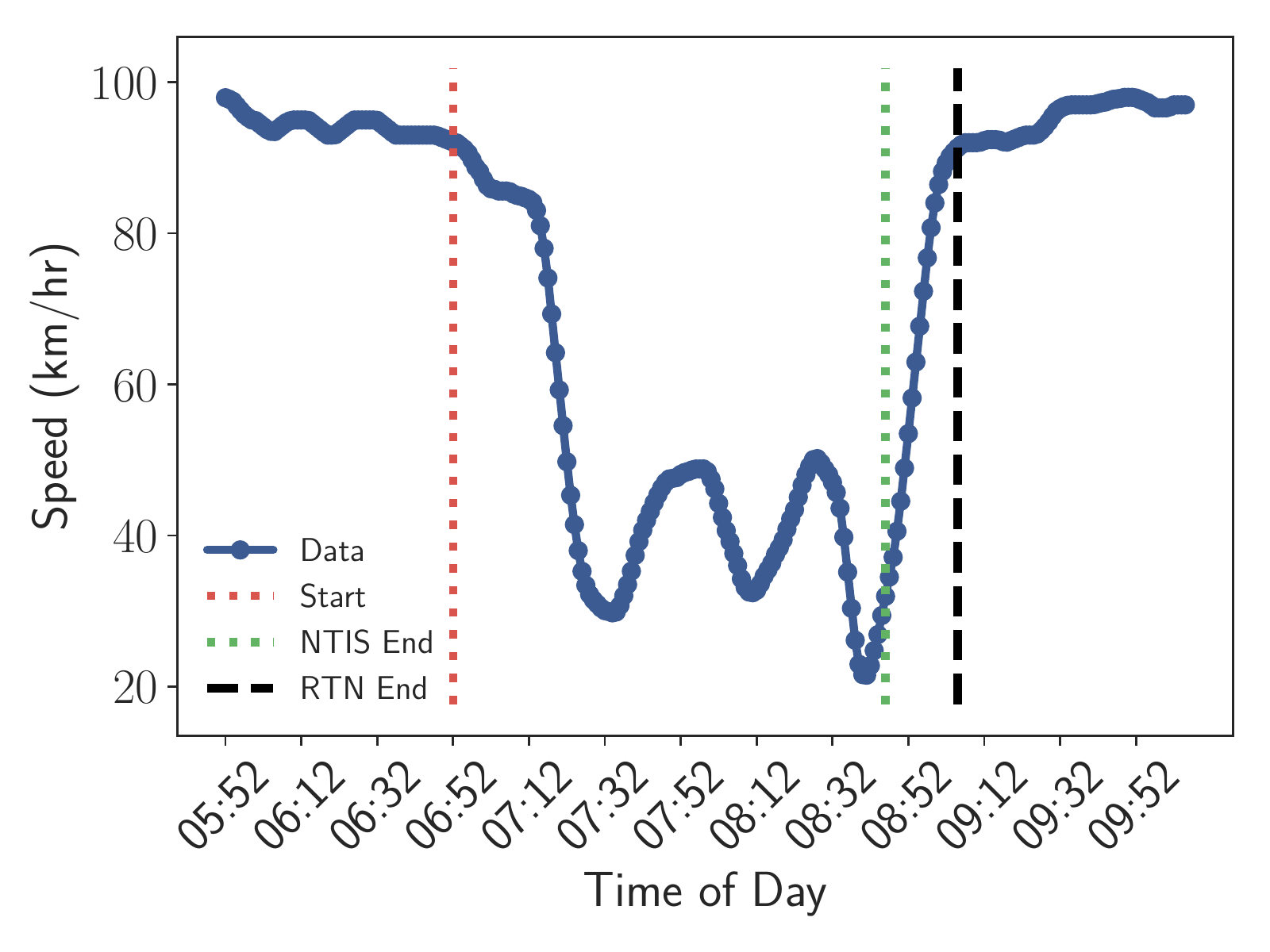}
		\centering \textbf{B}
	\end{minipage}
	\caption[Example Weekly Baseline]{The baseline for a single link and an example incident. The baseline captures clear seasonality in the data. On a much shorter time-scale, we see a drop in speed in the wake of an NTIS incident flag, and a recovery after a sustained period of low speed. \textbf{(A)}: An example weekly baseline, and residuals (data - baseline) for a single link, showing 4 weeks of data. \textbf{(B)} An example comparison between NTIS end times and the Return to Normal (RTN) end time. We are trying to predict the time until we return to normal operating conditions, rather than the time at which the NTIS flag is turned off. }\label{fig:ExampleIncidentWithFlags}
\end{figure}
Using this methodology, we process our dataset such that we have a set of records with the start time of each incident and the time at which the link returned to normal.
We include a safety margin in this baseline to account for any persistent but minor problems, shifting it down by 8km/hr ($\approx 5$mph).
A link is considered to have returned to normal when its speed is above this shifted baseline for at-least 3 consecutive minutes, acting as a persistence check.

\section{Methodology}\label{sec:Methodology}

Our modelling approach compares multiple methodologies to predict incident durations.
A concise summary with all models considered and the main points of note about them is available in the supplementary material. 

\subsection{Incident Features}\label{sec:EventFeatures}

As discussed, there is much work in the literature identifying the most influential features for incidents.
However, we are restricted in two ways. 
The first is that at any time-step of a prediction, we only incorporate features that would be known at that step. 
The second is that our dataset does not provide as comprehensive an overview of some features that are likely to be highly informative of the duration, for example we do not have in-depth injury reports or arrival time of police forces.
The features we do use are separated into time invariant and time varying categories.
The time varying features are derived from time-series of speed, flow and travel-time provided at a 1-minute resolution, which we take for the link an incident occurs on.
We first remove the seasonality from each of these series by determining `typical weeks' just as in the case of the speed baseline, then subtracting this from the time-series to generate a set of residual series.
We then hand engineer features that may be of use for some simple dynamic models, computing the gradient of the residuals at some time $t$ using the previous 5-minutes of data from $t$, as-well as simply recording the value of the series.
These are used as initial features from the time-series as they provide a sensible and intuitive summary of an incident from the sensors, however of-course more complex features might be derived from the series.
We consider models that do just this by applying temporal convolutions across the residual series, and compare the modelling results in section \ref{sec:Results}. 

The time invariant features on the other hand are detailed in \textbf{Table 1}, and are a combination of what an operator might know using existing camera and phone coverage of the SRN.
For completeness we also detail the time varying features there also.
\begin{table}[ht!]
\centering
\caption[Incident Features]{Overview of the features considered for incident duration modelling. These would be recorded by an NTIS operator when an incident is declared in the system and observed on the network. The time-series features are recorded by inductive loop sensors along the road, and reported each minute. We further consider machine learning models that engineer time-series features automatically. }\label{table:EventFeatures}
\resizebox{\textwidth}{!}{%
\renewcommand{\arraystretch}{2}
\begin{tabular}[t]{|c|c|c|}
\hline
Feature                   & Variable Type &  Description \\
\hline
\makecell{Binned daily \\ time}         & Categorical   & \makecell[l]{An indicator of the time of day. Bins: Morning Rush (6a.m.-9a.m.), \\ Afternoon (9a.m.-3p.m.), Evening Rush (3p.m.-6p.m.) and Night (6p.m.-6a.m.). } \\
\hline
\makecell{Capacity \\ reduction}        & Categorical   & \makecell[l]{The fraction of lanes that are blocked due to the incident, binned into 0-25\%,\\  25-50\%, 50-75\% and 75-100\%.} \\
\hline
Incident type             & Categorical   & \makecell[l]{Specified incident type, coded as Accident, Vehicle Obstruction, Non-Vehicle \\ Obstruction, and Abnormal Traffic.} \\
\hline
Link length               & Continuous    & \makecell[l]{Length of the link the incident occurred on in metres} \\
\hline 
\makecell{Link downstream \\ atypical?} & Binary        & \makecell[l]{Is the link downstream perturbed to some atypical state at the time of the \\ incident flag?} \\
\hline
\makecell{Link upstream \\ atypical?}   & Binary        & \makecell[l]{Is the link upstream perturbed to some atypical state at the time of the incident \\ flag?} \\
\hline
\makecell{Number of \\ vehicles}        & Categorical   & \makecell[l]{How many vehicles are involved in the incident?} \\ 
\hline
Has Cascade?              & Binary        & \makecell[l]{Did the incident occur immediately after another incident nearby in space \\ and time?} \\ 
\hline
Has roadworks?            & Binary        & \makecell[l]{Did the incident occur on a link with roadworks active?} \\ 
\hline
Spatial location          & Categorical   & \makecell[l]{Network is split into 8 sections (North, North East, East, \dots, North West) and \\ the incident location is specified with this encoding} \\
\hline
Season                    & Categorical   & \makecell[l]{What season did the incident occur during?} \\
\hline
\makecell{Vehicle types \\ involved}    & Binary        & \makecell[l]{Binary variables indicating if an incident involved a car, motorcycle, lorry, \\ trailer and articulated vehicle.} \\
\hline
Weekend indicator         & Binary        & \makecell[l]{1 if incident occurs on a weekend, 0 otherwise.} \\
\hline
Time-Series Residual      & Continuous    & \makecell[l]{Speed, flow and travel time residuals from their weekly baselines (used only \\ in landmarking models)} \\
\hline
\makecell{Gradient of Time \\ Series Residual}      & Continuous    & \makecell[l]{Gradient of speed, flow and travel time residuals from their weekly baselines \\ (used only in landmarking models)} \\
\hline
\end{tabular}%
}
\end{table}
Whilst not exhaustive, the features in \textbf{Table 1} offer a combination of contextual information in time, space and specific to the incident.
Our choice of bins for time of day reflects typical commuting patterns in the UK.
Some authors use day and night time as separation, as in \cite{application_of_finite_mixture_models_for_analysing_freeway_incident_clearance_time}, where as others account for peak times in their binning \cite{an_exploratory_hazard_based_analysis_of_highway_incident_duration} and \cite{a_simple_time_sequential_procedure_for_predicting_freeway_incident_duration}, as we have.

\subsection{Survival Analysis Methods}\label{sec:SurvivalAnalysisMethods}

We first offer more detail on models in the vein of classic survival analysis that we will consider for our problem. We remind the reader that, following the convention in the survival analysis literature, we use the word ``event" to mean the occurrence of the outcome of interest. In our case, this is the end of a traffic incident as determined by the return of the speed to within some threshold difference from the the profile value.
Survival analysis methods aim to model some property of the duration distribution.
Let $f(t)$ be the PDF of incident durations, and $F(t)$ be the cumulative distribution function (CDF).
A key component in survival analysis is the survival function $S(t)$, which in our context describes the probability an incident has not ended by time $t$, where time is measured from the start of the incident.
Denoting the event time (the incident end time) by $T$, we formally write:
\begin{equation}
\begin{split}
S(t) &= \mathbb{P}\left( T \geq t \right) \\
     &= 1 - F(t) \\
     &= \int_{t}^{\infty} f(x) dx.
\end{split}
\end{equation}
Further, many survival analysis methods are concerned with the hazard function $\lambda(t)$, describing the instantaneous rate of occurrence of events.
One can show that:
\begin{equation}
\begin{split}
\lambda(t) &= \lim_{dt \to 0} \left( \frac{\mathbb{P}\left( t \leq T < t + dt \, \, | \, \, T \geq t \right)}{dt} \right) \\
           &= \frac{f(t)}{S(t)}.
\end{split}
\end{equation}
In practice, this means that the instantaneous rate of events is equal to the density of events at that time divided by the probability of surviving to that time.
A final concept of note is the cumulative hazard function $\Lambda(t)$, which is the integral of the hazard function between time 0 and $t$:
\begin{equation}
\Lambda(t) = \int_{0}^t \lambda(s) ds.
\end{equation}

Using these concepts, the first model we apply is a Cox regression model, reviewed in \cite{proportional_hazards_model_a_review}.
Suppose some `individual' $i$ (incident in this application) has covariate vector $\bm{x_i}$.
A Cox model specifies the hazard function for individual $i$ as:
\begin{equation}\label{equ:CoxHazard}
\lambda_i(t \, \, | \, \, \bm{x}_i) = \lambda_0(t)e^{\bm{x}_i'\bm{\beta}}
\end{equation}
where $\lambda_0(t)$ is some baseline hazard at time $t$, and $\bm{\beta}$ is vector of regression coefficients.
The baseline hazard describes the hazard function for an individual with covariates all equal to $0$, and then it is adjusted for a particular individual with the exponential of the regression term. 
In this original formulation, the covariate effect is constant in time, but the baseline hazard varies in time.
Various methods exist for estimating a baseline hazard function, with more details found in \cite{handbook_of_survival_analysis}.
In short, the baseline hazard is assumed to be piecewise constant and determined without any distributional assumptions, allowing the data to construct an approximation.
One can determine $\bm{\beta}$ by optimizing the partial likelihood: 
\begin{equation}\label{equ:PartialLikelihoodBeta}
PL(\bm{\beta}) = \prod_{i=1}^N \left[ \frac{e^{\bm{x}_i'\bm{\beta}}}{\sum_{j=1}^N e^{\bm{x}_j'\bm{\beta}}Y_j(\tau_i)} \right]^{\delta_i}
\end{equation}
where $\delta_i$ is 1 if the event time is observed and 0 if it is censored and $Y_l(\tau_i)$ is 1 if individual $l$ is still at risk at time $\tau_i$.
Here $\tau_i$ represents the recorded incident duration for incident $i$.
The baseline hazard function can be determined using the Breslow estimator:
\begin{equation}
\lambda_0(\tau_i) = \frac{d_i}{\sum_{j \in \mathcal{R}(\tau_i)}e^{\bm{x}_j'\bm{\beta}}}
\end{equation}
where $\mathcal{R}(\tau_i)$ is the set of at-risk individuals at time $\tau_i$ and $d_i$ is the number of events that have occurred at the $i-th$ time.
We use the implementation of Cox models provided in the R package 'survival' \cite{survival_package_R}. 
Ties in event times are handled using the `Efron' method, detailed in \cite{the_efficiency_of_coxs_likelihood_function_for_censored_data}, altering the likelihood in Eq.~(\ref{equ:PartialLikelihoodBeta}).

The next model we apply is an accelerated failure time model (recall AFT from the literature review).
Such a model supposes relationships between the survival and hazard functions of the form:
\begin{equation}
s_i(t) = s_0\left(te^{\bm{x}_i\bm{\beta}}\right), \, \, \, \lambda_i(t) = \lambda_0\left(te^{\bm{x}_i\bm{\beta}}\right)e^{\bm{x}_i\bm{\beta}}.
\end{equation}
Here, $s_0(t)$ and $\lambda_0(t)$ represent assumed baseline survival and hazard forms, and covariates `accelerate' or `decelerate' the survival time of particular individuals.
Given an assumed form, for example Weibull, Log-normal or so on with parameters $\bm{\theta}$, one can then fit this model through maximum likelihood, optimizing:
\begin{equation}\label{equ:LikelihoodAFT}
L(\bm{\theta}) =  \prod_{i=1}^N \left[ f(\tau_i) \right]^{\delta_i}s(\tau_i)^{1-\delta_i}.
\end{equation}
A common way to interpret the AFT model is as a regression on the log of the durations:
\begin{equation}\label{equ:AFTEquation1}
log(\tau_i) = \tau_0 + \bm{x}_i'\bm{\beta} + \epsilon_i
\end{equation}
where $\epsilon_i$ is noise, with some assumed form.
We implement the models using the R package `flexsurv' detailed in \cite{flexsurv_package}.

As Cox and AFT models involve, in some way, a linear regression on covariates of interest, they are unable to account for potential non-linear, complex interactions and effects of variables without manual investigation and specification.
One way to account for this is to instead use random survival forests (recall RSF from the literature review), which are non-linear models based on an ensemble of individual tree models.
The basic idea is as follows. 
Firstly, one takes a training set and generates $B$ bootstrap samples from it, that is samples with replacement.
Each of these samples is used to grow a decision tree, however randomness is introduced in the growing of the tree, by selecting a set of potential split variables at each point in which the tree needs to be split.
The optimal split variable from this set is chosen to optimise some survival criterion. One of the most commonly used is the log-rank splitting rule \citep{random_survival_forests}.
The tree is then grown until some criteria is met, either a maximal size or minimum number of cases remaining, and the output at the end of any branch is the cumulative hazard function for all data-points that are placed into that branch when passed through the tree.
This process is repeated several times and the collection of trees is referred to as a forest.
Each decision tree is a non-linear mapping from input covariates to the output cumulative hazard function, and the collection of many trees acts as an ensemble learner.
Ensemble models are known to show promising performance in a range of tasks, and this in addition to the non-linear decision tree models suggests such models may improve upon Cox and AFT models for certain datasets.
For our work, we use the R implementation found in \citet{random_survival_forests_for_R}.

\subsection{Deep Learning Methods}\label{sec:DeepLearningMethods}

Alternative approaches in survival analysis have focused on applying methods from deep learning to incorporate non-linear covariate effects and behaviours.
One of the first such methods was in \cite{a_neural_network_model_for_survival_data}, where a Cox model was considered, but the term $\bm{x}_i'\bm{\beta}$ was replaced with $g\left( \bm{x}_i \right)$, which was the output from a neural network given input $\bm{x}_i$.
Similarly, \cite{deep_surv_personalized_treatment_recommender_system_using_a_cox} and \cite{deep_learning_for_patient_specific_kidney_graft_survival_analysis} extended the cox model to a neural network setting, however fundamentally such models are still somewhat restrictive in that they assume a form of the hazard function.
More recently, \cite{deephit_a_deep_learning_approach_to_survival_analysis_with_competing_risks} suggested to make far fewer assumptions, and instead train a network to directly model the function $F(t \, \, | \, \, \bm{x}) = 1 - \mathbb{P}\left( T > t \right)$, referred to as the failure function.
To avoid specifying any particular form of this function, the output space was treated as discrete, defined on times $\{ t_1, t_2, \dots t_{max} \}$.
We suppose a single output value in this discrete space at time $t_j$ gives $\mathbb{P}\left( t_j \, \, | \, \, \bm{x}_i \right)$ and hence we derive $F(t_j \, \, | \, \, \bm{x})$ as:
\begin{equation}
F(t_j \, \, | \, \, \bm{x}_i) = \sum_{t = t_1}^{t_j} \mathbb{P}\left( t \, \, | \, \, \bm{x}_i \right).
\end{equation}
However, we still need to enforce that the output vector actually defines a discrete probability distribution.
A natural way to enforce this is apply a softmax function on the output layer, normalizing the sum of the values to 1 though:
\begin{equation}
\sigma\left( \bm{z} \right)_j = \frac{e^{z_j}}{\sum_{k=1}^{t_{max}}e^{z_k}}.
\end{equation}
In particular, \cite{deephit_a_deep_learning_approach_to_survival_analysis_with_competing_risks} considered an application with competing risks, where individuals experienced one of many possible events.
Here we consider a simpler case, having only one event (traffic state returning to normal), however the methodology remains consistent in this application.
As we are only able to measure if an event has occurred each minute from our sensor data, the discrete nature of the model is not restrictive in our context, yet the non-parametric output is appealing as we have seen in an exploratory analysis (described in the supplementary material) that the data does not appear to be generated from any particular, simple closed form distribution.
We adapt our implementation from the implementation found in \citet{deephit_github}.

For our implementation, we specify a $t_{max}$ value equal to the longest duration, plus a 20\% margin as in the original implementation, and define the output grid at a 1 minute resolution.
However doing so leads to a very large output space for the model, and could potentially lead to over-fitting.
To combat this, we apply dropout after every full connected layer in the network, elastic net regularization on the weights and early stopping based on holdout data.
We further consider if a parametric distribution may be sufficiently flexible when attached to a neural network model to perform well in the prediction task.
To test this, we build another model as described above, but remove the softmax output layer and replace with with a mixture distribution layer, influenced by \cite{application_of_finite_mixture_models_for_analysing_freeway_incident_clearance_time}.
We choose our mixture components to be log-normal, avoiding the other specified distributions for numerical stability.
This alters the output size to be $3 \times N_m$ where $N_m$ is the number of mixtures, a hyper-parameter to tune.

A final alternative that compromises between the full non-parametric discrete output and the parametric mixture is to allow the output layer of the network to define a set of weights, and construct a probability distribution from these weights using kernel smoothing.
Kernel smoothing is a non-parametric technique that aims to construct a distribution by summing a set of kernel functions, evaluated at given data-points.
Formally, we can write a kernel smoothed result for some desired point $z$, with kernel centres $Z_i$ and weights $\omega_i$ as: 
\begin{equation}\label{equ:KernelSmoothWithWeights}
\hat{\nu}(z) = \frac{1}{h_{bw}} \sum_{i=1}^N \omega_iK\left(\frac{z-Z_i}{h_{bw}}\right).
\end{equation}
Here $h_{bw}$ is the smoothing bandwidth and the kernel $K(x)$ is often taken to be Gaussian, and the resulting estimate essentially builds a distribution as a weighted smoothed sum over all kernel centres.
A point with high weighting will result in a significant amount of mass near this location, and a wide bandwidth will smooth this mass out to the surrounding area.
Applying this to our problem, it allows us to avoid treating the output space as discrete, and instead we place a kernel centre at each point in the formerly discrete grid, and treat the neural network output (including having a softmax applied) as definitions of the weights $\omega_i$.
Doing this also enforces some amount of smoothness in the output distribution, determined by the choice of $h_{bw}$. 
We choose to use a bandwidth of 3 minutes, which still allows significant freedom to the distribution.

The actual function proposed in \cite{deephit_a_deep_learning_approach_to_survival_analysis_with_competing_risks} to optimize in order to train the network is a combination of two loss values, the first accounting for the likelihood of the observed data and the second enforcing ordering.
The likelihood loss function is given as: 
\begin{equation}
\mathcal{L}_1 = - \sum_{i=1}^N \left[ \delta_i \log\left( \hat{f}(\tau_i \, \, | \, \, \bm{x}_i) \right) + (1 - \delta_i) \log( 1 - \hat{F}(\tau_i \, \, | \, \, \bm{x}_i) ) \right]
\end{equation} 
where $\hat{f}$ is the PDF (or PMF in the discrete case) implied by the model output, and $\hat{F}$ is the CDF or (cumulative mass function in the discrete case) implied by the model output, given a particular input $\bm{x}_i$.
This is exactly as in Eq.~(\ref{equ:LikelihoodAFT}), but taking logs, describing the likelihood of survival data.
The second loss function is written:
\begin{equation}
\begin{split}
\mathcal{L}_2 &= \sum_{i \neq j} \mathbbm{1}\left( \tau_i < \tau_j \right) \eta\left( \hat{F}(\tau_i \, \, | \, \, \bm{x}_i), \hat{F}(\tau_i \, \, | \, \, \bm{x}_j) \right) \\
\eta(x,y)     &= e^{-\frac{x-y}{\eta_\sigma}}. \\
\end{split}
\end{equation} 
This loss penalizes the incorrect ordering of pairs in terms of the cumulative probability at their event time. 
If an incident $i$ ends before $j$, then we would expect $\hat{F}(\tau_i \, \, | \, \, \bm{x}_i)$ to be larger than $\hat{F}(\tau_i \, \, | \, \, \bm{x}_j)$, and if so this pair is considered correctly ordered.
Large deviations from correct ordering are penalized by $\eta(x,y)$.
The total loss function is then the sum of $\mathcal{L}_1$ and $\mathcal{L}_2$.
The hyper-parameter grids used for all machine learning models can be found in the supplementary material, and all models are trained using 100 instances of random search.

\subsection{Dynamic Methods}\label{sec:DynamicModels}

To this point, all models discussed in this section have been static, that is an individual has a covariate vector $\bm{x_i}$, it is passed through some model, and an estimate of its hazard function, failure function or alternative is attained.
However, in-practice our specific application contains a significant amount information that may be useful in determining the duration that is not available at the start of the incident.
Such information in the traffic domain is a police report made on the scene, recovery information, and specifically of use to us, the time-series provided by the sensors along the road.
A significant incident on a road network could lead to speed drops, flow breakdown and travel time spikes, all of which will be evident when we inspect the time-series as the incident progresses. 
However, the recovery of the link to normal operating conditions is closely tied to these time-series, firstly through the level of the speed series (as this defines how far from a baseline we are), but one could imagine much richer indicators of traffic state can be mined from them.
Recall \textbf{Figure 2B}, we see significant structure in the series, having a linear drop near the start of the incident, an unstable oscillation period at lower speeds, then a recovery to normal conditions. 
Further examples of these series for incident periods can be found in the supplementary material. 
A number of methods have been suggested to handle dynamic predictions in a survival analysis setting. 

\subsubsection{Landmarking}\label{sec:Landmarking}

With any dynamic prediction approach, the goal is to provide estimates of a hazard function, survival function or similar at some time $t$, conditioned on the fact that the individual has survived to time $t$ and any covariates they provide.
A simple method to do this is known as `landmarking' and is discussed in \cite{landmark_analysis_at_the_25_year_landmark_point}.
We note from the outset that landmarking is similar to truncated regression discussed in section \ref{sec:LitReview}, however this terminology is consistent with the wider survival analysis literature.
To carry out landmarking, one first specifies a set of `landmark times' $\{ t_{LM_1}, t_{LM_2}, \dots, t_{LM_K} \}$ at which we want to make dynamic predictions.
One then chooses some survival model, for example a Cox model, and the hazard function at landmark time $t_{LM_j}$ becomes:
\begin{equation}
\lambda_i(t \, \, | \, \, \bm{x}_i(t_{LM_j}), t_{LM_j}) = \lambda_0(t \, \, | \, \, t_{LM_j})e^{ \bm{x}_i(t_{LM_j})'\beta(t_{LM_j})}
\end{equation}
with $t_{LM_j} \leq t < t_{LM_j} + \Delta t$, for some $\Delta t$ defining how far ahead we are interested in looking.
Notice how, compared to Eq.~(\ref{equ:CoxHazard}), the covariate values $\bm{x}_i$ are replaced with those known at time $t_{LM_j}$ and the regression coefficients and baseline hazard can vary based on landmark time.
At each landmark time, only incidents that are still ongoing are retained, so the model is therefore conditioned on surviving up to this landmark time.
To account for potential time-varying effects and avoid misspecification of the regression parameters, events that occur after $t_{LM_j} + \Delta t$ are administratively censored, that is they are marked as censored if they survived past the look-ahead time of interest.
Such a model is simple to implement as one can refine the dataset at different times to produce dynamic models.  However, as the landmark time grows and less data becomes available, some power may be lost when drawing statistical conclusions.
To implement these models, we choose landmark times $t_{LM}$ of $\{ 0, 15, 30, 45, 60, 120 \}$ minutes and horizons $\Delta t$ of $\{ 5, 15, 30, 45, 60, 120, 180, 240 \}$ minutes and display results throughout section \ref{sec:Results}.
Finally, the landmarking framework can be applied with models other than a Cox model, so we consider both Cox and RSF landmarking models as two candidate dynamic prediction models, with RSF offering a non-linear alternative.  
The same was done to compare to dynamic models in \cite{dynamic_deephit_a_deep_learning_approach_for_dynamic_survival_analysis_with_competing_risks_based_on_longitudinal_data}.

\subsubsection{MATCH-Net Based Sliding Window Model}\label{sec:SlidingWindows}

The models considered so far require us to manually engineer features from the time-series variables to incorporate them as covariates.
As discussed, we use the level and gradient of the residual series, as these will indicate both how close the link is to reaching standard behaviour, and if the situation is getting better or worse.
However, gradients computed in short windows can be noisy, yet computed on large windows may lead to significant delay in identifying features.
Instead, we would like some automated method that, given a time-series, is able to learning meaningful features from them and incorporate them into predictions.
Such a method is proposed in \cite{dynamic_prediction_in_clinical_survival_analysis_using_temporal_convolutional_networks}, where the authors detail a sliding window model which they name MATCH-Net.
We note that the algorithm is designed to make dynamic predictions accounting for missing data, although in our application we do not have missing data so are interested in the dynamic prediction aspect only.
In this model, a window  of longitudinal measurements is fed through a convolutional neural network (CNN), with the convolutions learning features from the time-series that are then used for prediction of risk in some look-ahead window.
The model slides across the data, shifting the window each time, meaning features are updated as time progresses and predictions are also shifted forward.
It is upon this we base our sliding window methodology.

Specifically, we take a historical window of length $w$, and at time $t$ feed the time-series from time $t-w$ to $t$ through a CNN, where the filters in the CNN aim to derive features from the series without any manual specification of what they should be.
We then concatenate the features output by the CNN with the time-invariant features, and then pass these through a series of fully connected layers.
In \cite{dynamic_prediction_in_clinical_survival_analysis_using_temporal_convolutional_networks}, the output layer was a discrete space upon which a softmax activation was applied, and we again consider this, a mixture of log-normal distributions and a kernel smoothed output distribution.  
At each input time, we consider a window ahead of the the same length as in the fixed case, and treat $w$ as a hyper-parameter to optimize.
Since this model is more complex and has far more parameters than in the former case, we consider the discrete distribution to be piecewise constant for 5 minute intervals.
As a result, the output space decreases in size by 80\% without sacrificing too much freedom.
A schematic of the network architecture is given in \textbf{Figure 3}, with the output layer left intentionally vague to be clear that we consider multiple different forms of output.
\begin{figure}[ht!]
	\centering
	\includegraphics[width=\textwidth]{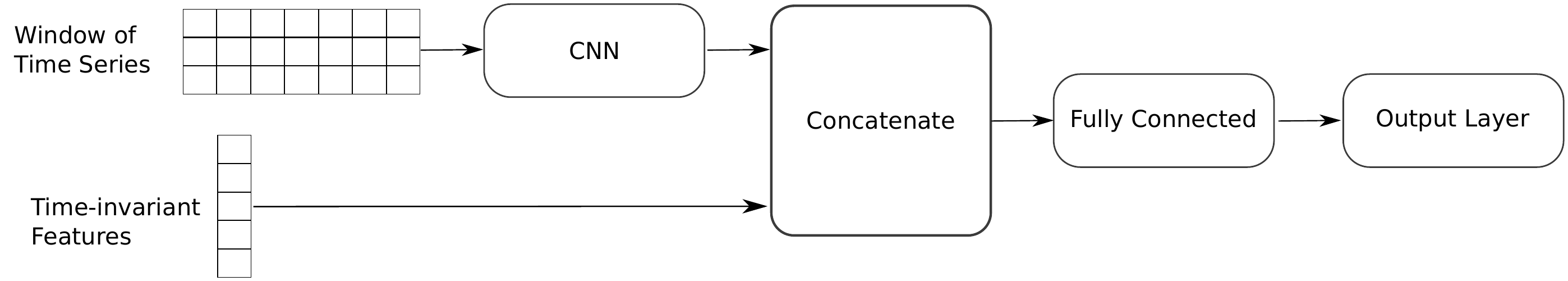}
	\caption{Network schematic for sliding window model. We pass filters across the residual time-series to engineer features from each time-series, then concatenate these with the time invariant features to create a feature vector, which is passed through a series of fully connected layers, and some output layer is applied to the result. The example shown is for a single traffic incident being passed through the network. The number of boxes for features is not to scale. The window of time-series represents 3 variables and a window size of 7 in this simple example. }\label{fig:NetworkDiagram}
\end{figure}

\section{Results}\label{sec:Results}

A point infrequently discussed in the context of traffic incidents,  is that there are multiple criteria that define a `good' hazard model, and multiple ways to measure this in the dynamic setting.
We discuss some of these ways in the text below.
We note also that elastic net regularization is applied to all deep learning methods, and the optimal Cox and AFT models are selected though inspection of sample-size adjusted Akaike information criterion (AIC) to avoid over-fitting.

\subsection{Discriminative Performance - Concordance Index}

The concordance index (C-index) has different definitions in the static and dynamic setting.
In the static setting, we write it as:
\begin{equation}\label{equ:CIndexFixed}
C = \mathbb{P}\left( \hat{F}(\tau_i \, \, | \, \, \bm{x}_i ) > \hat{F}(\tau_i \, \, | \, \, \bm{x}_j ) \, \, | \, \, \tau_i < \tau_j \right).
\end{equation}  
Eq.~(\ref{equ:CIndexFixed}) is the so called `time dependent' definition used in \cite{deephit_a_deep_learning_approach_to_survival_analysis_with_competing_risks}, accounting for the fact that we care about the entire function $\hat{F}$ and not a single point value.
In the dynamic setting, it is written given prediction time $t$ and evaluation time $\Delta t$ as:
\begin{equation}\label{equ:CIndexVaried}
C(t, \Delta t) = \mathbb{P}\left( \hat{F}(t+\Delta t \, \, | \, \, \bm{x}_i(t)) > \hat{F}(t+\Delta t \, \, | \, \, \bm{x}_j(t)) \, \, | \, \, \tau_i < \tau_j, \tau_i < t + \Delta t \right)
\end{equation}  
The only difference here is now we are specifically computing the C-index at a given prediction time and horizon rather than over the entire dataset, and this is the definition given in \cite{dynamic_deephit_a_deep_learning_approach_for_dynamic_survival_analysis_with_competing_risks_based_on_longitudinal_data}.
In computing this, we are taking the $\hat{F}$ values at some time $t + \Delta t$ and compare pairs where an incident $i$ actually ended in the horizon.

As described in \cite{dynamic_deephit_a_deep_learning_approach_for_dynamic_survival_analysis_with_competing_risks_based_on_longitudinal_data}, such a measure compares the ordering of pairs.
If individual $i$ experienced an event before individual $j$, then we would expect a good model to correctly assign more chance of an event to individual $i$ than $j$.
A model with perfect C-index, given $N$ traffic incidents, will perfectly predict the order in which the incidents will end. 
This idea stems from viewing survival analysis as a ranking problem, and since we compare the CDF for two events, we see that it incorporates the entire history from a prediction time up to an evaluation time, not just a single point measurement.
A random model will achieve a C-index on average of 0.5, and a perfect model will attain a value of 1.0, so these are reference values to consider when interpreting this measure.

We formally compute Eq.~(\ref{equ:CIndexVaried}) given our dataset by evaluating:
\begin{equation}\label{equ:CIndexDynamic}
\begin{split}
C(t, \Delta t) &\approx \frac{ \sum_{i \neq j}A_{i,j} \cdot \mathbbm{1}\left( \hat{F}(t+\Delta t \, \, | \, \, \bm{x}_i(t)) > \hat{F}(t+\Delta t \, \, | \, \, \bm{x}_j(t)) \right) }{ \sum_{i \neq j} A_{i,j} } \\
A_{i,j}          &= \mathbbm{1}\left( \tau_i < \tau_j, \tau_i < t + \Delta t \right)
\end{split}
\end{equation}  
where we simply evaluate empirically how often the ordering is correct conditioned on the requirements.
The same is true for the static case.
If two incidents happen to give exactly the same CDF values when evaluating, we take the convention of adding 0.5 to the total rather than 0 or 1, following the convention in \cite{multivariable_prognostic_models_issues_in_developing_models}.

\subsection{Calibration Performance - Brier Score}

The brier score measures how well calibrated a model is, and compares the binary label (1 if an event has happened at some time, 0 otherwise) with the model prediction at that time.
Formally, we write:
\begin{equation}
BS(t, \Delta t) = \frac{1}{N}\sum_{i=1}^N \left( \mathbbm{1}\left( \tau_i < t + \Delta t \right) - \hat{F}(t+\Delta t \, \, | \, \, \bm{x}_i(t)) \right)^2
\end{equation} 
where we sum over all events still active at time $t$, and ask did incident $i$ end before $t + \Delta t$.
If it did, then we would expect a good model to have a high CDF value at this point, with 1 being perfect (i.e. predicting the incident would end by $t + \Delta t$ for certain).
On the other hand, if the incident did not end before $t + \Delta t$, then we would expect a low CDF value.
This definition is that proposed in the supplementary material of \cite{dynamic_deephit_a_deep_learning_approach_for_dynamic_survival_analysis_with_competing_risks_based_on_longitudinal_data}.
In a sense, this measures the mean square error of a probabilistic forecast of a binary outcome.
In terms of reference values, a model that outputs a survivor function value equal to 0.5 at a particular time will have a Brier score of 0.25, so lower values than this are desirable, and a perfect model will achieve a score of 0.

\subsection{Point-Wise Performance - Mean Absolute Percentage Error}

Whilst C-index and Brier score are used throughout the survival literature, we also note that mean absolute percentage error (MAPE) is used throughout the traffic literature and has some practical relevance to our application.
Since we have no censoring, we know the true duration for each traffic incident.
As such, we can evaluate for every data-point, what is the error between a point-prediction, and the true duration.
A natural choice for such a point prediction is the median of each models output distribution, \cite{predictiion_performance_of_survival_models}, \cite{an_information_based_time_sequential_approach_to_online_incident_duration_prediction}, \cite{competing_risk_mixture_model_and_text_analysis_for_sequential_incident_duration_prediction} as the distribution of traffic incident durations is known to be heavy tailed.
We can then ask what is the point-wise error for each model.
Note that, C-index and Brier score asked about the accuracy of the output distribution, where as this asks about a single point taken from the distribution.
Highways England currently states that NTIS are measured on their prediction of the `likely delay associated with an event.' 
Specifically, NTIS is scored as follows.
One aggregates all incidents that have a predicted return to profile time made at their half way point and lasted over 1 hour.
The MAPE between these predictions at the half-way point and the true values is computed.
The target for NTIS predictions is for this to be below 35\%, and it is stated in \cite{highways_englands_provison_of_information_to_road_users} that the current value in practice is 35.49\%.

There is of course a problem with this criterion, in that a `perfect' model by this standard would just always predict double the current duration, which would optimise the prediction at the mid-point, but be of no practical use aside from this.
Regardless, this rough measure allows us to frame our work in the context of the practical considerations traffic operators are currently working towards.

\subsection{Static Prediction Models}

We begin by considering how a range of models perform in the static sense.
For this, we use only the fixed covariate information available at the start of the incident to fit the models.
We apply each of the discussed models, and show results for all metrics in \textbf{Table 2}.
\begin{table}[ht!]
  \centering
  \caption{Performance measures for models in a static setting, where we make only a single prediction per incident using a set of time-invariant covariates. Optimal values are highlighted in bold. AFT (LN) - An accelerated failure time model assuming a log-normal distribution of incident durations. AFT (W) - An accelerated failure time model assuming a Weibull distribution of incident durations. Cox - A linear Cox regression model. RSF - Random survival forest. NN (LN) - A feed-forward neural network model with an output layer that parametrises a mixture of log-normal distributions. NN (NP) - A feed-forward neural network model with a non-parametric output layer. NN (Kernel) - A feed-forward neural network with a kernel smoothed output.}\label{table:StaticResults}
  \resizebox{\textwidth}{!}{%
  \renewcommand{\arraystretch}{1.25}
  \begin{tabular}{|c|c|c|}
  \hline
  \multirow{3}{*}{Model} & \multirow{3}{*}{\makecell{C-\\Index}} & Point \\
  						 & 						      	         & -Wise \\
                         &                                       & MAPE  \\
  \hline
  AFT (LN)      & 0.624 & 40.677 \\
  \hline
  AFT (W)       & 0.624 & 38.543 \\
  \hline
  Cox           & 0.626 & 38.545 \\
  \hline
  RSF           & \textbf{0.676} & 39.961 \\
  \hline
  NN (LN)       & 0.666 & 41.401 \\
  \hline
  NN (NP)       & 0.647 & \textbf{37.416} \\
  \hline
  NN (Kernel)   & 0.659 & 39.332  \\
  \hline
  \end{tabular}
  ~
  \begin{tabular}{|c|c|c|c|c|c|c|c|c|}
  \hline
  \multicolumn{9}{|c|}{Brier Score} \\
  \hline
  \multicolumn{8}{|c|}{Prediction Horizon (minutes)} & \multirow{2}{*}{Mean} \\
  \cline{1-8}
  5 & 15 & 30 & 45 & 60 & 120 & 180 & 240 & \\
  \hline
  \textbf{0.000} & 0.052 & 0.110 & 0.146 & 0.185 & 0.231 & 0.168 & 0.102 & 0.124 \\
  \hline
  \textbf{0.000} & 0.052 & 0.113 & 0.148 & 0.186 & 0.226 & 0.164 & 0.100 & 0.124 \\
  \hline
  \textbf{0.000} & 0.052 & 0.113 & 0.149 & 0.186 & 0.226 & 0.164 & 0.100 & 0.124 \\
  \hline
  \textbf{0.000} & \textbf{0.048} & \textbf{0.103} & \textbf{0.134} & \textbf{0.167} & \textbf{0.210} & \textbf{0.149} & \textbf{0.093} & \textbf{0.113} \\
  \hline
  \textbf{0.000} & \textbf{0.049} & 0.106 & 0.141 & 0.178 & 0.221 & 0.157 & \textbf{0.094} & 0.118 \\
  \hline
  \textbf{0.000} & 0.050 & 0.108 & 0.142 & 0.179 & 0.223 & 0.163	& 0.101 & 0.121 \\
  \hline
  \textbf{0.000} & 0.049 & 0.104 & 0.137 & 0.173 & 0.218 & 0.157 & 0.097 & 0.117 \\
  \hline
  \end{tabular}%
  }
\end{table}
We see the ordering of incident durations, measured by the C-index, attains values between 0.624 and 0.676.
All models are informative, beating the 0.5 reference value, and the biggest gains in C-index are seen when we go from the linear to non-linear modelling frameworks.
The RSF achieves the optimal C-index, followed by the neural network with a mixture of log-normals.

In terms of point-wise error, we do not make predictions at the half-way point of incidents, we only make them at the start of an incident in this setting.
Doing so and measuring for all incidents longer than 60 minutes, we see that all models achieve a MAPE between 37\% and 41\%, with the best model being the non-parametric neural network.
No model achieves an MAPE of less than 35\%. 
Finally, the optimal Brier score is always achieved by the RSF method, with the most noticeable differences observed at horizons of 120 and 180 minutes.
There is not much to distinguish many of these models, and ultimately one might suggest that in a static setting, a RSF offers a good compromise between performance measured by C-index, Brier score and MAPE, however if MAPE is the single desired criterion, a non-parametric neural network model would be preferred.

\subsection{Dynamic Prediction Models}

We now consider the models in a dynamic setting.
We consider C-index as defined in Eq.~(\ref{equ:CIndexDynamic}), and show results for it at various prediction times and horizons in \textbf{Table 3}.
Initially, the RSF achieves optimal C-index across all horizons when predicting at $t=0$.
As time of prediction increases, we see a strong favouring of neural network models, specifically the sliding window model with a kernel smoothed output achieves the optimal C-index in most cases.
There is a systematic preference for the non-parametric sliding window models compared to others at all prediction horizons when considering prediction times of 30 minutes or greater.
Even at a prediction time of 15 minutes, the non-parametric sliding window models are preferred when considering 180 and 240 minute horizons.
As a general summary of \textbf{Table 3}, one should see that out of all $47$ prediction time, prediction horizon pairs considered, the optimal model in terms of C-index is the RSF roughly 34\% of the time, the sliding window neural network with kernel output 43\% of the time, and the sliding window neural network with non-parametric output the remainder of times.

Averaged over all horizons, we see that one would prefer the RSF model when initially making predictions, but all prediction times after 15 minutes favour the kernel smoothed output, with the non-parametric neural network often similar in performance.
The neural network model parametrising a mixture of log-normals achieved the highest C-index of the neural network models in the static case, closely following the RSF model, however it never wins in the dynamic case, suggesting that when we provide the time-series features, the RSF makes better use of them initially, and then the other neural network models make better use of them as time-passes.
All models achieve C-index values higher than the reference value of 0.5, across all prediction times and horizons showing their predictions remain informative.

One point of note from \textbf{Table 3} is that the Cox model has quite poor C-index compared to the alternatives considered when predicting at a horizon of 5 minutes. 
We believe this is due to the amount of administrative censoring introduced at such a short horizon.
If we look to \cite{dynamic_prediction_in_clinical_survival_analysis}, an assumption of the Cox landmarking model is that there is not too much censoring at the horizon time. 
For a very short horizon of 5 minutes, almost all incidents last longer than this, and hence, when we are applying our administrative censoring, this assumption becomes invalid, and we suspect this is why the Cox model has poor results at this horizon.

\begin{table}[ht!]
\centering
  \caption{C-Index values for considered models, across a range of different prediction times (when predications are made) and prediction horizons (at what time after the prediction time they are evaluated). Higher values show a better model. Optimal values for each prediction time - prediction horizon pair are shown in bold. $Cox$ - A linear Cox landmarking model. RSF - Random survival forest landmarking model. SW (LN) - Sliding window with log-normal mixture output. SW (NP) - Sliding window with non-parametric output. SW (Kernel) - Sliding window with kernel smoothed output.}\label{table:DynamicCIndexScores}
  \resizebox{\textwidth}{!}{\renewcommand{\arraystretch}{1.25}\begin{tabular}{|c|c|c|c|c|c|c|c|c|c|c|}
    \hline
    Prediction Time            & \multirow{2}{*}{Model} & \multicolumn{8}{c|}{Prediction Horizon (minutes)} & Mean Over \\
    \cline{3-10} 
    (minutes)				   &             & 5 & 15 & 30 & 45 & 60 & 120 & 180 & 240 & Horizons \\
    \hline
    \multirow{6}{*}{$t = 0$}   & Cox         & - & 0.851 & 0.799 & 0.754 & 0.712 & 0.654 & 0.642 & 0.638 & 0.721 \\
                               & RSF         & - & \textbf{0.870} & \textbf{0.832} & \textbf{0.803} & \textbf{0.774} & \textbf{0.698} & \textbf{0.667} & \textbf{0.651} & \textbf{0.756} \\ 
                               & SW (LN)     & - & 0.766 & 0.709 & 0.673 & 0.650 & 0.606 & 0.599 & 0.597 & 0.657 \\ 
							   & SW (NP)     & - & 0.798 & 0.743 & 0.705 & 0.682 & 0.642 & 0.637 & 0.634 & 0.692 \\ 
							   & SW (Kernel) & - & 0.823 & 0.757 & 0.717 & 0.689 & 0.641 & 0.634 & 0.630 & 0.699 \\ 
	\hline
    \multirow{6}{*}{$t = 15$}  & Cox         & 0.513 & 0.864 & 0.784 & 0.732 & 0.698 & 0.648 & 0.639 & 0.637 & 0.689 \\
                               & RSF         & \textbf{0.956} & \textbf{0.893} & \textbf{0.826} & \textbf{0.788} & \textbf{0.751} & \textbf{0.686} & 0.662 & 0.653 & \textbf{0.777} \\ 
                               & SW (LN)     & 0.927 & 0.851 & 0.773 & 0.732 & 0.694 & 0.644 & 0.633 & 0.632 & 0.736 \\ 
							   & SW (NP)     & 0.947 & 0.884 & 0.811 & 0.772 & 0.731 & 0.678 & \textbf{0.669} & \textbf{0.666} & 0.770 \\ 
							   & SW (Kernel) & 0.953 & 0.891 & 0.815 & 0.774 & 0.733 & 0.679 & 0.667 & 0.662 & 0.772 \\ 
	\hline
    \multirow{6}{*}{$t = 30$}  & Cox         & 0.529 & 0.861 & 0.770 & 0.731 & 0.702 & 0.662 & 0.652 & 0.648 & 0.664 \\
                               & RSF         & 0.921 & 0.880 & 0.795 & 0.760 & 0.738 & \textbf{0.699} & 0.676 & 0.662 & 0.766 \\ 
                               & SW (LN)     & 0.947 & 0.867 & 0.774 & 0.735 & 0.707 & 0.662 & 0.653 & 0.653 & 0.750 \\ 
							   & SW (NP)     & 0.960 & 0.905 & 0.803 & 0.761 & 0.733 & 0.689 & \textbf{0.681} & \textbf{0.677} & 0.776 \\ 
							   & SW (Kernel) & \textbf{0.971} & \textbf{0.907} & \textbf{0.810} & \textbf{0.768} & \textbf{0.739} & 0.691 & 0.679 & 0.675 & \textbf{0.780} \\ 
    \hline
    \multirow{6}{*}{$t = 45$}  & Cox         & 0.504 & 0.859 & 0.796 & 0.764 & 0.724 & 0.679 & 0.667 & 0.661 & 0.707 \\
                               & RSF         & 0.970 & 0.884 & 0.817 & 0.788 & \textbf{0.757} & \textbf{0.706} & 0.684 & 0.654 & 0.783 \\ 
                               & SW (LN)     & 0.950 & 0.852 & 0.778 & 0.750 & 0.718 & 0.673 & 0.663 & 0.662 & 0.756 \\ 
							   & SW (NP)     & 0.967 & 0.880 & 0.813 & 0.784 & 0.749 & 0.703 & \textbf{0.692} & \textbf{0.688} & 0.785 \\ 
							   & SW (Kernel) & \textbf{0.974} & \textbf{0.893} & \textbf{0.821} & \textbf{0.787} & 0.750 & 0.701 & 0.687 & 0.683 & \textbf{0.787} \\ 

	\hline
    \multirow{6}{*}{$t = 60$}  & Cox         & 0.578 & 0.872 & 0.796 & 0.755 & 0.723 & 0.684 & 0.670 & 0.668 & 0.718 \\
                               & RSF         & 0.954 & 0.887 & 0.811 & 0.782 & 0.742 & 0.700 & 0.674 & 0.654 & 0.776 \\ 
                               & SW (LN)     & 0.928 & 0.871 & 0.796 & 0.759 & 0.726 & 0.681 & 0.672 & 0.671 & 0.763 \\ 
							   & SW (NP)     & 0.953 & 0.903 & 0.830 & 0.787 & 0.755 & 0.711 & \textbf{0.702} & \textbf{0.699} & 0.793 \\ 
							   & SW (Kernel) & \textbf{0.969} & \textbf{0.912} & \textbf{0.834} & \textbf{0.793} & \textbf{0.758} & \textbf{0.712} & 0.701 & 0.698 & \textbf{0.797} \\ 
	\hline
    \multirow{6}{*}{$t = 120$} & Cox         & 0.522 & 0.850 & 0.804 & 0.781 & 0.750 & 0.706 & 0.692 & 0.683 & 0.724 \\
                               & RSF         & 0.961 & 0.889 & 0.839 & 0.807 & 0.777 & 0.731 & 0.697 & 0.688 & 0.799 \\ 
                               & SW (LN)     & 0.944 & 0.878 & 0.822 & 0.799 & 0.769 & 0.718 & 0.715 & 0.713 & 0.795 \\ 
							   & SW (NP)     & 0.968 & 0.896 & 0.852 & 0.824 & 0.791 & \textbf{0.744} & \textbf{0.739} & \textbf{0.735} & 0.819 \\ 
							   & SW (Kernel) & \textbf{0.986} & \textbf{0.904} & \textbf{0.853} & \textbf{0.825} & \textbf{0.793} & 0.743 & 0.737 & 0.732 & \textbf{0.822} \\ 
	\hline
  \end{tabular}}
\end{table}

We further show the Brier scores for each model in \textbf{Table 4}.
Again, we observe that initially, the random survival forest achieves optimal scores across all horizons, however as time of prediction increases we gradually see the sliding window neural network with kernel smoothed output start to achieve better Brier scores for short prediction horizons.
This is again systematic, and we see for a prediction time of 120 minutes that the optimal model is the sliding window neural network with kernel smoothed output at horizons up to and including 45 minutes, but for a prediction time of 45 minutes it is only optimal for horizons up to and including 15 minutes.
One could postulate that initially, time-series provide less useful information than the fixed features, that is at the very start of an incident, we see only the state of the link before the incident, which might have been reasonably seasonal.
However, as time progresses, we will attain more informative features specific to single incidents, and in this case the fact the sliding window method engineers its own features, rather than our noisy gradient and level values we manually input to the RSF model, may prove more useful.
Despite this, if a duration is very long, say 4 hours, and make a prediction 60 minutes into it, how much do we truly expect to gain from inspecting the time-series so far? It may be that there is just simply no sign of recovery and all we can really conclude is that speed has been slow for a long time, and shows no other clear features.

Another point of note when considering Brier score is that RSF appeared to perform well compared to a non-parametric neural network model in other works.
If we look to the supplementary material of \cite{dynamic_deephit_a_deep_learning_approach_for_dynamic_survival_analysis_with_competing_risks_based_on_longitudinal_data}, we see that a neural network with a non-parametric output did not consistently improve upon the Brier score achieved by a RSF model (see table VI in the supplementary material of the cited reference).
It is unclear therefore if there is some fundamental reason for this in the modelling framework, as two entirely different datasets and applications appear to have observed the same behaviour.
Despite this, all models achieve Brier scores below the reference value of 0.25.

\begin{table}[ht!]
\centering
  \caption{Brier scores for considered models, across a range of different prediction times (when predications are made) and prediction horizons (at what time after the prediction time they are evaluated). Lower values indicate a better model. Optimal values for each prediction time - prediction horizon pair are shown in bold. All keys are as defined in \textbf{Table 3}.}\label{table:DynamicBrierScores}
  \resizebox{\textwidth}{!}{\renewcommand{\arraystretch}{1.25}\begin{tabular}{|c|c|c|c|c|c|c|c|c|c|c|}
    \hline
    Prediction Time            & \multirow{2}{*}{Model} & \multicolumn{8}{c|}{Prediction Horizon (minutes)} & Mean Over \\
    \cline{3-10}
    (minutes)				   &             & 5 & 15 & 30 & 45 & 60 & 120 & 180 & 240 & Horizons \\
    \hline
    \multirow{6}{*}{$t = 0$}   & Cox         & \textbf{0.000} & 0.046 & 0.096 & 0.130 & 0.171 & 0.224 & 0.164 & 0.099 & 0.116 \\
                               & RSF         & \textbf{0.000} & \textbf{0.041} & \textbf{0.089} & \textbf{0.120} & \textbf{0.154} & \textbf{0.205} & \textbf{0.149} & \textbf{0.091} & \textbf{0.106} \\ 
                               & SW (LN)     & 0.006 & 0.096  & 0.197 & 0.260 & 0.315 & 0.317 & 0.204 & 0.115 & 0.189 \\ 
							   & SW (NP)     & 0.007 & 0.082  & 0.160 & 0.213 & 0.262 & 0.290 & 0.195 & 0.113 & 0.165 \\ 
							   & SW (Kernel) & 0.006 & 0.079  & 0.159 & 0.211 & 0.261 & 0.292 & 0.197 & 0.114 & 0.165 \\ 
	\hline
    \multirow{6}{*}{$t = 15$}  & Cox         & 0.027 & 0.062 & 0.113 & 0.152 & 0.192 & 0.218 & 0.149 & 0.088 & 0.125 \\
                               & RSF         & \textbf{0.018} & \textbf{0.059} & \textbf{0.107} & \textbf{0.141} & \textbf{0.177} & \textbf{0.204} & \textbf{0.138} & \textbf{0.083} & \textbf{0.116} \\ 
                               & SW (LN)     & 0.020 & 0.077 & 0.157 & 0.212 & 0.266 & 0.280 & 0.176 & 0.097 & 0.161 \\ 
							   & SW (NP)     & 0.019 & 0.069 & 0.134 & 0.180 & 0.230 & 0.259 & 0.170 & 0.096 & 0.145 \\ 
							   & SW (Kernel) & \textbf{0.018} & 0.070 & 0.136 & 0.181 & 0.230 & 0.259 & 0.171 & 0.097 & 0.145 \\ 
	\hline
    \multirow{6}{*}{$t = 30$}  & Cox         & 0.025 & 0.068 & 0.127 & 0.165 & 0.198 & 0.204 & 0.138 & 0.082 & 0.126 \\
                               & RSF         & 0.020 & \textbf{0.064} & \textbf{0.121} & \textbf{0.157} & \textbf{0.186} & \textbf{0.190} & \textbf{0.128} & \textbf{0.076} & \textbf{0.118} \\ 
                               & SW (LN)     & 0.019 & 0.074 & 0.154 & 0.204 & 0.250 & 0.255 & 0.161 & 0.090 & 0.151 \\ 
							   & SW (NP)     & 0.018 & 0.067 & 0.138 & 0.183 & 0.223 & 0.236 & 0.155 & 0.088 & 0.139 \\ 
							   & SW (Kernel) & \textbf{0.017} & 0.066 & 0.137 & 0.180 & 0.218 & 0.235 & 0.156 & 0.089 & 0.137 \\ 
	\hline
    \multirow{6}{*}{$t = 45$}  & Cox         & 0.028 & 0.073 & 0.131 & 0.162 & 0.196 & 0.194 & 0.132 & 0.082 & 0.125 \\
                               & RSF         & 0.018 & \textbf{0.069} & \textbf{0.125} & \textbf{0.153} & \textbf{0.185} & \textbf{0.185} & \textbf{0.125} & \textbf{0.077} & \textbf{0.117} \\ 
                               & SW (LN)     & 0.021 & 0.082 & 0.157 & 0.201 & 0.245 & 0.239 & 0.154 & 0.090 & 0.149 \\ 
							   & SW (NP)     & 0.019 & 0.074 & 0.138 & 0.176 & 0.217 & 0.221 & 0.148 & 0.089 & 0.135 \\ 
							   & SW (Kernel) & \textbf{0.016} & 0.070 & 0.136 & 0.173 & 0.213 & 0.219 & 0.149 & 0.089 & 0.133 \\ 
	\hline
    \multirow{6}{*}{$t = 60$}  & Cox         & 0.034 & 0.083 & 0.139 & 0.168 & 0.196 & 0.185 & 0.129 & 0.083 & 0.127 \\ 
                               & RSF         & 0.023 & 0.079 & \textbf{0.135} & \textbf{0.161} & \textbf{0.190} & \textbf{0.177} & \textbf{0.119} & \textbf{0.077} & \textbf{0.120} \\ 
                               & SW (LN)     & 0.025 & 0.082 & 0.154 & 0.199 & 0.240 & 0.228 & 0.145 & 0.090 & 0.145 \\ 
							   & SW (NP)     & 0.023 & 0.073 & 0.136 & 0.176 & 0.212 & 0.210 & 0.139 & 0.089 & 0.132 \\ 
							   & SW (Kernel) & \textbf{0.019} & \textbf{0.070} & \textbf{0.135} & 0.172 & 0.208 & 0.208 & 0.140 & 0.090 & 0.130 \\ 
	\hline
    \multirow{6}{*}{$t = 120$} & Cox         & 0.034 & 0.090 & 0.146 & 0.168 & 0.184 & 0.169 & 0.117 & 0.083 & 0.124 \\
                               & RSF         & 0.024 & 0.084 & 0.132 & 0.157 & \textbf{0.174} & \textbf{0.159} & \textbf{0.109} & \textbf{0.078} & \textbf{0.115} \\ 
                               & SW (LN)     & 0.026 & 0.087 & 0.148 & 0.179 & 0.213 & 0.199 & 0.137 & 0.091 & 0.135 \\ 
							   & SW (NP)     & 0.023 & 0.082 & \textbf{0.128} & 0.157 & 0.189 & 0.183 & 0.130 & 0.089 & 0.123 \\ 
							   & SW (Kernel) & \textbf{0.018} & \textbf{0.076} & \textbf{0.128} & \textbf{0.155} & 0.186 & 0.183 & 0.132 & 0.090 & 0.121 \\ 
	\hline
  \end{tabular}}
\end{table}

Finally, we show the error in a point prediction made at various times throughout incidents in \textbf{Table 5}.
For reference we also include the value achieved by the fixed model to get an idea of what we are gaining from making dynamic predictions.
From \textbf{Table 5}, we see that when making a prediction after 30\% of the duration of an incident has passed, we can expect between 30\% and 33\% MAPE in that prediction.
This is around an 5-10\% improvement from the prediction made from the corresponding static models at the start of the incident.
If we predict half way through an incident, we see that the neural network models all now achieve quite a significantly better MAPE value than the landmarking models, with an optimal MAPE of 21.6\% achieved by the mixture of log-normals model, followed by 22\% for the non-parametric model. 
The discrepancy between the sliding window and landmarking models grows as we make predictions later and later, with the sliding window models achieving an MAPE of between 16.5\% and 17.3\% compared to a value of 26.5\% for the optimal landmarking model (RSF).
Additionally, the prediction error shows very little improvement moving beyond the 50th percentiles of an incidents duration for the RSF model, and actually increases for the Cox model, suggesting that they are not sufficiently capturing signs in the time-series that indicate the end is near. 
A key point of practical interest is that with the dynamic models, we do indeed achieve a MAPE value below 35\% as desired by Highways England.
Of-course, we would need to attain data for all incidents across the UK to truly ensure that we are able to maintain this on a wider scale, but as far as we are able to measure we achieve what would be considered industrially satisfactory error rates with the dynamic models.

Whilst the landmarking models appear to plateau in point-wise performance here, we note that it is partly due to plateauing or noisy error they appear to exhibit when making predictions at large landmarking times when only very few incidents remain active. 
We visualize the error per minute into incidents in the supplementary material, which shows this.
The plots within the supplementary material are more akin to something one can find in other dynamic works, for example Figure 2 in \cite{competing_risk_mixture_model_and_text_analysis_for_sequential_incident_duration_prediction}. 

\begin{table}[ht!]
\centering
  \caption{MAPE at various points for incidents, all of which are at-least 60 minutes long. The optimal model for each prediction point is shown in bold. The point prediction is generated as the median of the output distribution from each model. All keys are as defined in \textbf{Table 3}.}\label{table:DynamicErrorAtPercentiles}
  \renewcommand{\arraystretch}{1.25}\begin{tabular}{|c|c|c|c|c|c|}
    \hline
    \multirow{3}{*}{Model} & \multirow{3}{*}{\makecell{MAPE \\ Static Model}} & \multicolumn{4}{c|}{MAPE Dynamic Model} \\
    \cline{3-6}
                & & \multicolumn{4}{c|}{Percentile Into Incident Prediction Made at}   \\
    \cline{3-6}
                & & 30th   & 50th   & 70th   & 90th \\
    \hline
    Cox         & 38.545 & 32.513 & 31.002 & 32.180 & 35.923 \\
    \hline 
    RSF         & 41.607 & \textbf{30.286} & 27.707 & 26.478 & 25.319 \\ 
    \hline
    SW (LN)     & 37.416 & 32.839 & \textbf{21.576} & \textbf{16.506} & 11.319 \\ 
    \hline
    SW (NP)     & 41.401 & 31.660 & 21.998 & 17.069 & 10.399 \\ 
    \hline
    SW (Kernel) & 39.332 & 31.056 & 22.432 & 17.319 & \textbf{10.040} \\ 
	\hline
  \end{tabular}
\end{table}

\subsection{Do Temporal Convolutions Improve Predictions?}

As we have applied methods from \cite{dynamic_prediction_in_clinical_survival_analysis_using_temporal_convolutional_networks} to formulate a sliding window model, we have naturally included temporal convolutions to generate information from the time-series.
However, a valid question one could ask is are these required in our application, or do the simple levels and gradients retain enough information to make informative predictions?
We test this by implementing a model without the CNN structure and instead feeding an input vector consisting of the time-invariant features and the level and gradients computed as in the landmarking case to a feed forward network.
Doing so results in a model that achieves a worse Brier score and C-index across all prediction time, horizon pairs and a worse error at the half way point. 
Exact results are given in the supplementary material.

\section{Feature Importance and Model Interpretability}\label{sec:VariableImportance}

Variable importance is a topic often addressed in the literature, and we offer some discussion of it here for both the RSF model and the non-parametric neural network model, chosen for simplicity compared to the kernel model.

\subsection{Random Survival Forest Variable Importance}\label{sec:RSFVariableImportance}

As RSF are adaptations of random forest methods, standard variable importance metrics are well explored and readily implemented.
Recall that trees are trained based on a bootstrap sampled dataset, meaning a set of observations remains for each tree that are `out of bag' which will be used for measuring variable importance.
Given a trained forest and some variable of interest $x$, we drop the out of bag data for each tree down the tree and whenever a split on $x$ is encountered, we assign a daughter node at random instead of evaluating based on the value of $x$. 
We then compute the estimates from the model doing this, and the variable importance for $x$ is the prediction error for the original ensemble subtracted from the prediction error for the new ensemble ignoring the $x$ value.
A large variable importance suggests that a variable is highly useful in accurately predicting the output.
We compute the importance for all features, then scale the importance values by diving by the largest.
This yields variable importance on a scale from 0 to 1 and we plot particularly important variables in \textbf{Figure 4}.
\begin{figure}[ht!]
	\centering
	\begin{minipage}[t]{.46\textwidth}
		\includegraphics[width=\textwidth]{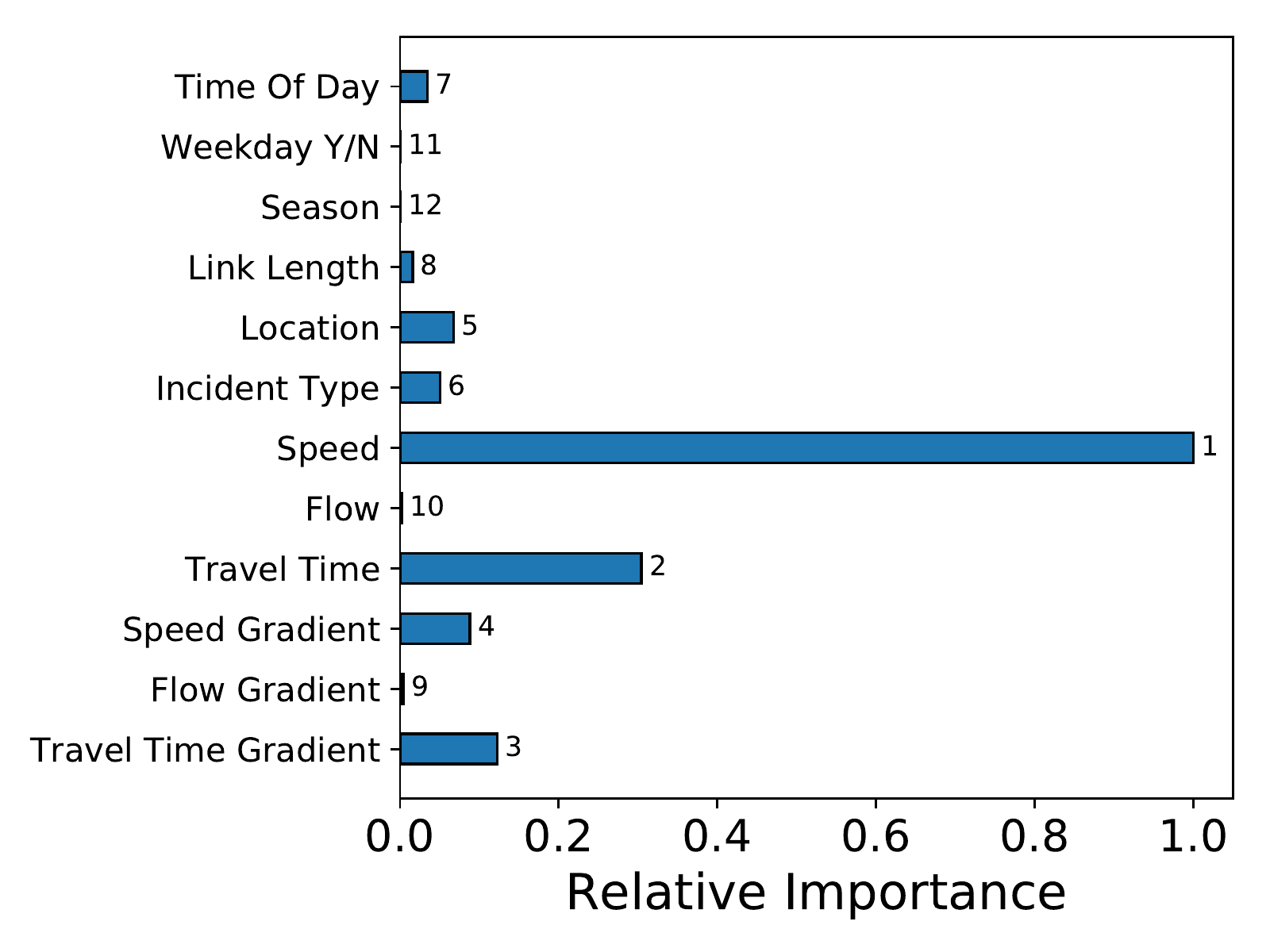}
		\centering \textbf{A}
	\end{minipage}
	~
	\begin{minipage}[t]{.46\textwidth}
		\includegraphics[width=\textwidth]{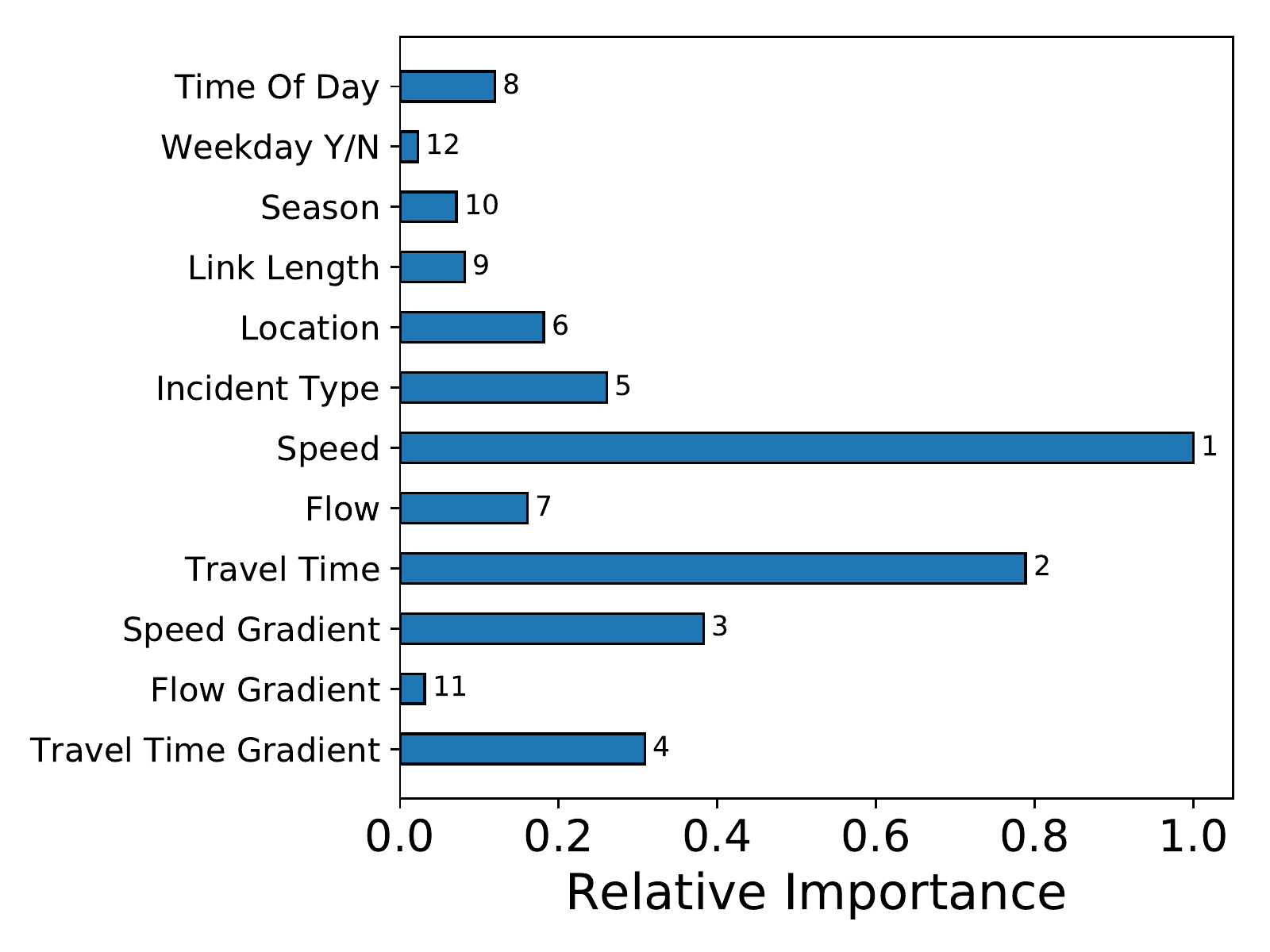}
		\centering \textbf{B}
	\end{minipage}
	
	\begin{minipage}[t]{.46\textwidth}
		\includegraphics[width=\textwidth]{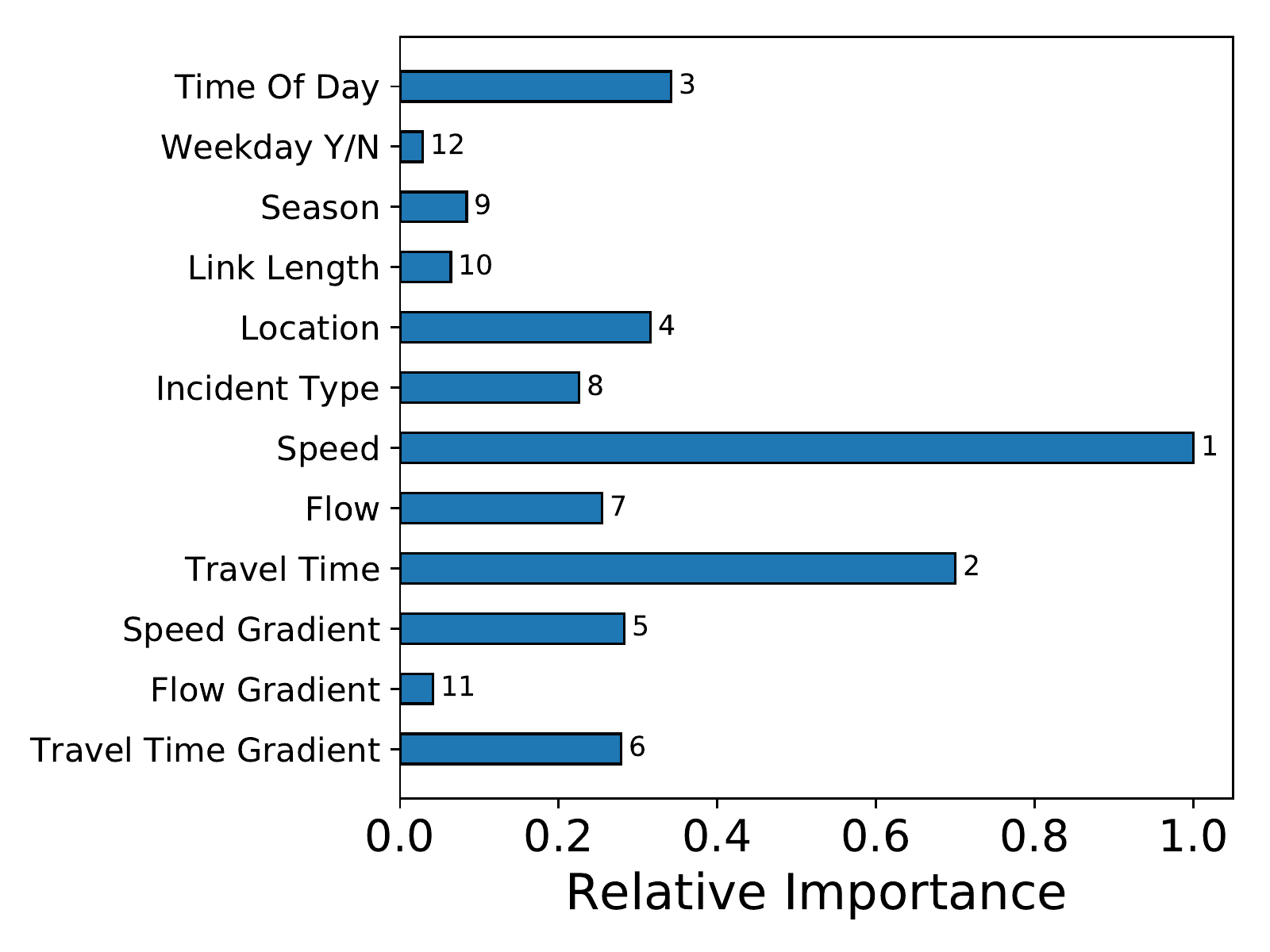}
		\centering \textbf{C}
	\end{minipage}
	~
	\begin{minipage}[t]{.46\textwidth}
		\includegraphics[width=\textwidth]{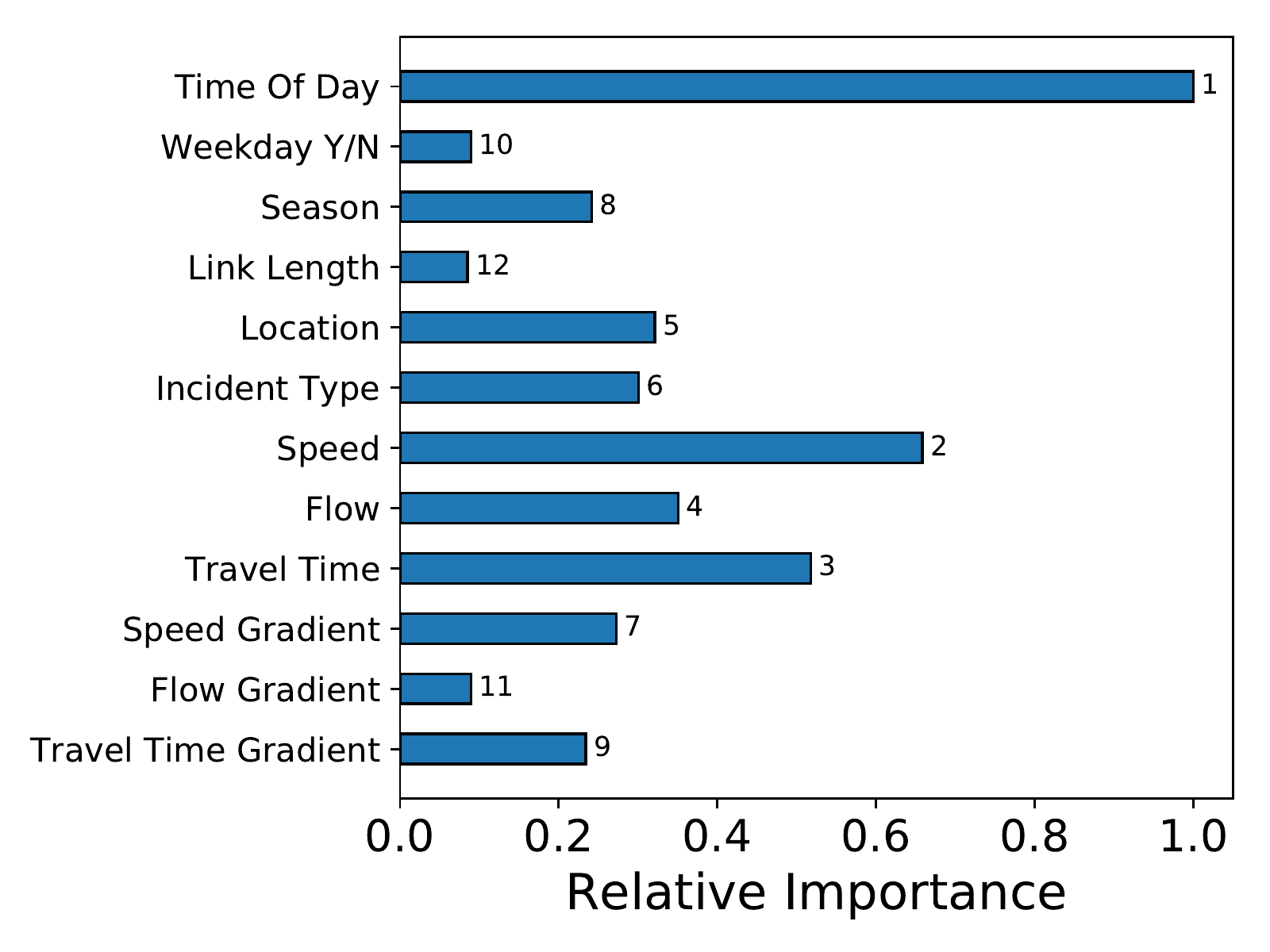}
		\centering \textbf{D}
	\end{minipage}
	\caption{Variable importance, as measured for the random survival forest model, for a subset of the important variables. Each plot is a different prediction horizon $h$, all shown are at a prediction time of $t=30$. The importance at each prediction horizon has been normalized such that the largest importance at any time is 1, and all others are relative to this. The rank of each variable at a given time is written beside each bar. Due to the scaling, one should focus on the the ranking and relative difference between bars in each plot. \textbf{(A)}: $h=5$. \textbf{(B)}: $h=30$. \textbf{(C)}: $h=60$. \textbf{(D)}: $h=180$. }\label{fig:VariableImportanceRSF}
\end{figure}

From \textbf{Figure 4}, we see that the speed value is the most important variable for horizons of 5, 30 and 60 minutes. 
At 180 minutes, the most important variable becomes the time of day.
This makes intuitive sense, as we expect the time-varying features to provide more useful information about the immediate future rather than times far into the future.
Similar reasoning applies to the importance of travel time and flow. 
It is interesting to note that flow is often less important than other, time-invariant features, across all horizons. 
The gradients are always less important than the residual values themselves, which may be a consequence of noise when estimating them, or the fact that the we can see short term rises and falls in the traffic variables that do not indicate the incident is actually near ending, but rather traffic state is just unstable as in \textbf{Figure 4}.
The location of an incident is always somewhat important, ranking between sixth and fourth across all horizons. 
This suggests clear heterogeneity in durations by location.
Note that with a location and length, a model should be able to identify single links in the network, so predictions can be specific to these if it improves performance.
However, the importance of length decays over horizons, and is always less than the location itself, so the coarse segmentation we have introduced for location seems more important than the specific link an incident occurs on.
We see that the season becomes increasingly important as horizon increases.
The type of incident varies between the sixth, fifth, eighth and sixth most important variable going from horizons of 5, 30, 60 and 180 minutes respectively. 

\subsection{Neural Network Variable Importance}\label{sec:NNVarImportanceSHAP}

Recently, there has been a significant effort to improve the interpretability of prediction models, both those involving neural networks and more general frameworks.
Examples of this include \cite{why_should_I_trust_you_explaining_the_predictions_of_any_classifier} and \cite{learning_important_features_through_propagating_activation_differences}.
In the first, the general idea is to build a simpler `explainer' model $g$ that locally approximates some complex model $f$
One optimizes $g$ by penalizing complexity, and assigning more weight to data-points near the one we wish to explain the prediction of, hence resulting in local accuracy.
It was then shown in \cite{a_unified_approach_to_interpreting_model_predictions} that many existing model interpretability methods could be phrased in-terms of a concept from game-theory known as Shapley values.
In short, they proposed explainer models of the form:
\begin{equation}\label{equ:SHAPExplainerModel}
g(z') = \phi_0 + \sum_{i=1}^M\phi_iz_i'
\end{equation}
where $z_i'$ is a binary value indicating the inclusion or exclusion of a particular feature and we have $M$ features.
They then specified three properties that one might desire in a feature attribution method: local accuracy (predictions of the same input give the same output), missingness (no attributed impact of missing features) and consistency (for two models, if ones output is more sensitive to a particular feature change than another, then it achieves a higher attribution value).
The authors showed that under these properties and Eq.~(\ref{equ:SHAPExplainerModel}), the $\phi_i$ values actually coincide with Shapley values from game theory. 
This approach unified many existing methods, including \cite{why_should_I_trust_you_explaining_the_predictions_of_any_classifier} and \cite{learning_important_features_through_propagating_activation_differences}. 
They denoted $\phi_i$ a `SHAP value' and they are appealing as they are additive, showing how particular features shift a models predictions away from some mean $\phi_0$ to the final result for a particular data-instance.
Their absolute value shows the size of a particular features importance, however one can go deeper and ask for a given feature, is this value increasing or decreasing the final prediction of the model?
One actually computes the SHAP value for feature $i$ as: 
\begin{equation}\label{equ:MainTexShapValue}
\phi_i =\sum_{S \subseteq \mathcal{M} \text{\textbackslash} {i} }  \frac{|S|!(|M| - |S| - 1)!}{M!} \left[ F(S \cup \{i\}) - F(S) \right]
\end{equation} 
where $\mathcal{M}$ is the set of all features.
In Eq.~(\ref{equ:MainTexShapValue}), we sum over all subsets of feature vectors that do not include feature $i$.
For any one of these sets $S$, we compute the difference between the model output using the features in $S$ and feature $i$, and the model output using only the features in $S$, shown by $F(S \cup \{i\}) - F(S)$. 
The remaining term $\frac{|S|!(|M| - |S| - 1)!}{M!}$ accounts for all possible orderings of the feature vector.

It is upon this that we base our feature importance exploration for the neural network model.
We compute the SHAP values of the network, for the fixed features and the features output from the CNN deriving information from the time-series.
We use the implementation provided by the original authors\footnote{Implementation and examples given in \url{https://github.com/slundberg/shap}}, specifically the permutation method for computational speed and the incorporation of structured inputs.
More details on SHAP values are given in the supplementary material.

A point of note for this method of feature importance is that we are computing values for each output neuron in the network that correspond to particular horizons, and how this output value changes, not how some performance metric changes.
As a result, we can question if different features have more or less impact on different parts of the output distribution for a single input data-point.
First, we consider raw feature importance, that is does a variable have a large or small impact the output of the model at particular horizons, showing results in \textbf{Figure 5}.
\begin{figure}[ht!]
	\centering
	\begin{minipage}[t]{.46\textwidth}
		\includegraphics[width=\textwidth]{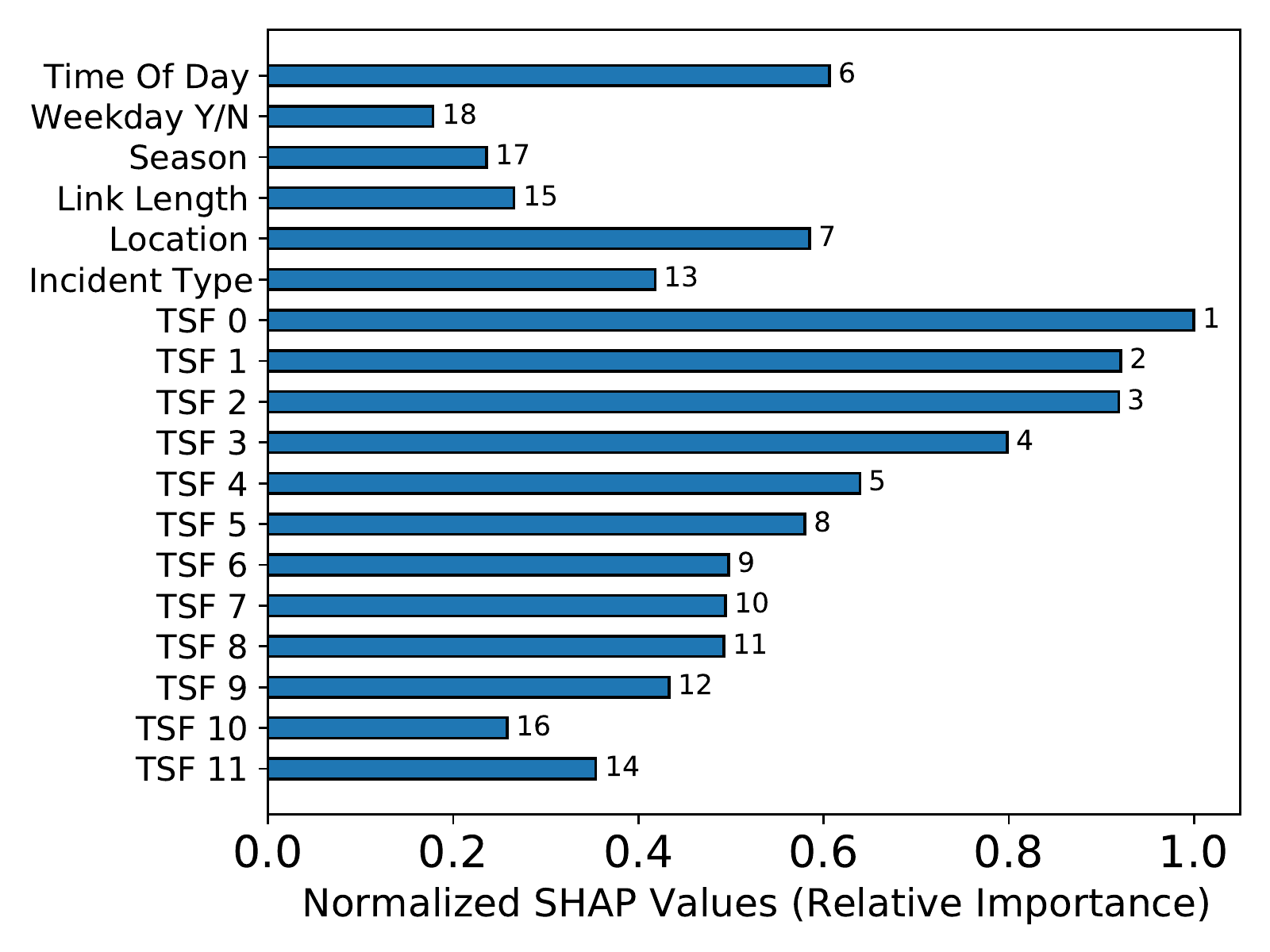}
		\centering \textbf{A}
	\end{minipage}
	~
	\begin{minipage}[t]{.46\textwidth}
		\includegraphics[width=\textwidth]{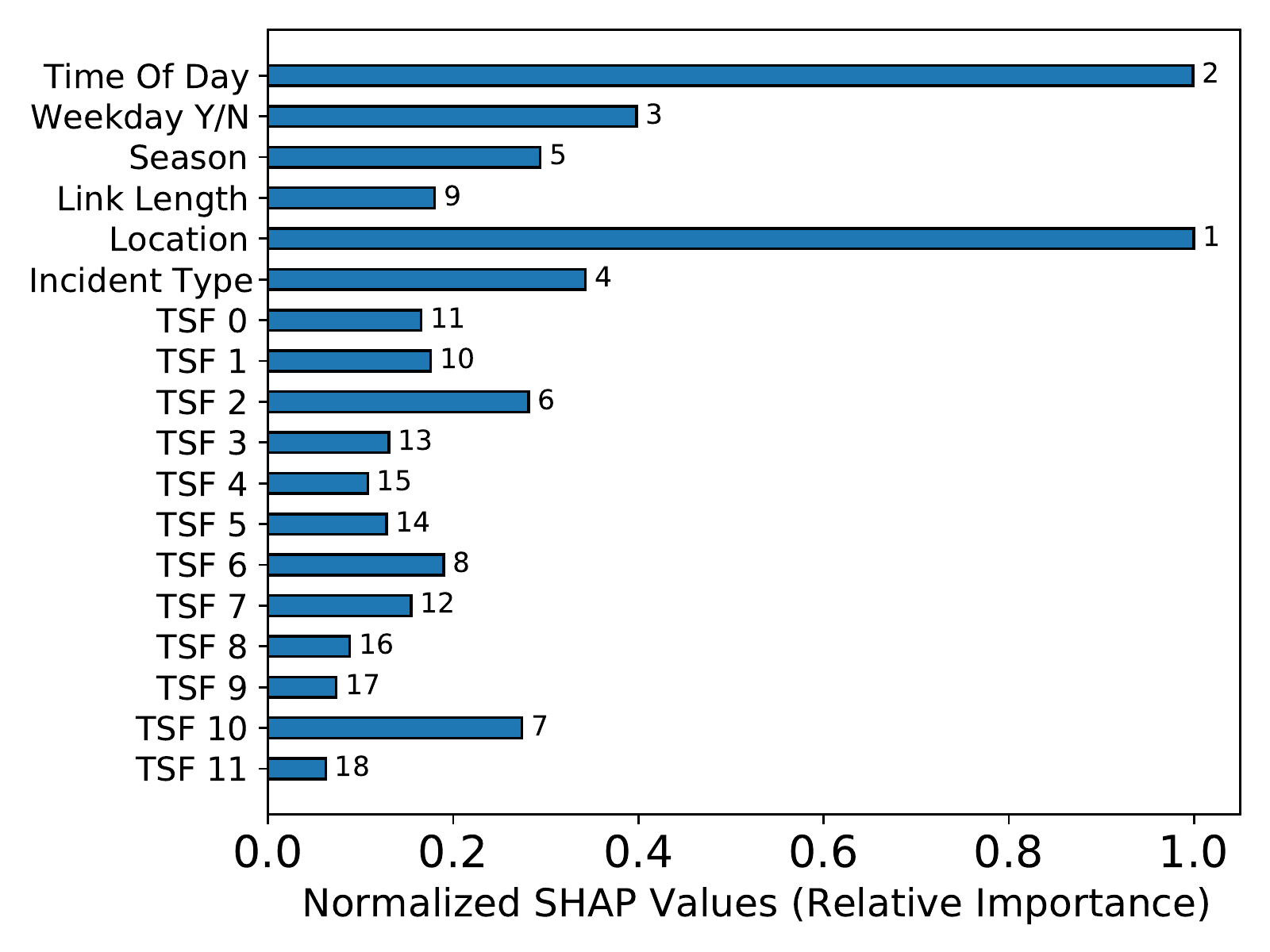}
		\centering \textbf{B}
	\end{minipage}
	
	\begin{minipage}[t]{.46\textwidth}
		\includegraphics[width=\textwidth]{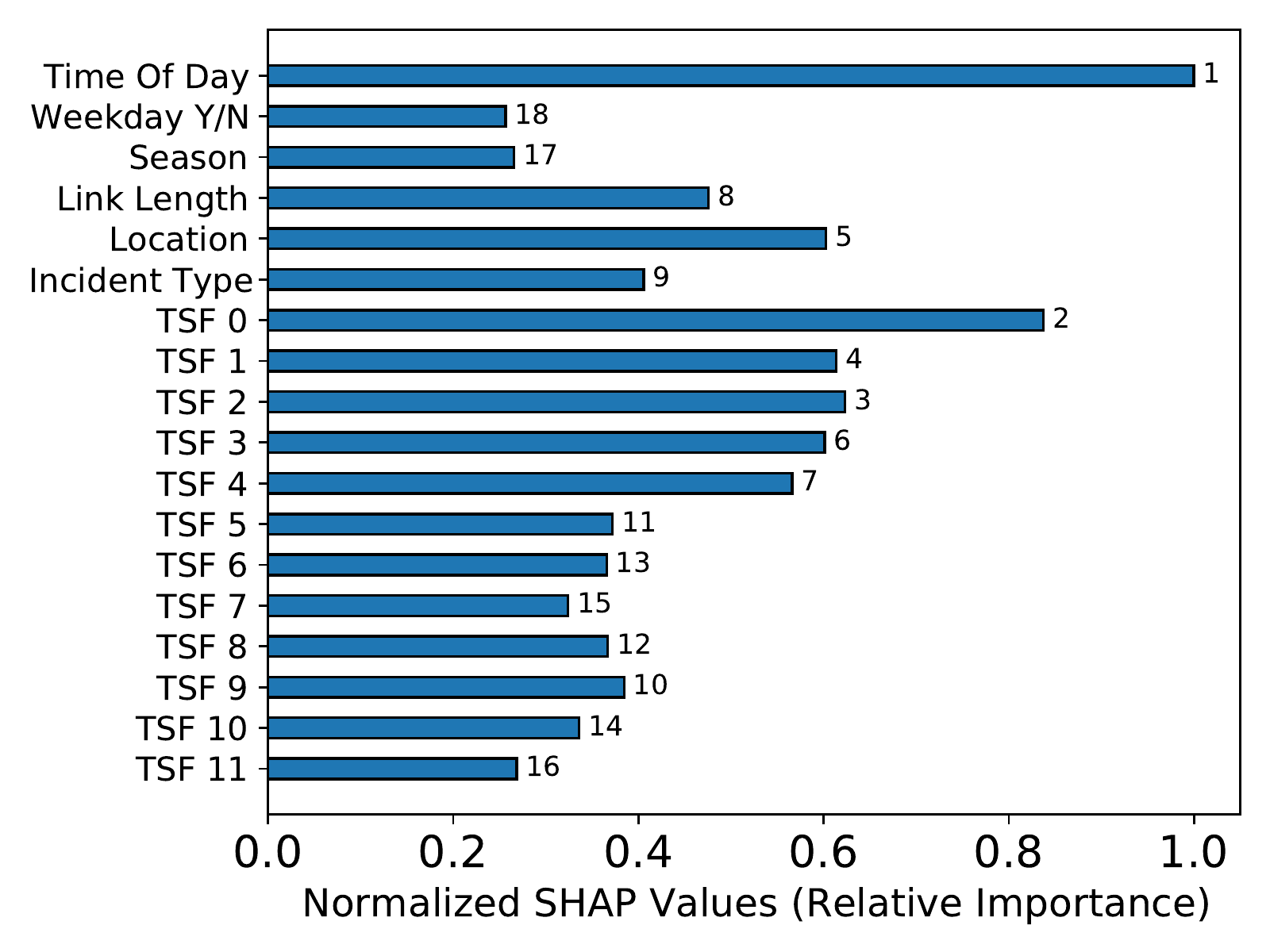}
		\centering \textbf{C}
	\end{minipage}
	~
	\begin{minipage}[t]{.46\textwidth}
		\includegraphics[width=\textwidth]{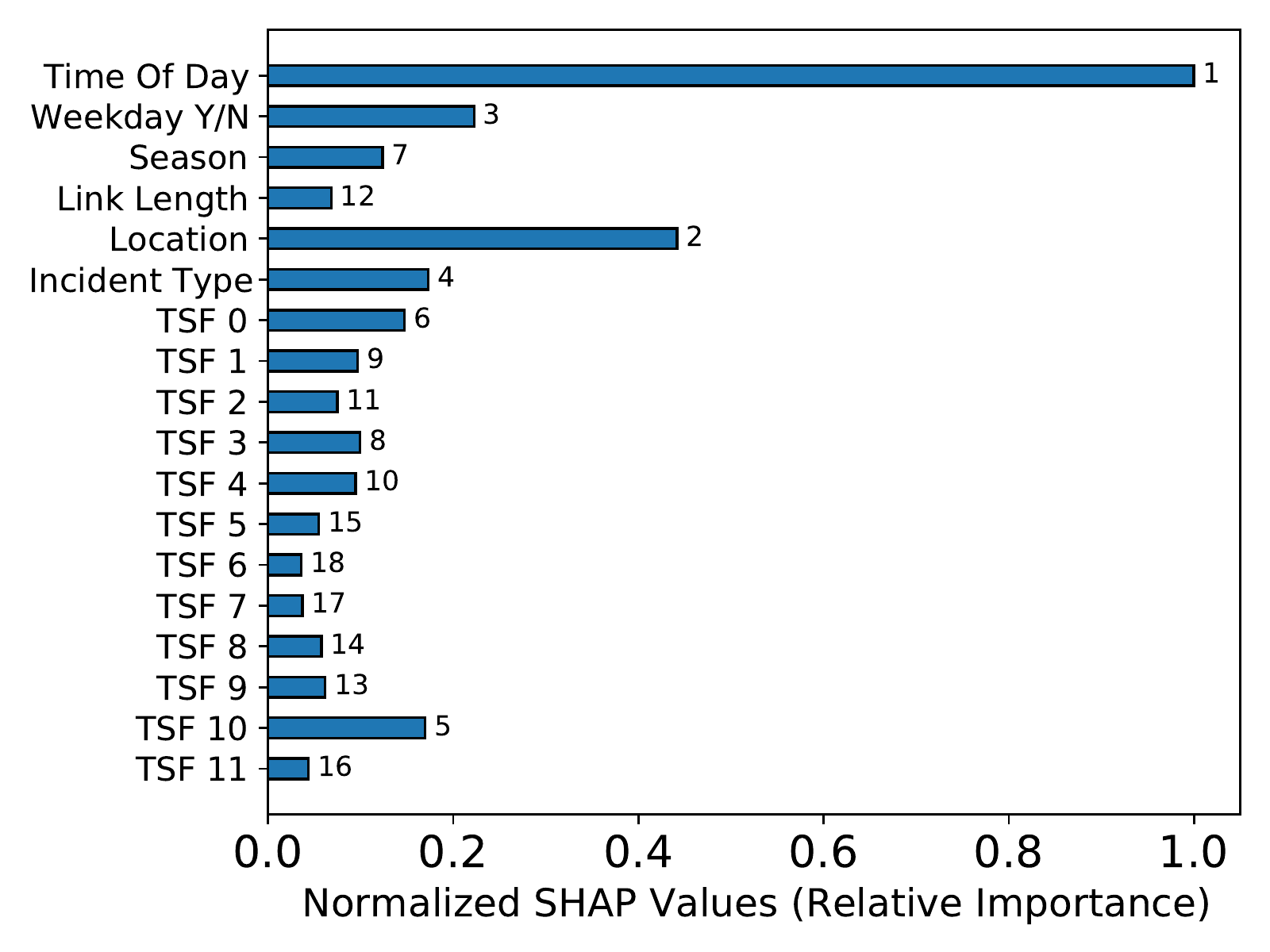}
		\centering \textbf{D}
	\end{minipage}
	\caption{Variable importance, as measured for the sliding window neural network model, for a subset of the important variables. Each plot is a different prediction horizon, all shown are at a prediction time of $t=30$. The importance at each prediction time is computed by taking the average of the absolute SHAP values for each feature across all query instances, and then has been normalized such that the largest importance at any horizon is 1, and all others are relative to this. The rank of each variable at a given time is written beside each bar. Due to the scaling, one should focus on the the ranking and relative difference between bars in each plot. TSF stands for time-series feature, that have been extracted through passing temporal convolutions across the data. \textbf{(A)}: $h=5$. \textbf{(B)}: $h=30$. \textbf{(C)}: $h=60$. \textbf{(D)}: $h=180$. }\label{fig:VariableImportanceNN}
\end{figure}
From \textbf{Figure 5}, we see that for very short horizons (5 minutes) there are a large number of time-series features with high importance.
This makes intuitive sense for the same reasoning as in the RSF case.
After the time-series features, we see the time of day, location and incident type are the features with the highest impact.
Moving to a horizon of 30 minutes, we then see the time-series features become less important, and location and time of day dominate the other features.
Note here that in the 5 minute horizon, there were lots of features with quite high importance, showing quite a number influenced the model's output, but at a horizon of 30 minutes we see two with large importance and many others with far less.
At a horizon of 60 minutes, we again see the importance of the time-series features increase, but the time of day and location are still the two most important of the time-invariant features, ranking first and fifth respectively.
One might question why the time-series resurge in importance here, and we explore this further in \textbf{Figure 6} and the analysis of it.
At a long horizon of 180 minutes, the time of day is by far the most important feature, and the location is second, but is less important relatively than it was at a horizon of 30 minutes.
A natural interpretation of this might be that for long horizons into the future, knowing if an incident will overlap with rush hour or go into lunch time or the night is a good indication of if we believe it might last a long time.

Having inspected the magnitude of SHAP values, we now question how do actual feature values shift the network output, either increasing or decreasing it.
Visualizing this is more complex due to the fact that we want to plot the impact of various features, the value they attain, and if this shifted the prediction up or down.
A standard way to do this for SHAP values is to make a `beeswarm' plot, in which each data-instance is plotted as a single dot, once per each feature.
Examples of making such plots for our dataset are given in \textbf{Figure 6}.
One should read these plots as follows. Firstly, along the y-axis are features the model used, where categorical ones have been split into their one-hot encoded states.
Secondly, the x-axis displays the SHAP values, not normalized as they were in the previous analysis, showing how the particular feature shifts the model output either up or down.
Thirdly, the colour indicates the feature value. For binary features such as `Morning Rush' a high value indicates the data-point was in the morning rush. 
Fourth, where many data-points had similar SHAP values for the same feature, points are expanded outwards in the y-axis, so a large vertical strip of points for a single feature indicates a high density of points at that SHAP value. Horizon is indicated by $h$ in each sub-caption, which corresponds to looking at a particular output node of the network. We split the fixed and series features to aid readability. Note that pushing the output `up' (a feature with a positive SHAP value) indicates increasing the probability mass functions value at this time-horizon. 

\begin{figure}[ht!]
	\centering
	\begin{minipage}[t]{.30\textwidth}
		\includegraphics[width=\textwidth]{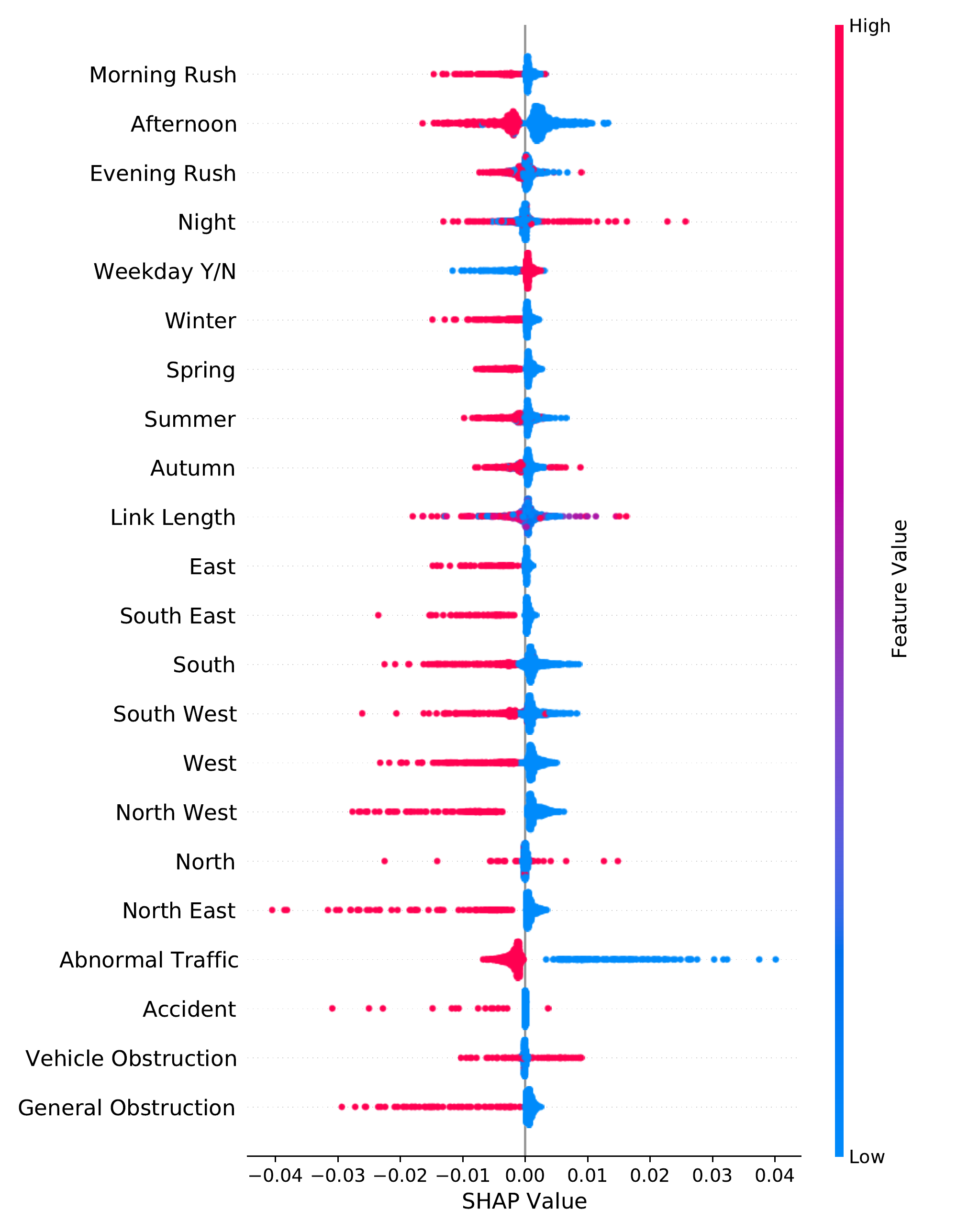}
		\centering \textbf{A}
	\end{minipage}
	~
	\begin{minipage}[t]{.30\textwidth}
		\includegraphics[width=\textwidth]{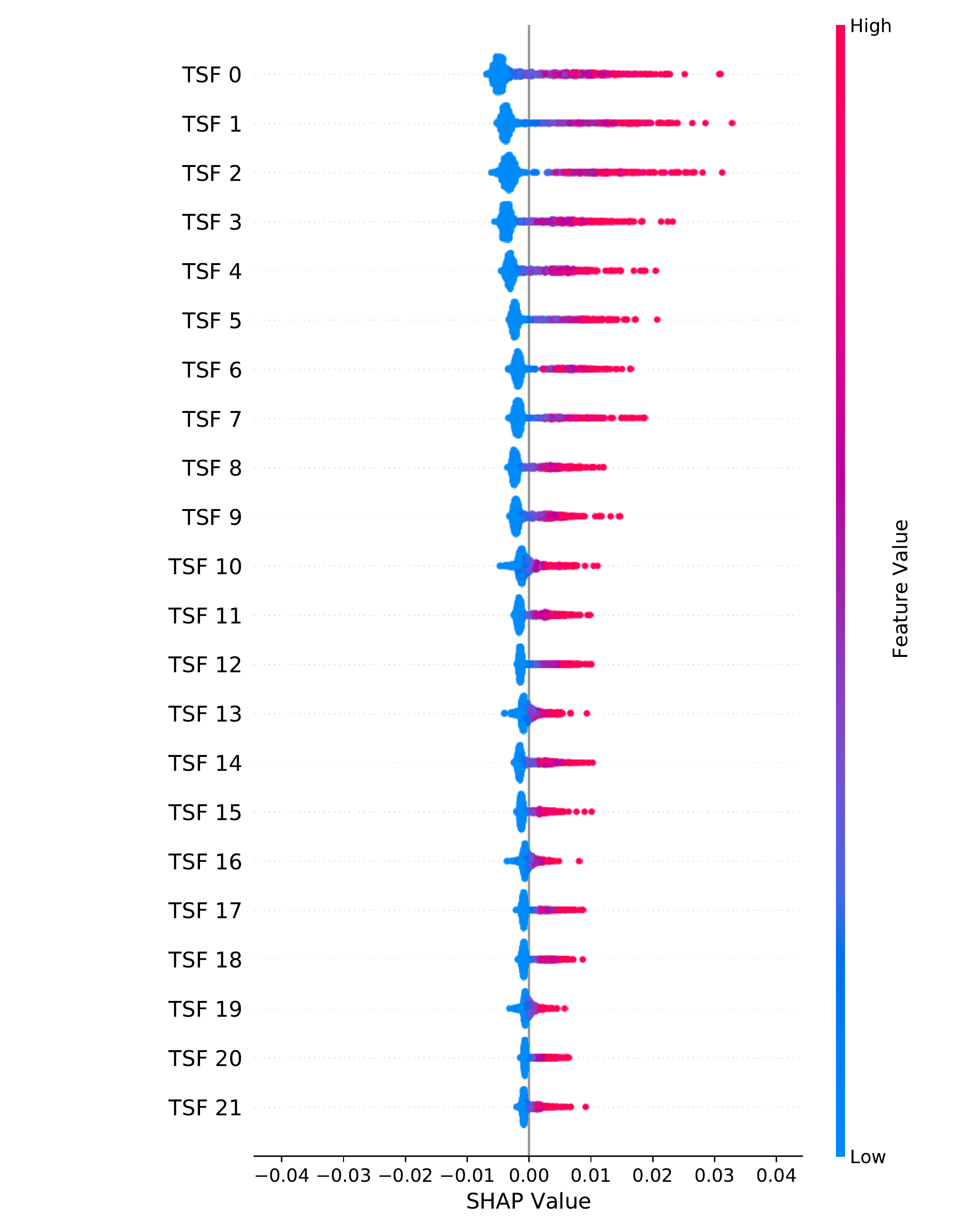}
		\centering \textbf{B}
	\end{minipage}
	~
	\begin{minipage}[t]{.30\textwidth}
		\includegraphics[width=\textwidth]{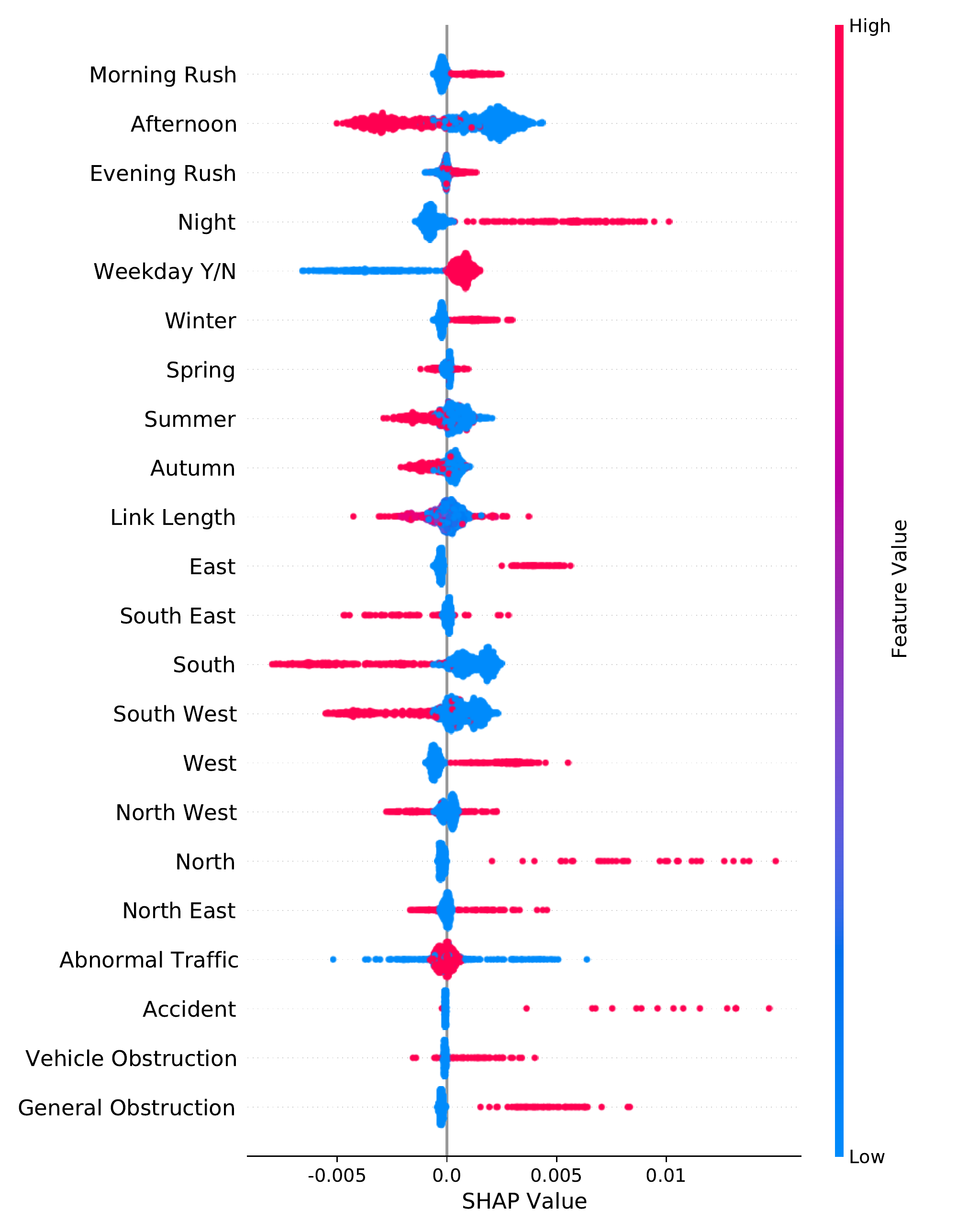}
		\centering \textbf{C}
	\end{minipage}
	
	\begin{minipage}[t]{.30\textwidth}
		\includegraphics[width=\textwidth]{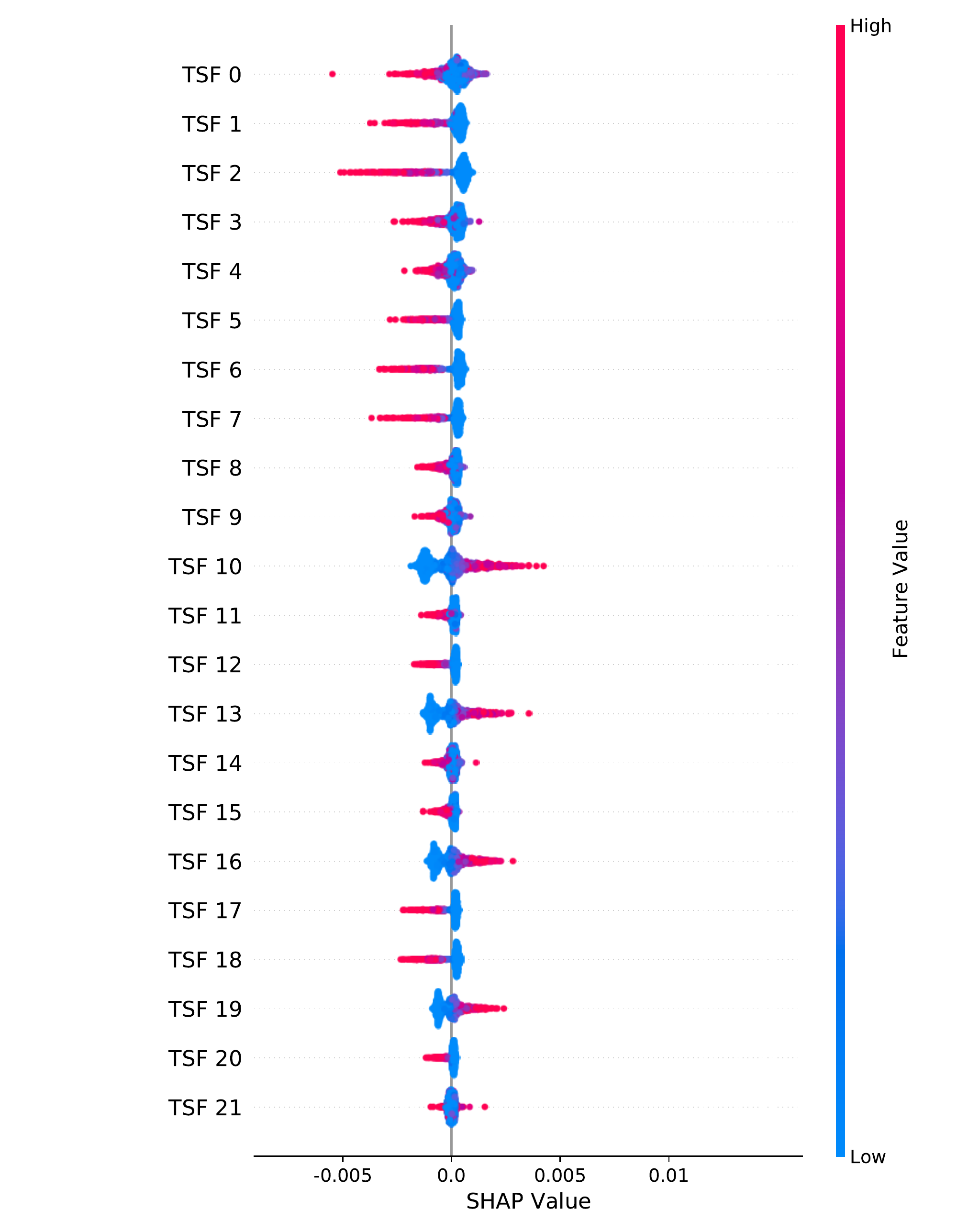}
		\centering \textbf{D}
	\end{minipage}
	~
	\begin{minipage}[t]{.30\textwidth}
		\includegraphics[width=\textwidth]{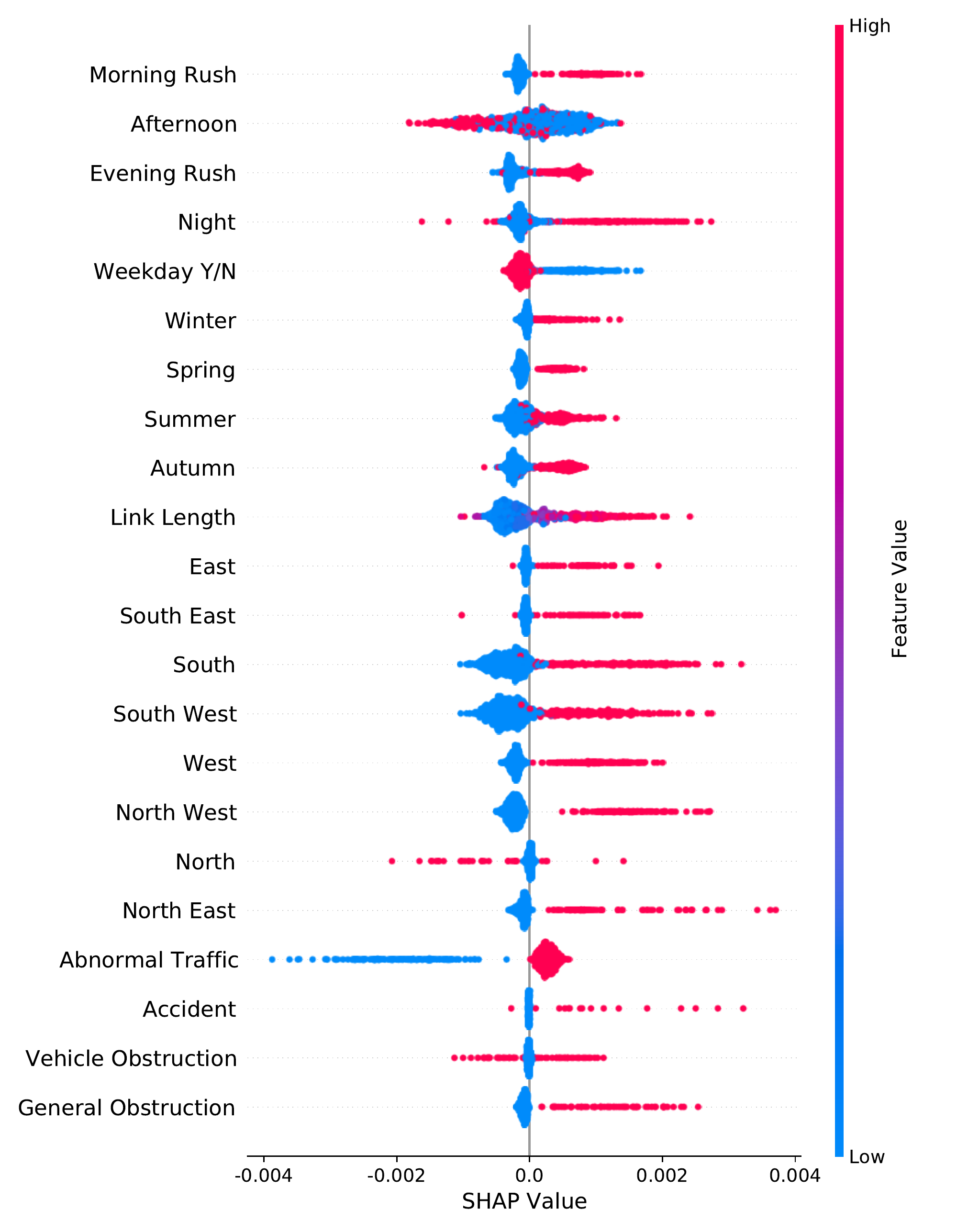}
		\centering \textbf{E}
	\end{minipage}
	~
	\begin{minipage}[t]{.30\textwidth}
		\includegraphics[width=\textwidth]{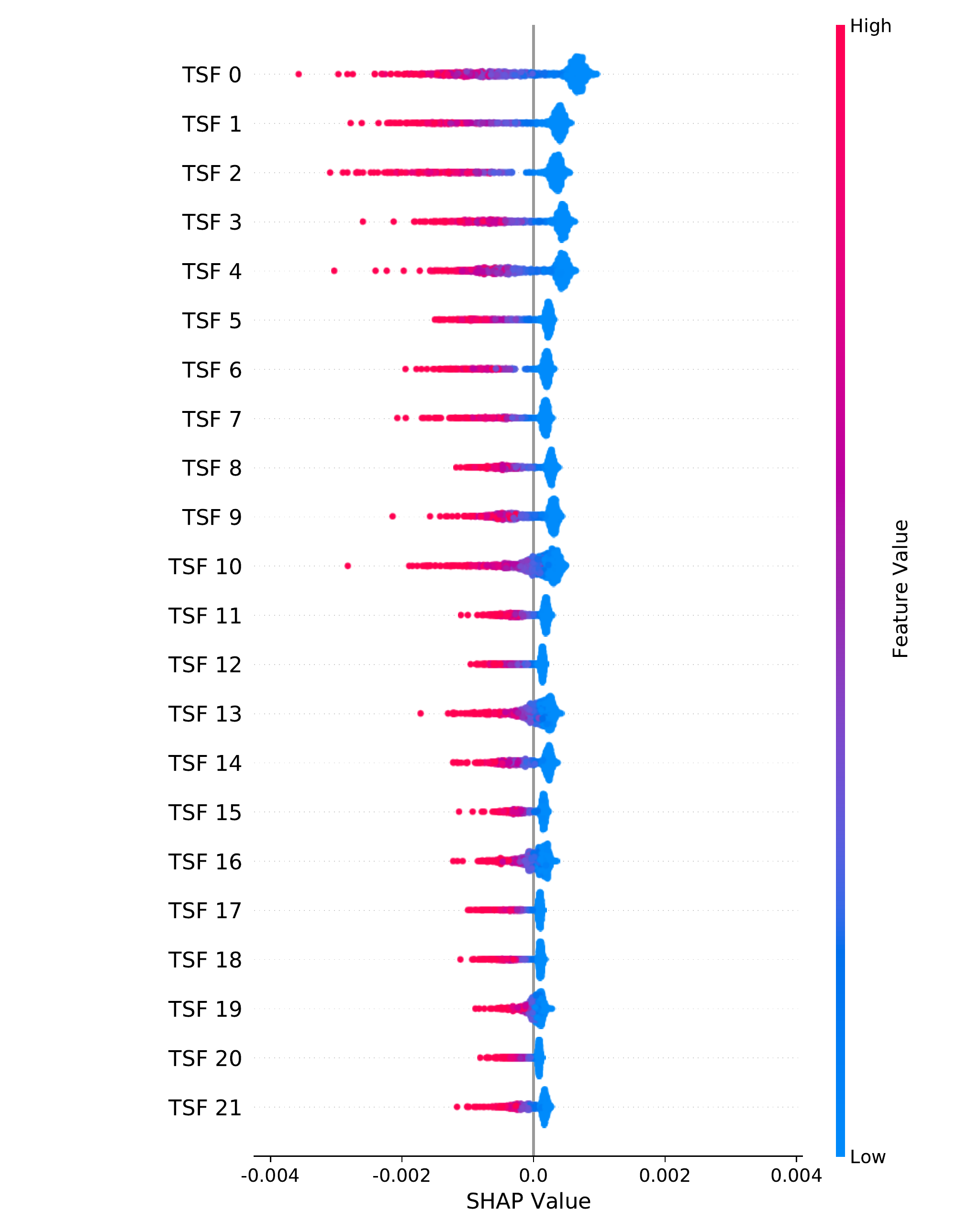}
		\centering \textbf{F}
	\end{minipage}
	
	\begin{minipage}[t]{.30\textwidth}
		\includegraphics[width=\textwidth]{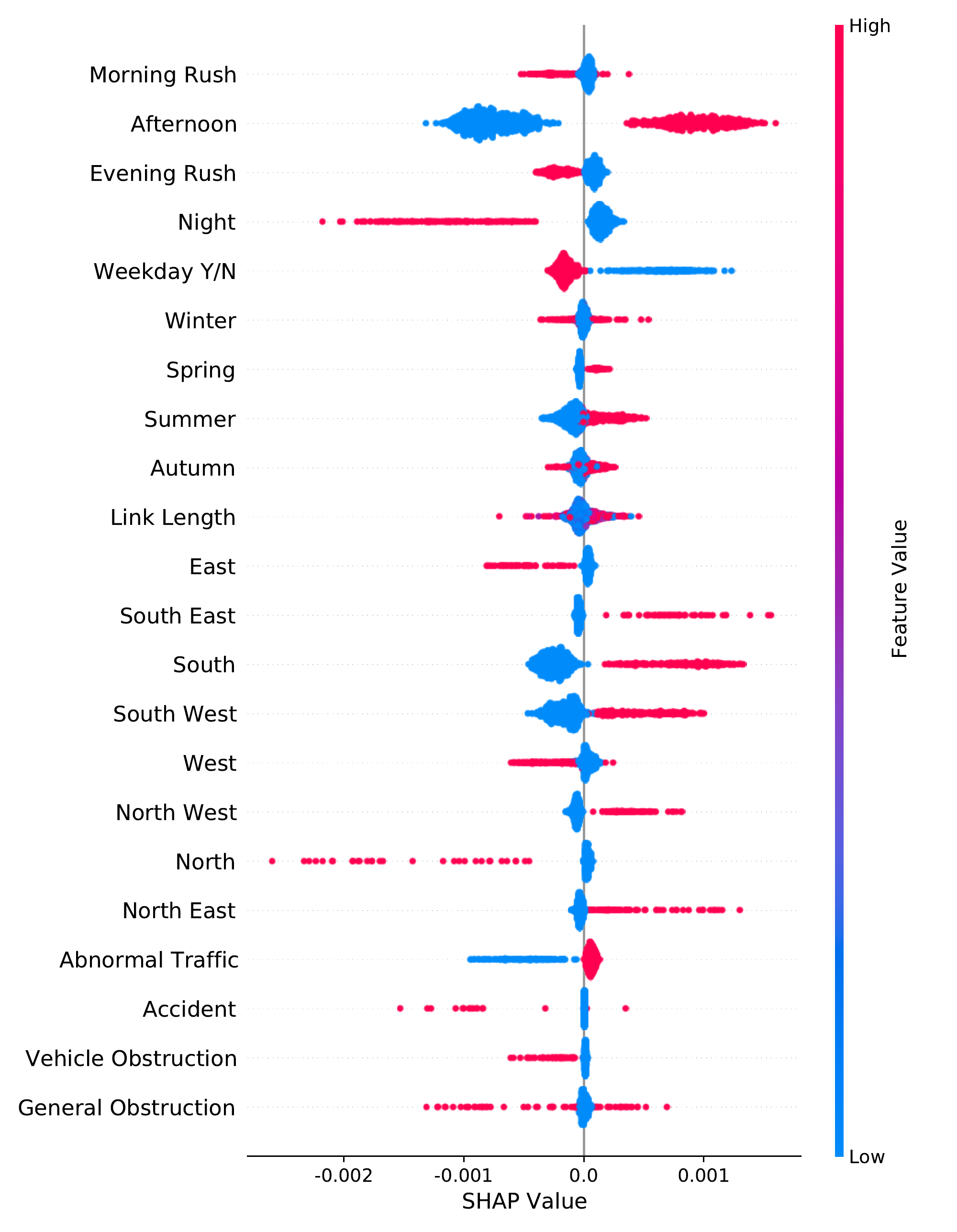}
		\centering \textbf{G}
	\end{minipage}
	~
	\begin{minipage}[t]{.30\textwidth}
		\includegraphics[width=\textwidth]{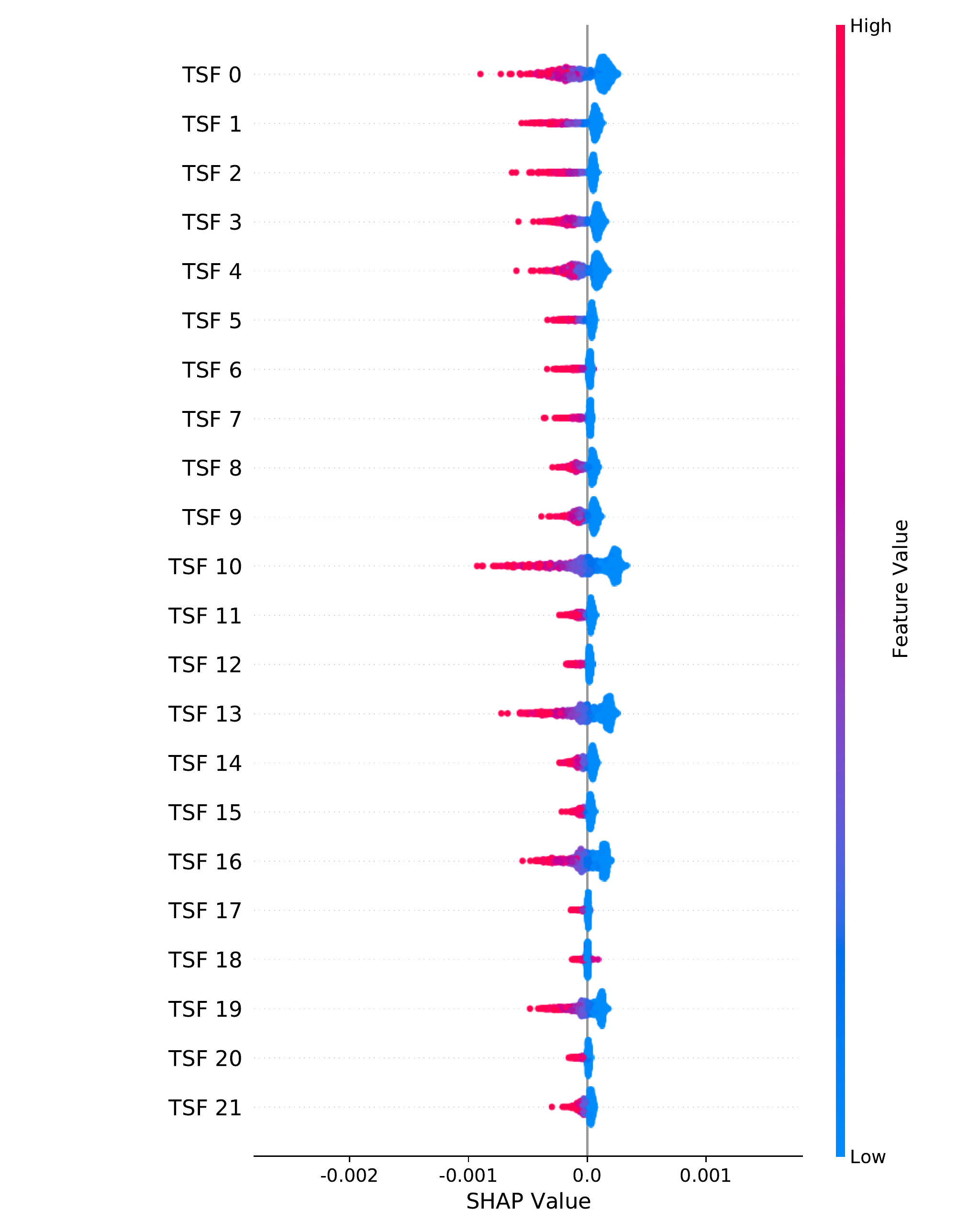}
		\centering \textbf{H}
	\end{minipage}
	
	\caption{SHAP values visualising how feature values shift the output of the network at particular horizons $h$. 
	\textbf{(A)}: Time-invariant features, $h=5$. \textbf{(B)}: Time-varying features, $h=5$. 
	\textbf{(C)}: Time-invariant features, $h=30$. \textbf{(D)}: Time-varying features, $h=30$.
	\textbf{(E)}: Time-invariant features, $h=60$. \textbf{(F)}: Time-varying features, $h=60$.
	\textbf{(G)}: Time-invariant features, $h=180$. \textbf{(H)}: Time-varying features, $h=180$. }\label{fig:SHAPImpact}
\end{figure}

Whilst there is a large amount of information in \textbf{Figure 6}, we breakdown the main points here.
Firstly, we see that at a horizon of 5 minutes (\textbf{Figures 6a, 8b}), there is clear coherence in the features, shown by there not being a random assortment of colours across single features.
Earlier, we saw time-series features, the time of day and location were influential factors at this horizon.
Considering the time of day more finely, we see from \textbf{Figure 6a} that when incidents are in the morning rush period, the value of the PMF at this time is decreased, suggesting the model believes it is unlikely incidents at this time of day will end very quickly. 
Inspecting horizons of 30 and 60 minutes, in \textbf{Figures 6c, 6d}, we see that that when incidents occur in the morning rush, this increases the output values at these horizons.
Finally, there appears to be more complex behaviour at a horizon of 180 minutes, as incidents occurring in the morning rush sometimes increase, and sometimes decrease the output at this time.
If we turn instead to view how a location impacts the result, for example inspecting the SHAP values for `West' we see that attaining a value of 1 here decreases the model output at a horizon of 5 minutes (\textbf{Figure 6a}), increases it at horizons of 30 and 60 minutes (\textbf{Figures 6c, 6e}) and then decreases it again at a horizon of 180 minutes (\textbf{Figure 6g}).
Note however that since some of these features are in-fact categorical and have been one-hot encoded for use with a neural network, care must be taken in interpreting the impact of such values.
In doing this analysis, we attain a SHAP value for each feature, however every data-point has as single location value equal to 1, and the rest equal to 0.
So each location feature here will alter the neural network output, but the total impact of a data-point having a particular location will be the sum of the SHAP values for that data-point over all encoded categories.
As such, we can better visualize the impact of categorical variables, for example location, by first summing the SHAP values for a data-point for all encoded groups of a particular feature, then visualizing how the overall feature impacted predictions.
We do so in \textbf{Figure 7}.
\begin{figure}[ht!]
	\centering
	\begin{minipage}[t]{.46\textwidth}
		\includegraphics[width=\textwidth]{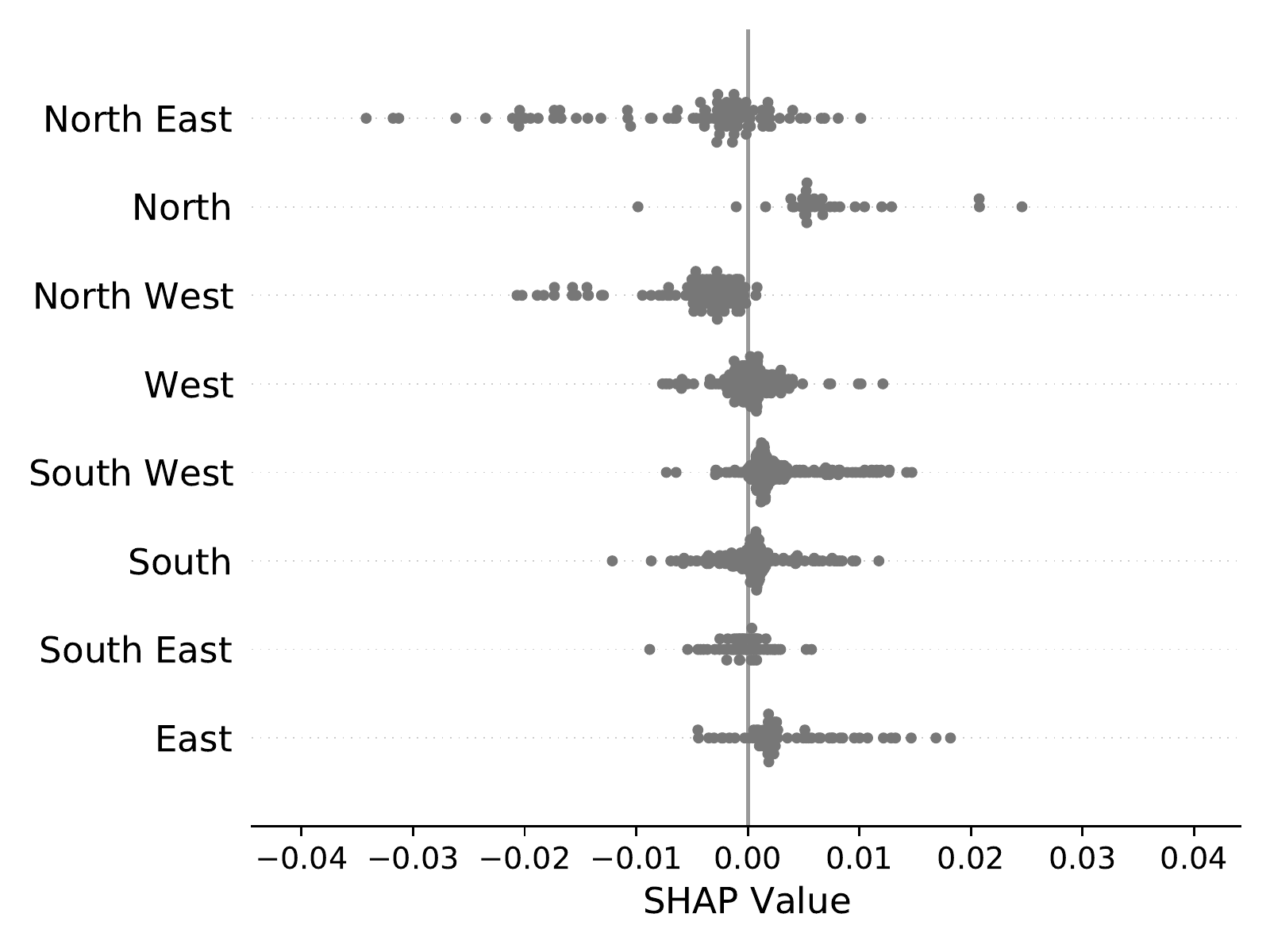}
		\centering \textbf{A}
	\end{minipage}
	~
	\begin{minipage}[t]{.46\textwidth}
		\includegraphics[width=\textwidth]{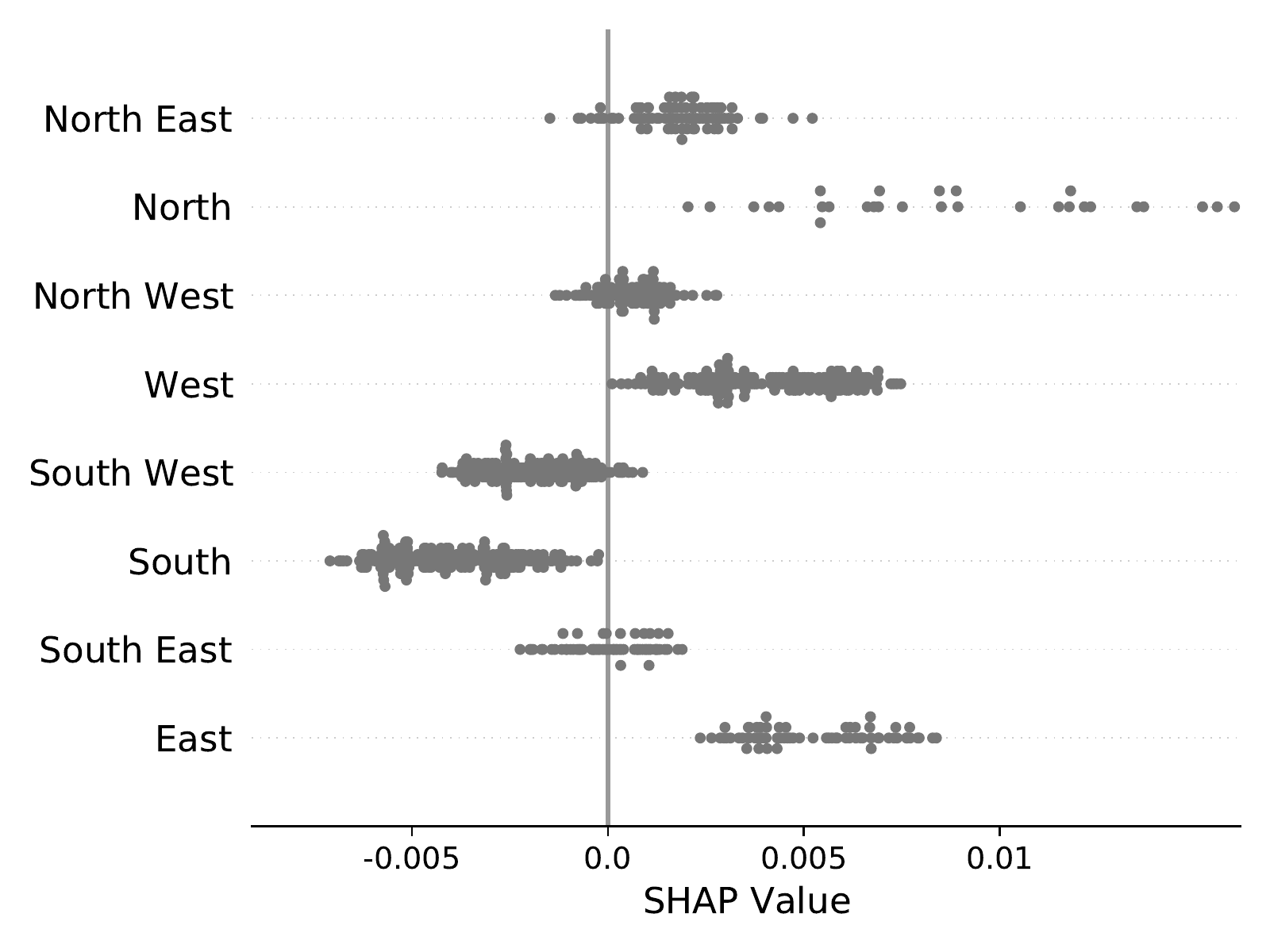}
		\centering \textbf{B}
	\end{minipage}
	
	\begin{minipage}[t]{.46\textwidth}
		\includegraphics[width=\textwidth]{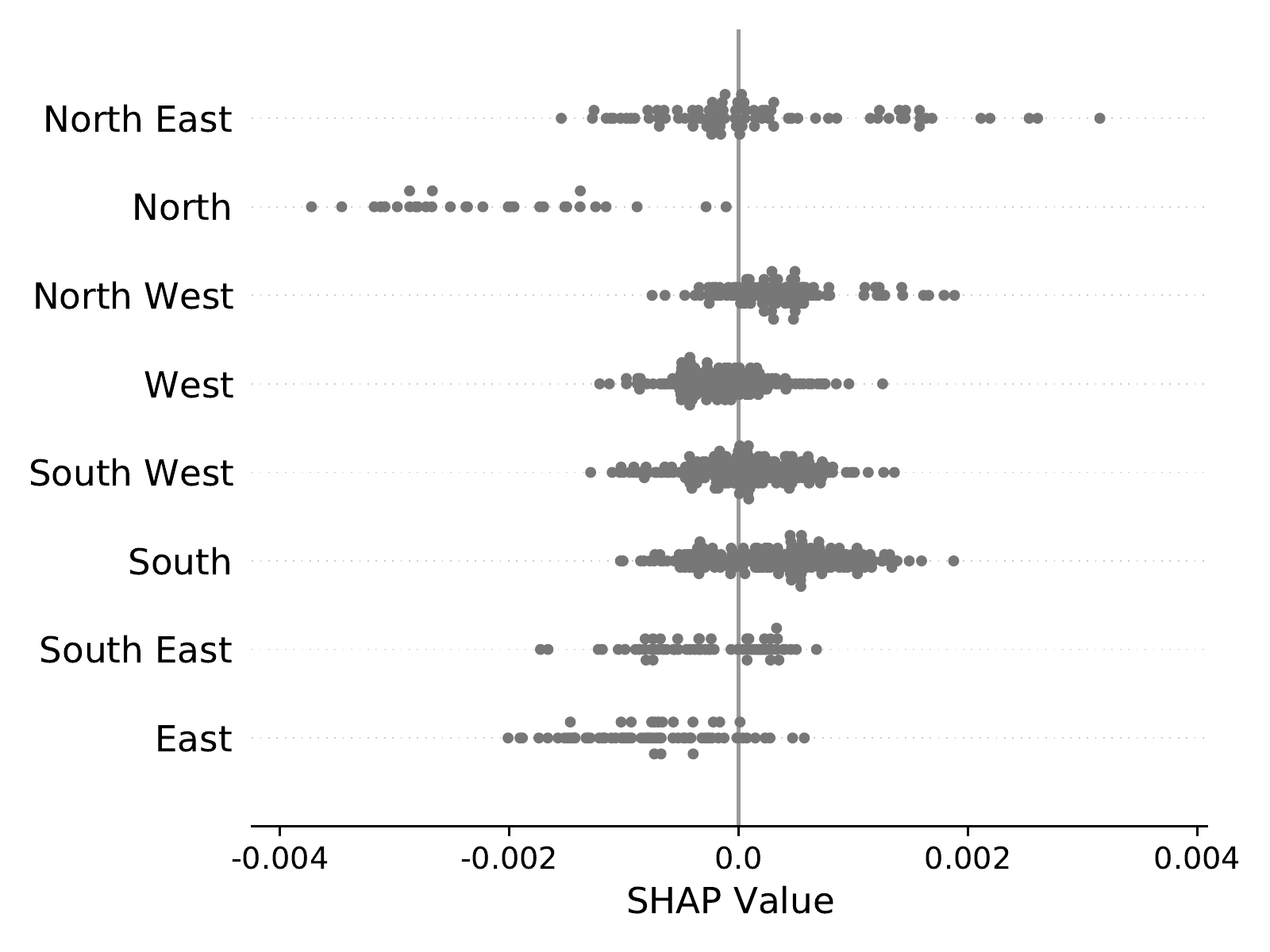}
		\centering \textbf{C}
	\end{minipage}
	~
	\begin{minipage}[t]{.46\textwidth}
		\includegraphics[width=\textwidth]{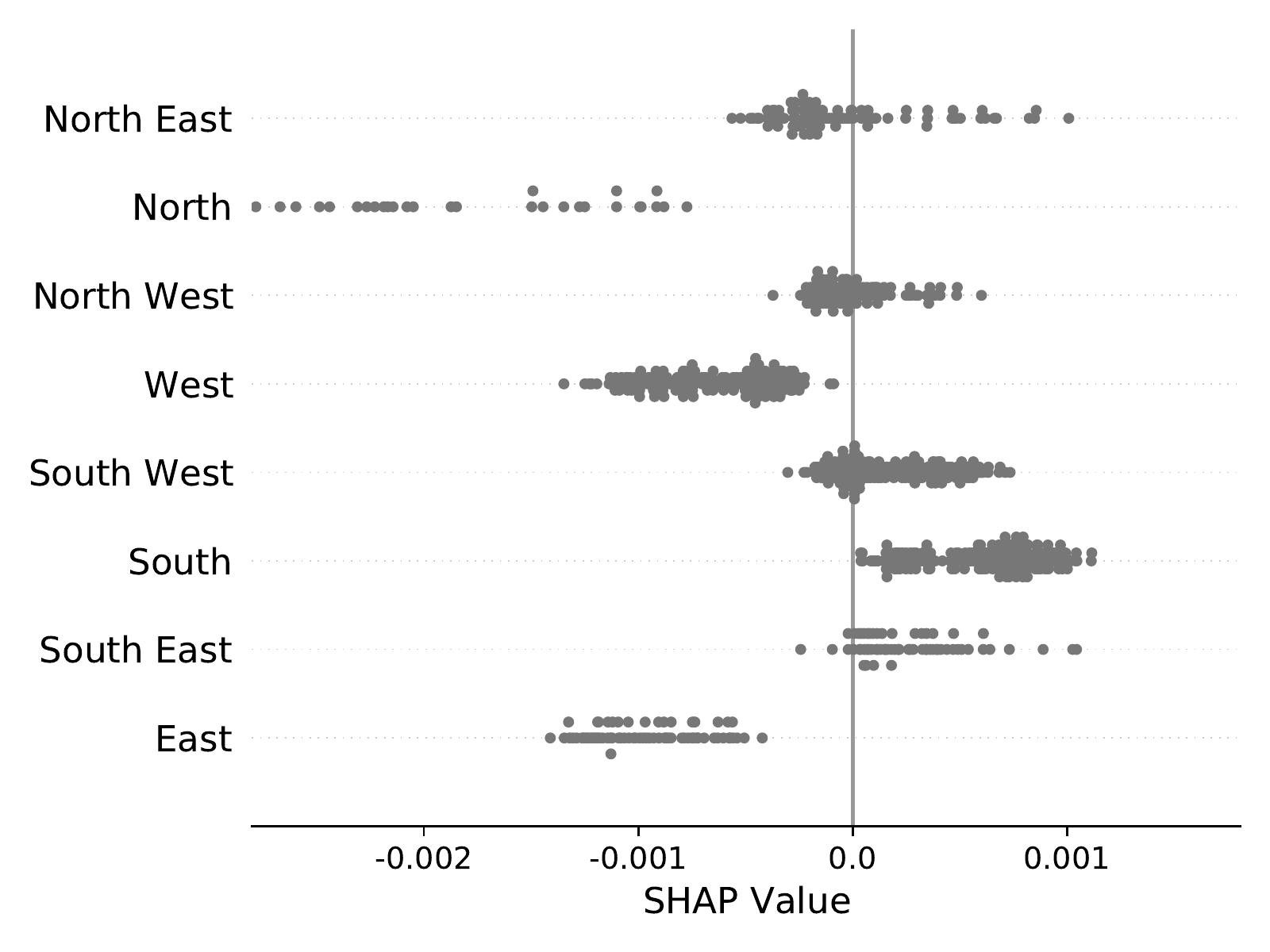}
		\centering \textbf{D}
	\end{minipage}
	\caption{SHAP values for the location feature. In each plot, we have summed the SHAP values for each one-hot encoded value, and then only plotted the resulting value in the row corresponding to each data-points true feature value. This shows the overall impact of the location feature, and allows one to view this impact separately for each value it attains. \textbf{(A)}: $h=5$. \textbf{(B)}: $h=30$. \textbf{(C)}: $h=60$. \textbf{(D)}: $h=180$. }\label{fig:SHAPLocationSplit}
\end{figure}

The more intuitive description offered in \textbf{Figure 7} allows one to see the overall impact of the location variable.
An example interpretation one could read from \textbf{Figure 7} is: for data-points where the spatial location was east, the overall impact of having this location on the model output was an increase in the output at horizons of 5 and 30 minutes, and a decrease at 60 and 180 minutes.
From this more refined view, we can see that incidents in the north east are quite varied, as their SHAP values sometimes shift up and sometimes shift down for all horizons.
Incidents in the north typically have the output increased at horizons of 5 and 30 minutes (\textbf{Figures 7a, 7b}) and then decreased at horizons of 60 and 180 minutes (\textbf{Figures 7c, 7d}).
This suggests that, for example, the model has learnt incidents in the north are of shorter duration than incidents in the south, however this can then be adjusted further by other observed features.
Locations can be compared in this way for all possible pairs.

A further question of interpretability relates to the features engineered from the time-series: what effect do these have on the model?
We previously saw that they were highly important at 5 and 60 minute horizons, but we can now use \textbf{Figure 6} to consider what impact they are having.
If we start with a horizon of 5 minutes, we see from \textbf{Figure 6b} that when the time-series features attain a high value, the model output at this horizon is increased.
We do not have an interpretable explanation of these features, but we see that they can provide quite significant shifts up in the output if their values are high.
Moving to a horizon of 30 minutes, we see from \textbf{Figure 6d} that the series features become less coherent.
Some features attain a high value and shift the output up, others down, and the overall impact is small for many features compared to the impact of the fixed features.
This does indicate that the features learnt from the series are distinct in some sense, providing different impacts on the model output.
At a horizon of 60 minutes, we see from \textbf{Figure 6f} that the features attaining a high value indicates a decrease in the model value at this point.
From this we can interpret that when high values of features from the are attained, they are making the model put more mass in the immediate future, and less at horizons of 60 minutes or longer.
This is perhaps an intuitive result, that the time-series are providing information that can significantly increase the model output at short term horizons, and attaining these same values shifts down the predictions at long horizons.

Of course, the sheer amount of data available here is somewhat overwhelming, however using SHAP values one can gain a significant understanding of why the machine learning model is outputting particular values.
We further show plots for the overall impact of categorical features in the supplementary material, for interested readers, omitted for brevity here.
However, care must always be taken in ensuring that feature importance and causality are not assumed, rather we are able to question why the model gives a set of outputs for a particular set of inputs.

\section{Conclusion}\label{sec:Conclusion}

In this work, we have addressed a number of issues raised in the literature regarding traffic incident duration prediction.
Firstly, we considered a method to determine incident duration as not only when an operator declared it cleared, but also when the traffic stated had returned to some typical behaviour.
This ensured that our predictions reflected when commuters could expect normal traffic conditions to resume if an incident had a significant impact even after it was cleared.
Secondly, we considered a range of models, some based on classic survival analysis and others based on machine learning and assessed how they performed on our dataset in both a static and dynamic setting.
In-particular, we took inspiration from work in the domain of healthcare and applied emerging methods used there to problems in traffic incident analysis.
We saw that in a static setting, there was little to choose between the models but generally either a neural network or random survival forest method would be preferred to the others considered.
We moved into the dynamic setting by utilising landmarking and sliding window neural networks that applied temporal convolutions to the time-series, both of which were inspired by success in healthcare applications.
We saw non-parametric methods that made no distributional assumptions were preferred to methods that parametrized mixture distributions, and saw some benefit in applying kernel smoothing to enforce some minimal structure on a non-parametric output.
Utilizing models with non-parametric distributional forms was influenced by the increasingly complex distributions being considered for this problem in the traffic literature, and showed significant promise in our results.

We assessed how each model performed using three different scoring criteria and in the dynamic sense, we saw clear structure in the results, with the kernel smoothed neural network model achieving an optimal C-index, Brier score depending on prediction time and horizon as to which model was preferred between a random survival forest and kernel smoothed neural network, and finally saw the neural network models showed much better performance in terms of point-wise error than the comparison models.
We saw that all score criteria were improved by engineering features from windows of sensor data through temporal convolutions, compared to feeding in local levels and gradients, suggesting future work should continue to explore methods of deriving features from sensor data rather than inputting raw values.

After, we considered variable importance for our model in the dynamic sense, assessing how the random survival forest and neural network models were influenced by the derived features.
Whilst we are aware of variable importance being studied in the traffic literature previously, we are not aware of SHAP values being applied to neural network models for incident duration analysis.
Time of day and location were generally important across models, particularly at long horizons, and the time-series features were shown to have significant impact on the neural network output at 5 and 60 minute horizons.

Finally, our suggestion to use the output of a neural network to define kernel weights, and from this construct a non-parametric distribution through smoothing was novel in that we are not aware of this being done before to the considered model.
It has relevance to other applications, specifically if one requires a continuous distribution to be output, but does not want to make strong parametric assumptions about what form this will take.
Further work could be done to consider alternative methods to select a bandwidth when applying this type of model, but the freedom it provides appeared promising on our data.
Other avenues for further work include incorporating more features in the dynamic prediction models.
In-particular, traffic operators will be able to report when recovery vehicles arrive, police involvement and details from on-site reports.
Further, social media and weather data could be collected in real-time.
It is likely these will have some predictive power for the duration of an incident, and considering how best to incorporate them is an interesting problem.
Finally, from our analysis in section \ref{sec:Results} we suspect that if one was able to derive robust, complex features from the time-series and feed them into a random survival forest model, we may see further improvements in the model.
One way to do this may be through an auto-encoder framework in which we pre-train a model to determine hidden representations of the series, however it is unclear if these will offer the same predictive power as we have observed from the time-series here.

\section*{Funding}
This work was supported by the EPSRC (grant number EP/L015374/1).

\section*{Acknowledgments}
We thank Dr. Steve Hilditch, Thales UK for sharing expertise on NTIS and UK transportation systems. 

\appendix

\section{Distribution Information}\label{sec:AppendixDistributionInfo}

In \textbf{Table 6} we summarize the various probability distributions discussed throughout this work.
\begin{table}[ht!]
  \centering
  \caption{Summary of distributions considered throughout this work. All distributions are defined for positive $x$. 
  Log-normal: Can be seeing as assuming the logarithm of $x$ is normally distributed. $erf(x)$ is the complementary error function. 
  Generalized F: $\tilde{B}_{\frac{d_1w}{d_1w+d_2}}\left( \frac{d_1}{2}, \frac{d_2}{2} \right)$ represents the regularized incomplete beta function. If $B(z; x,y)$ is the incomplete beta function and $B(x,y)$ is the beta function, then: $\tilde{B}_{\frac{d_1w}{d_1w+d_2}}\left( \frac{d_1}{2}, \frac{d_2}{2} \right) = \frac{B\left(\frac{d_1w}{d_1w+d_2}; \frac{d_1}{2}, \frac{d_2}{2}\right)}{B\left( \frac{d_1}{2}, \frac{d_2}{2} \right)}$. 
  Mixture: Each mixture component has some PDF $\tilde{f}_i(x; \bm{\theta}_i)$, CDF $\tilde{F}_i(x; \bm{\theta}_i)$ that depend on some parameter set $\bm{\theta}_i$, and a component weight $\pi_i$. The weighted sum of all components generates the final distribution. }\label{table:DistSummary}
  \resizebox{\textwidth}{!}{%
  \renewcommand{\arraystretch}{1.25}\begin{tabular}{|c|c|c|c|}
  \hline
  Distribution         & Parameters   &  PDF & CDF \\
  \hline
  Log-normal           & \makecell{$\mu \in (-\infty, \infty)$ \\ $\sigma \in (0, \infty)$} & $\frac{1}{x \sigma \sqrt{2\pi}}e^{-\frac{\left( \log(x) - \mu \right)^2}{2\sigma^2}}$ & $\frac{1}{2}\left( 1 + erf\left( \frac{\log(x) - \mu}{\sigma\sqrt{2}}\right)\right)$ \\

  \hline
  Log-logistic         & \makecell{$\alpha \in (0, \infty)$ \\ $\beta \in (0, \infty)$} & $\frac{\left(\frac{\beta}{\alpha}\right)\left(\frac{x}{\alpha}\right)^{\beta-1}}{\left( 1 + \left(\frac{x}{\alpha}\right)^\beta \right)^2}$ & $\frac{1}{1 + \left(\frac{x}{\alpha}\right)^{-\beta} }$ \\
  \hline
  Weibull              & \makecell{$\lambda \in (0, \infty)$ \\ $k \in (0, \infty)$} & $\frac{k}{\lambda}\left( \frac{x}{\lambda} \right)^{k-1} e^{-\left( \frac{x}{\lambda} \right)^k}$ & $1 - e^{-\left( \frac{x}{\lambda} \right)^k}$ \\
  \hline
  Generalized F        & \makecell{$\mu \in (0, \infty)$ \\ $\sigma \in (0, \infty)$ \\ $d_1 \in (0, \infty)$ \\ $d_2 \in (0, \infty)$ } & \makecell{ Let $w = \frac{\log(x) - \mu}{\sigma}$, then: \\
  $\frac{1}{wB\left( \frac{d_1}{2}, \frac{d_2}{2} \right)}\left(\frac{d_1}{d_2}\right)^{\frac{d_1}{2}}w^{\frac{d_1}{2}-1}\left( 1 + \frac{d_1}{d_2}w \right)^{-\frac{d_1 + d_2}{2}}$ } & $\tilde{B}_{\frac{d_1w}{d_1w+d_2}}\left( \frac{d_1}{2}, \frac{d_2}{2} \right)$ \\
  \hline
  Mixture              & \makecell{ $\bm{\theta}_i \forall i \in \{1, \dots, k\}$ \\ $\pi_i \forall i \in \{1, \dots, k\}$ } & $\sum_{i=1}^k \pi_i\tilde{f}_i\left( x; \bm{\theta}_i \right)$ & $\sum_{i=1}^k \pi_i \tilde{F}_i\left( \bm{x; \theta}_i \right)$ \\
  \hline
  \end{tabular}%
  }
\end{table}

\section{Incident Visualization}\label{sec:AppendixIncidentVisulization}

We saw in the main text that some structure clearly existed in the time-series.
One could question what features might we aim to discover in the series, and how varied is their behaviour during incidents? 
To aid in this, we visualize a number of windows with incidents in them, plotting the speed again and seeing some representative series of incidents in \textbf{Figure S1}.
\begin{figure}[ht!]
	\centering
	\begin{minipage}[t]{.32\textwidth}
		\includegraphics[width=\textwidth]{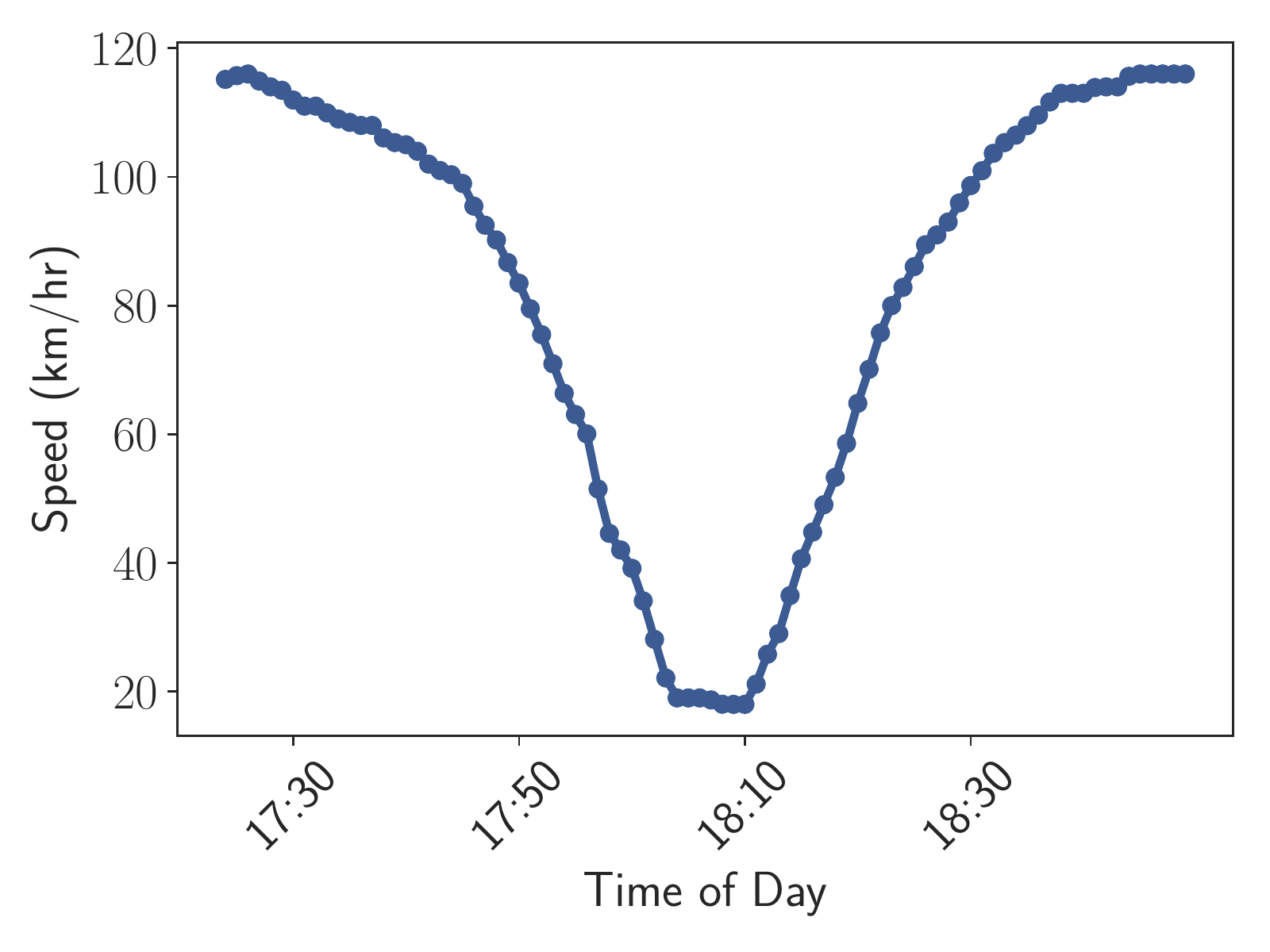}
		\centering \textbf{A}
	\end{minipage}
	~
	\begin{minipage}[t]{.32\textwidth}
		\includegraphics[width=\textwidth]{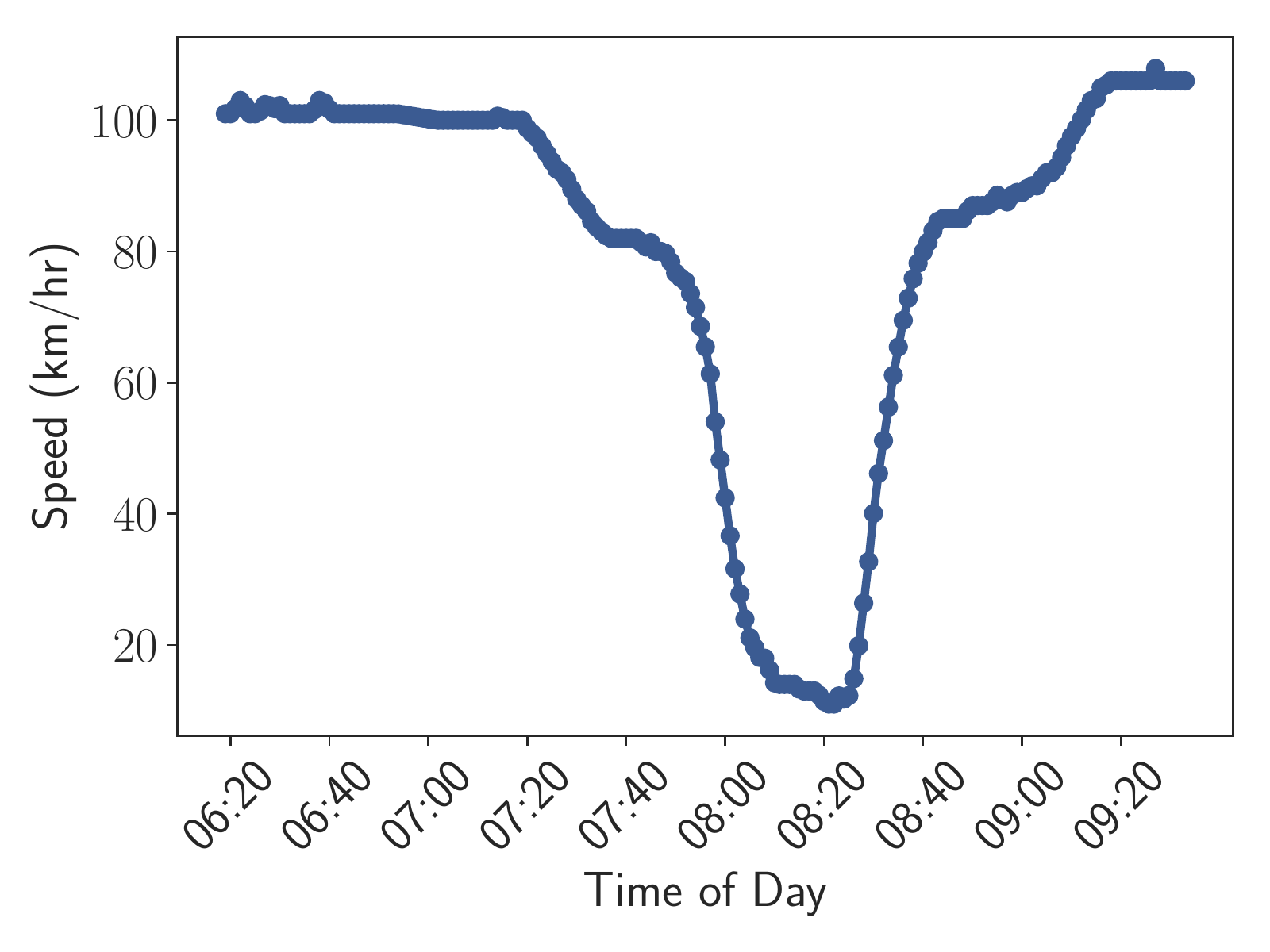}
		\centering \textbf{B}
	\end{minipage}
	~
	\begin{minipage}[t]{.32\textwidth}
		\includegraphics[width=\textwidth]{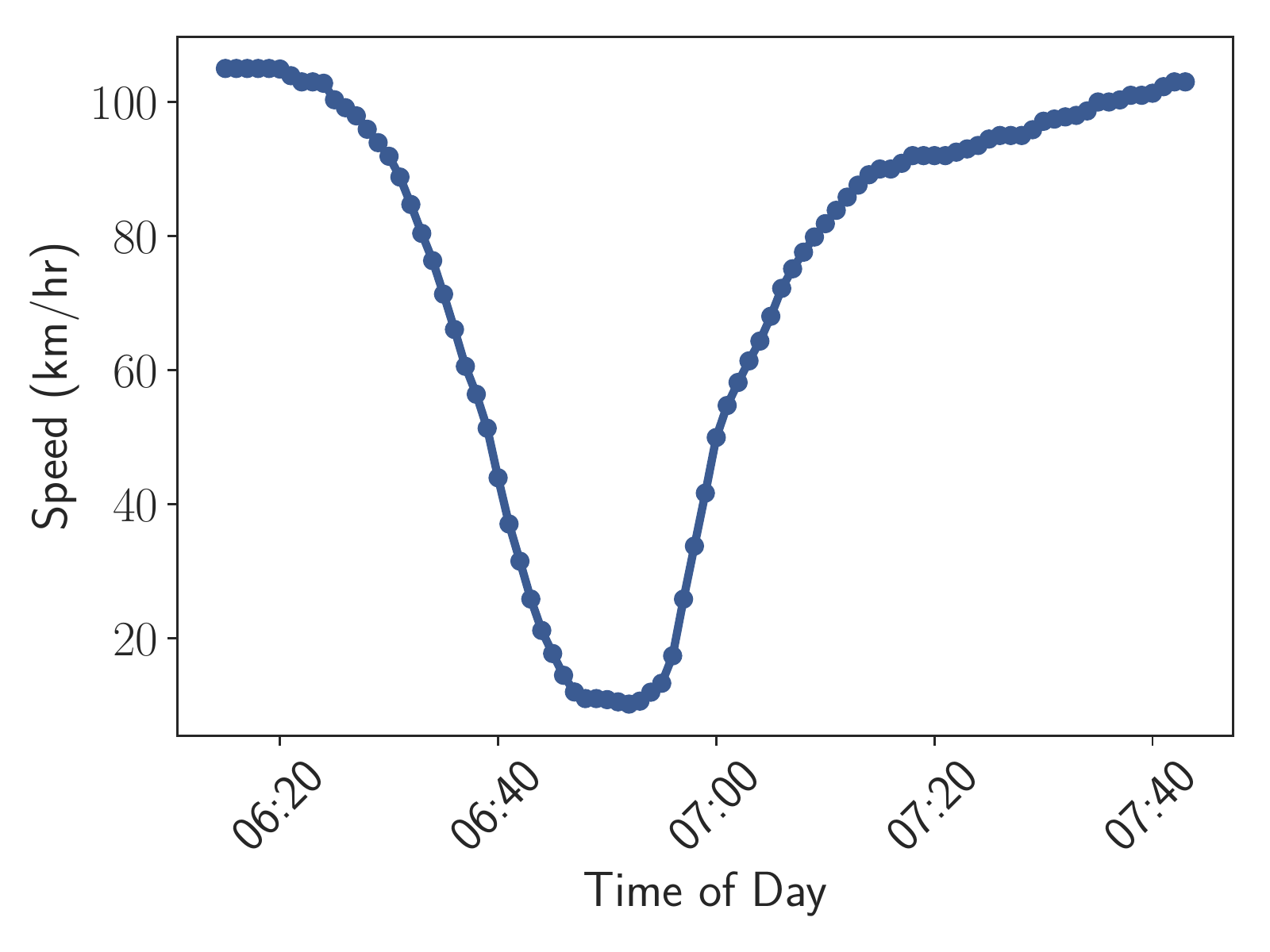}
		\centering \textbf{C}
	\end{minipage}

	\begin{minipage}[t]{.32\textwidth}
		\includegraphics[width=\textwidth]{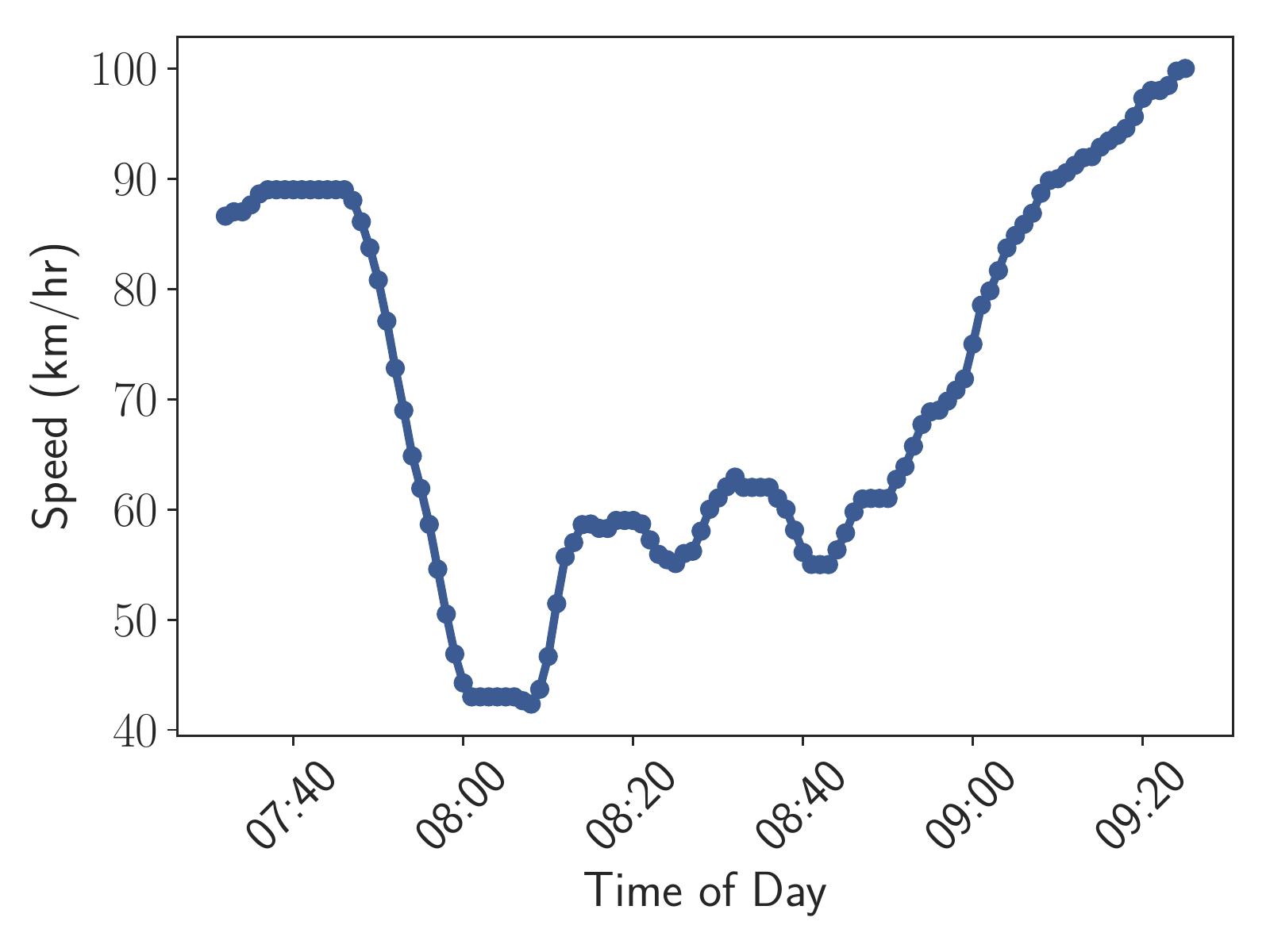}
		\centering \textbf{D}
	\end{minipage}
	~
	\begin{minipage}[t]{.32\textwidth}
		\includegraphics[width=\textwidth]{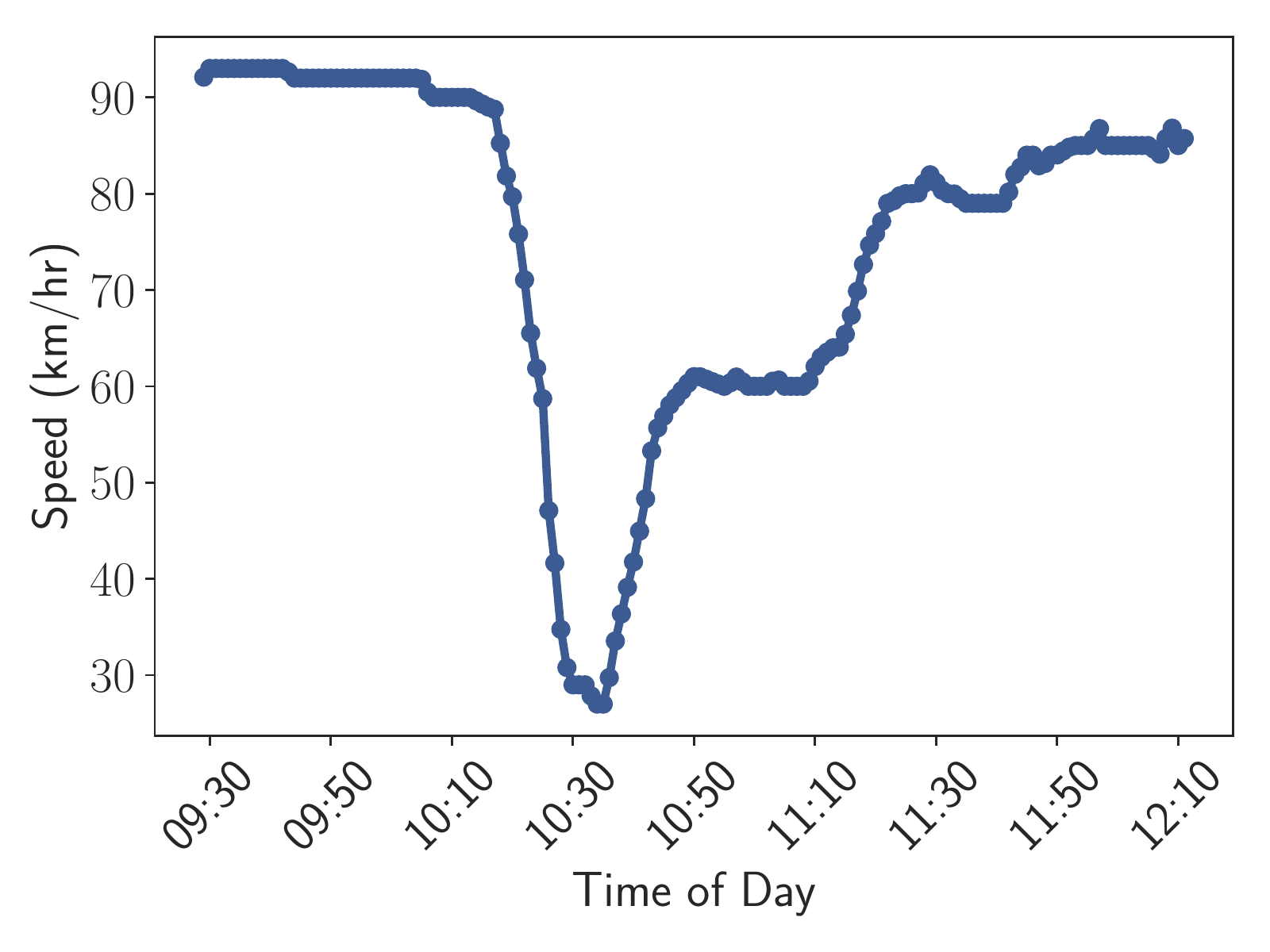}
		\centering \textbf{E}
	\end{minipage}
	~
	\begin{minipage}[t]{.32\textwidth}
		\includegraphics[width=\textwidth]{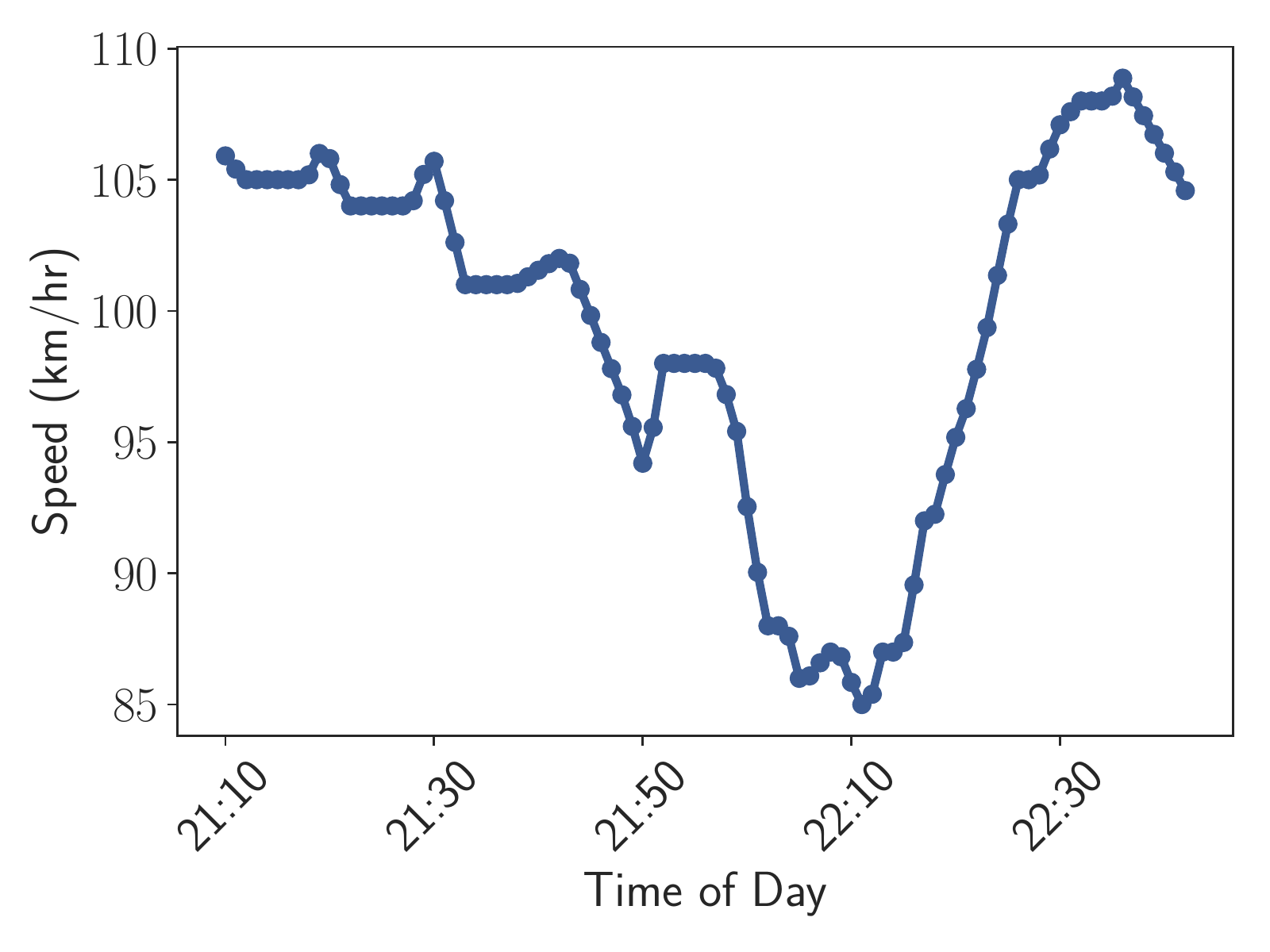}
		\centering \textbf{F}
	\end{minipage}

	\begin{minipage}[t]{.32\textwidth}
		\includegraphics[width=\textwidth]{./Supp_Images/Fig_1_d_Supp_Material.pdf}
		\centering \textbf{G}
	\end{minipage}
	~
	\begin{minipage}[t]{.32\textwidth}
		\includegraphics[width=\textwidth]{./Supp_Images/Fig_1_e_Supp_Material.pdf}
		\centering \textbf{H}
	\end{minipage}
	~
	\begin{minipage}[t]{.32\textwidth}
		\includegraphics[width=\textwidth]{./Supp_Images/Fig_1_f_Supp_Material.pdf}
		\centering \textbf{I}
	\end{minipage}
	\caption{Various example speed time-series from some incidents in the dataset. }\label{fig:EventsExamplesForCompare}
\end{figure}
From \textbf{Figure 8}, we see that different incidents show quite different representations in the speed time-series.
\textbf{Figures 8A, 8B, 8C} show short lived but large drops in speed associated with incidents, each at a different time of day. 
The speed profile is reasonably symmetric in time and the traffic state starts to recover soon after reaching its minimal speed.
\textbf{Figures 8D, 8E, 8F} on the other hand show very asymmetric incidents. 
In \textbf{Figures 8D, 8E} there is quite a sudden drop in speed, but a comparatively slow recovery.
In particular \textbf{Figure 8E} shows two periods where the speed recovers somewhat, but then stalls at particular values, but later recovering again.
\textbf{Figure 8F} shows the opposite case, a slow fall in speed to 85 km/hr and then a comparatively rapid recovery.
Finally \textbf{Figures 8G, 8H, 8I} show extreme incidents, with a long period of very low speed.
In the case of \textbf{Figure 8H}, we see the speed recovers somewhat to 40 km/hr however it remains around this for an extended period of time, where as in \textbf{Figures 8G, 8I} the speed remains around 20 km/hr for a significant period of time.

\section{Exploratory Analysis}\label{sec:ExploratoryAnalysis}

\subsection{Incident Duration Analysis}\label{sec:ExploratoryAnalysisDuration}

An initial exploratory analysis of our data reveals a number of properties.
As discussed, an initial step for many works in this field is to take all of the durations and determine if they are well modelled by a particular statistical distribution.
Our review of the literature suggested fitting log-normal, log-logistic, Weibull and generalized F distributions, so we do exactly this and show our results in \textbf{Figure 9}.
Specifically, \textbf{Figure 9A} shows the probability density functions (PDFs) compared to the data and \textbf{Figure 9B} shows a quantile-quantile plot.
An overview of each distribution considered is given in the supplementary material. 
\begin{figure}[ht!]
	\centering
	\begin{minipage}[t]{.47\textwidth}
		\includegraphics[width=\textwidth]{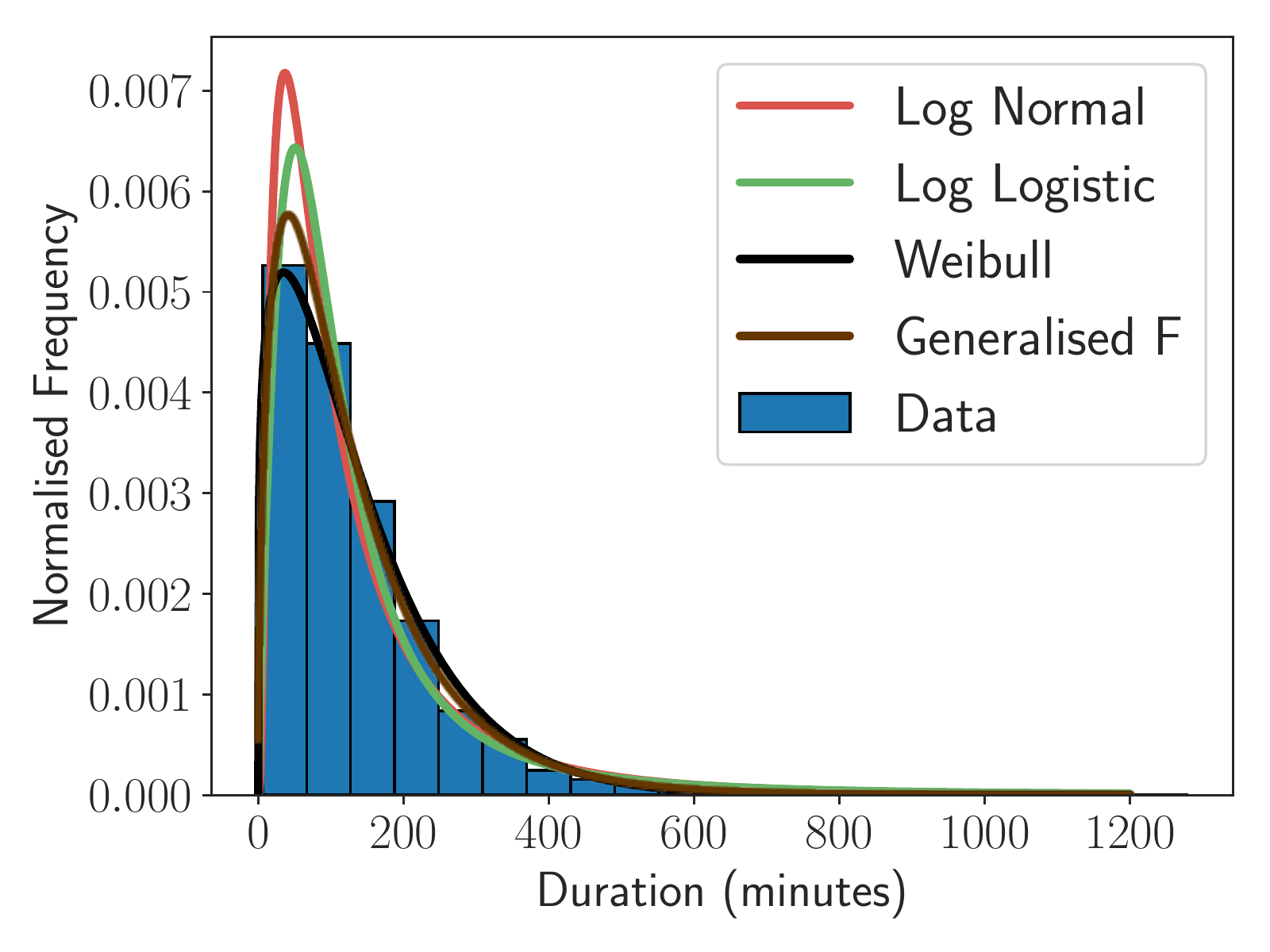}
		\centering \textbf{A}
	\end{minipage}
	~
	\begin{minipage}[t]{.47\textwidth}
		\includegraphics[width=\textwidth]{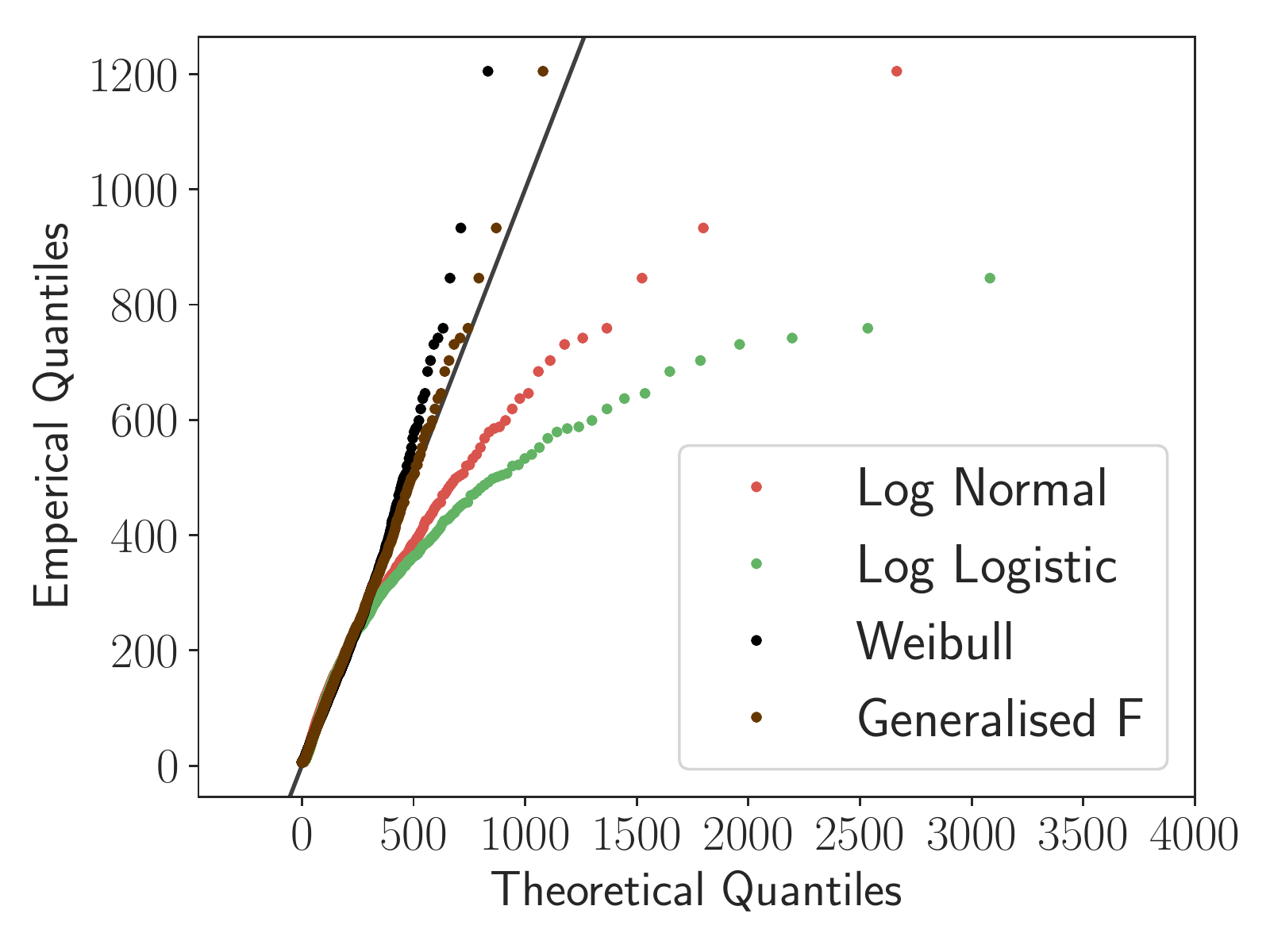}
		\centering \textbf{B}
	\end{minipage}
	\caption[Example Weekly Baseline]{Analysis of various distributions fit to incident duration data across the M25. Overall, the Weibull distribution fits the bulk of the data most closely, but the heavy tail is best captured by the generalized F distribution. \textbf{(A)}: Comparison of probability density functions. \textbf{(B)}: Quantile-Quantile plot. }\label{fig:DurationDistribution}
\end{figure}
From \textbf{Figure 9A}, we see that the Weibull PDF follows the data most closely, but inspecting the quantiles we see only the generalized F distribution has reasonable estimates of the extreme values.
When we apply a Kolmogorov-Smirnov with estimated parameters \cite{goodness_of_fit_tests_when_parameters_are_estimated}, we see no significant evidence that any of these candidate distributions are statistically valid fits.
Note that in \cite{forecasting_the_clearance_time_of_freeway_accidents}, a chi-squared test was applied to determine the statical significance of distributional fits, however doing so requires binning the data, which we avoid here.
These results suggest that to properly describe the data, we may want to consider more complex distributional representations, either non-parametric or mixtures of many components.
Note also that the tail of the durations contains around 6 outliers, suggesting very rare incidents where the link was at a reduced speed for almost an entire day, however the bulk of the incidents attain a duration of less than 300 minutes.

\subsection{Incident Clustering Analysis}\label{sec:ExploratoryAnalysisClustering}

We saw throughout the literature that time-invariant features were sometimes clustered to reveal different sub-populations of incidents, and provided useful features in modelling.
This was particularly evident in \cite{cluster_based_lognormal_distribution_model_for_accident_duration} and \cite{a_comparative_study_of_knn_and_hazard_based_models}, for example.
We question if these clusters are reflected also in the time-series for the data.
To do so, we require a distance metric between time-series.
A common choice is dynamic time warping (DTW) discussed in \cite{using_dynamic_time_warping_to_find_patterns_in_time_series} and used in, for example, \cite{fuzzy_clustering_of_time_series_data_using_dynamic_time_warping_distance}.
We specifically use the implementation in \cite{tsclust_package}.
The idea is to stretch or squeeze a pair of time-series such that they are as similar as possible. 
If one time-series has the same shape as another, it will have a low DTW distance, where as if two are fundamentally different, they will have a large distance.
We first standardize all link data by it's mean and standard deviation, then compute the DTW distance between all pairs of series that represent a window where an incident occurred.
From this, we attain a distance matrix, and we use hierarchical agglomerative clustering (HAC) with a ward linkage function \cite{hierarchical_grouping_to_optimize_an_objective_function} to construct a dendrogram to visualize distances between elements. 
The results are shown in \textbf{Figure 10}, where \textbf{Figure 10A} clusters the data by their time-invariant features and \textbf{Figure 10B} clusters the data by their time-series.
\begin{figure}[ht!]
	\centering
	\begin{minipage}[t]{.47\textwidth}
		\includegraphics[width=\textwidth]{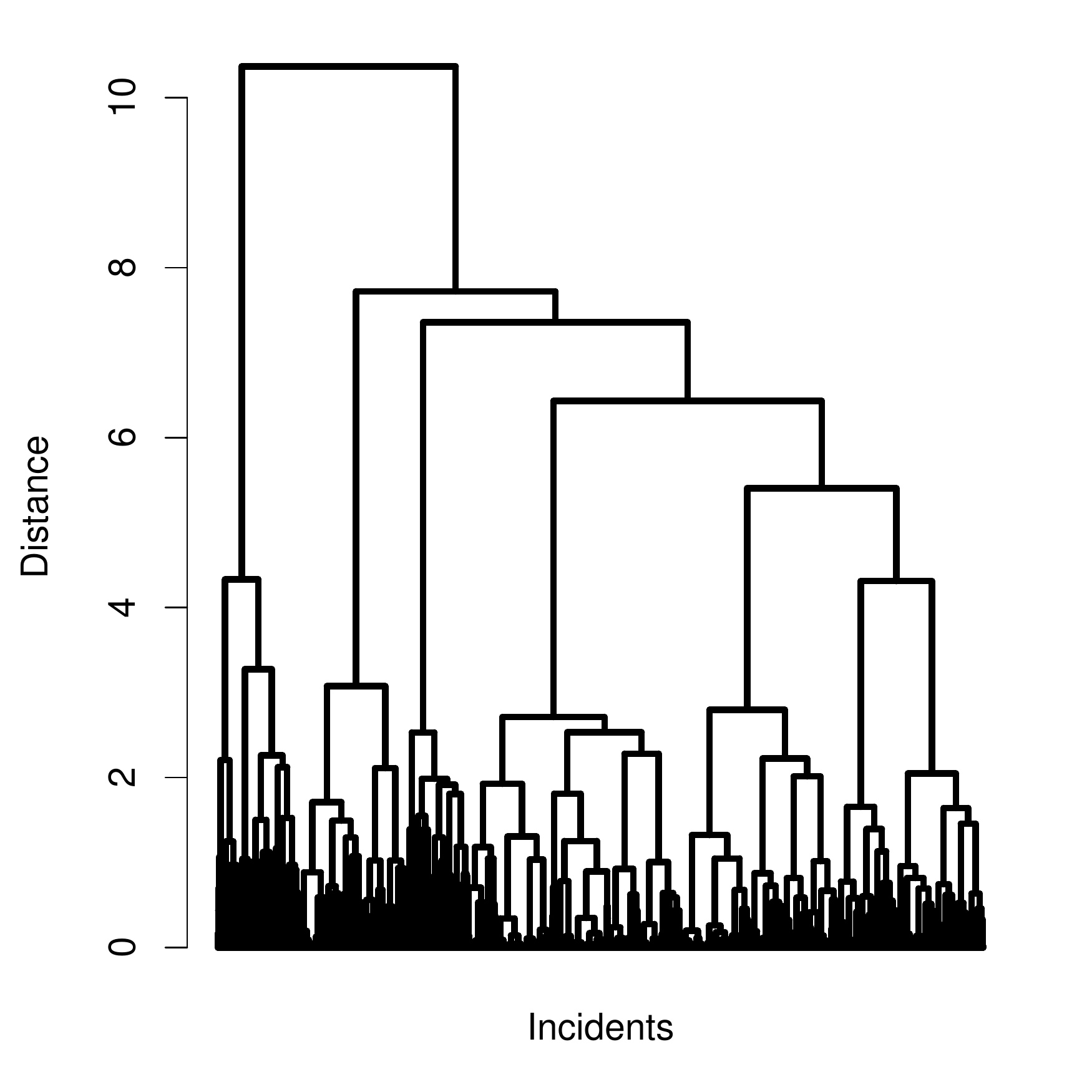}
		\centering \textbf{A}
	\end{minipage}
	~
	\begin{minipage}[t]{.47\textwidth}
		\includegraphics[width=\textwidth]{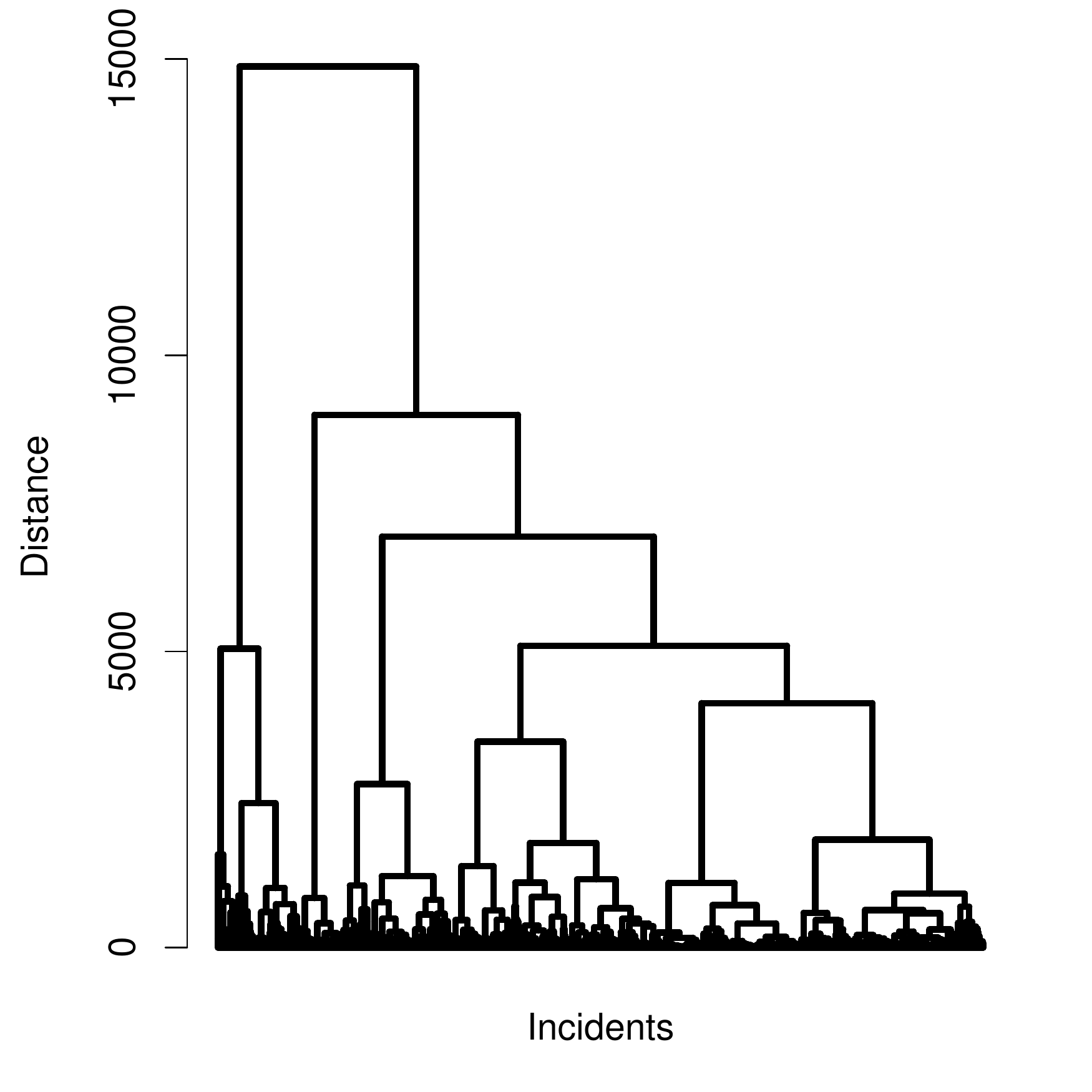}
		\centering \textbf{B}
	\end{minipage}
	\caption[Example Weekly Baseline]{Clustering Dendrogram using all incidents in the NTIS data. We see clear structure, perhaps suggesting 4 distinct groups exist. \textbf{(A)}: Clustering of incidents using the time-invariant features. \textbf{(B)}: Clustering of incidents using the speed time-series. }\label{fig:ClusteringDendrogram}
\end{figure}
From \textbf{Figure 10}, there is clear structure in the distance matrix, suggesting that incidents cluster by both time invariant features, but also by the observed traffic metrics on the links.
The number of clusters is somewhat subjective, but if we inspect the average within cluster sum of squared distances, we see that the reduction in this begins to diminish after around 6 clusters are identified for both clustering instances. 
Examples of different series observed in incident windows are given in the supplementary material, where we observe differences in the severity, symmetry, time-scales and stability of speed behaviour during incidents, and more details on the clustering discussed.

\subsection{Clustering Details}\label{sec:AppendixClustering}

Whilst we saw that the dataset appeared to cluster on both the time-series and the time-invariant features, we offer more explanation of the clustering here.
When clustering the data by their time-invariant features, we use the `daisy' dissimilarity measure \cite{finding_groups_in_data} as it is able to handle features of varying types.
To do so, it uses the Gower dissimilarity coefficient \cite{a_general_coefficient_of_similarity_and_some_of_its_properties}. 
Inspecting the average within cluster sum of squares distance values on both clustering results, we construct a so called `elbow plot' shown in \textbf{Figure 11}.
\begin{figure}[ht!]
	\centering
	\begin{minipage}[t]{.48\textwidth}
		\includegraphics[width=\textwidth]{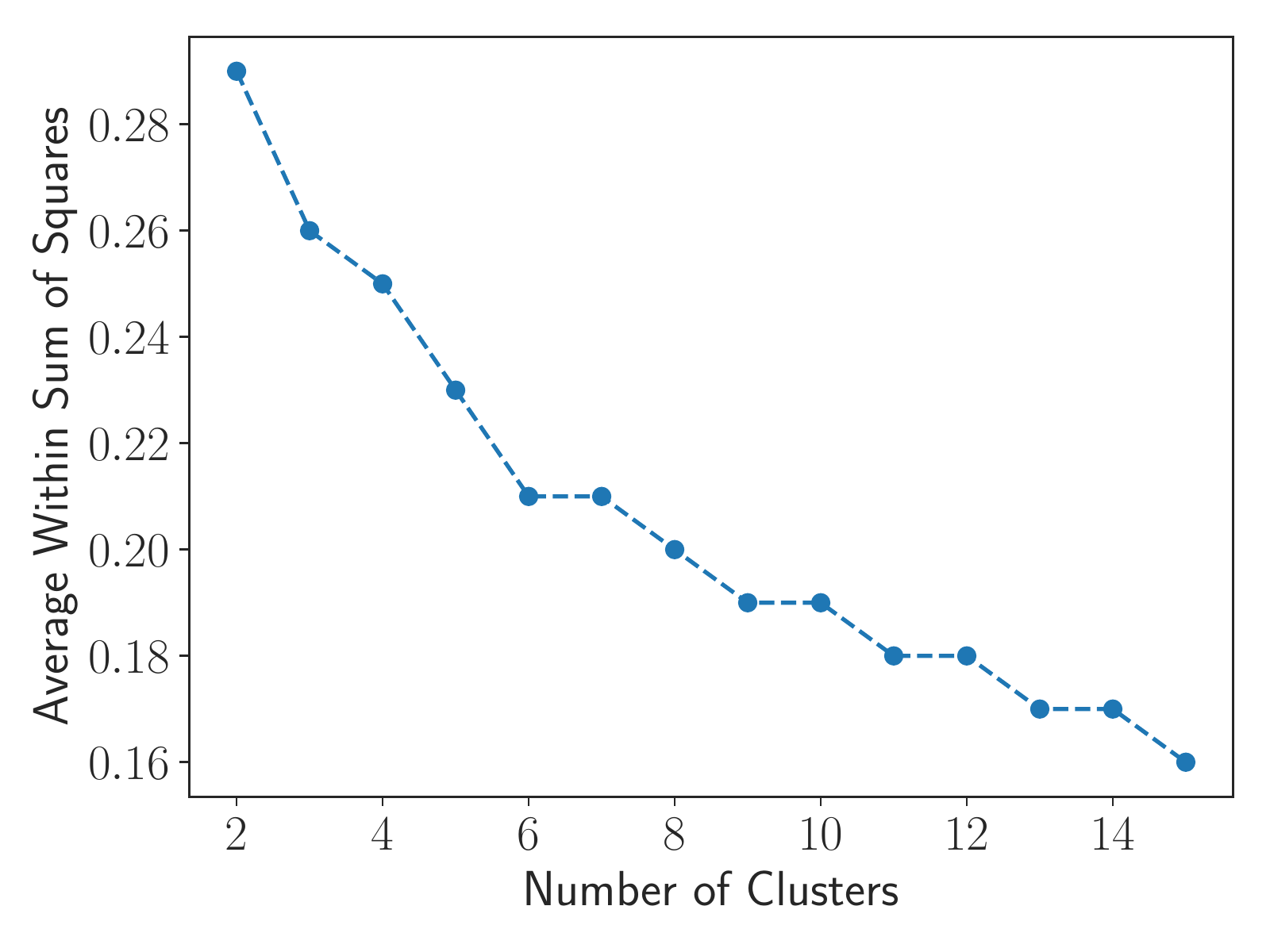}
		\centering \textbf{A}
	\end{minipage}
	~
	\begin{minipage}[t]{.48\textwidth}
		\includegraphics[width=\textwidth]{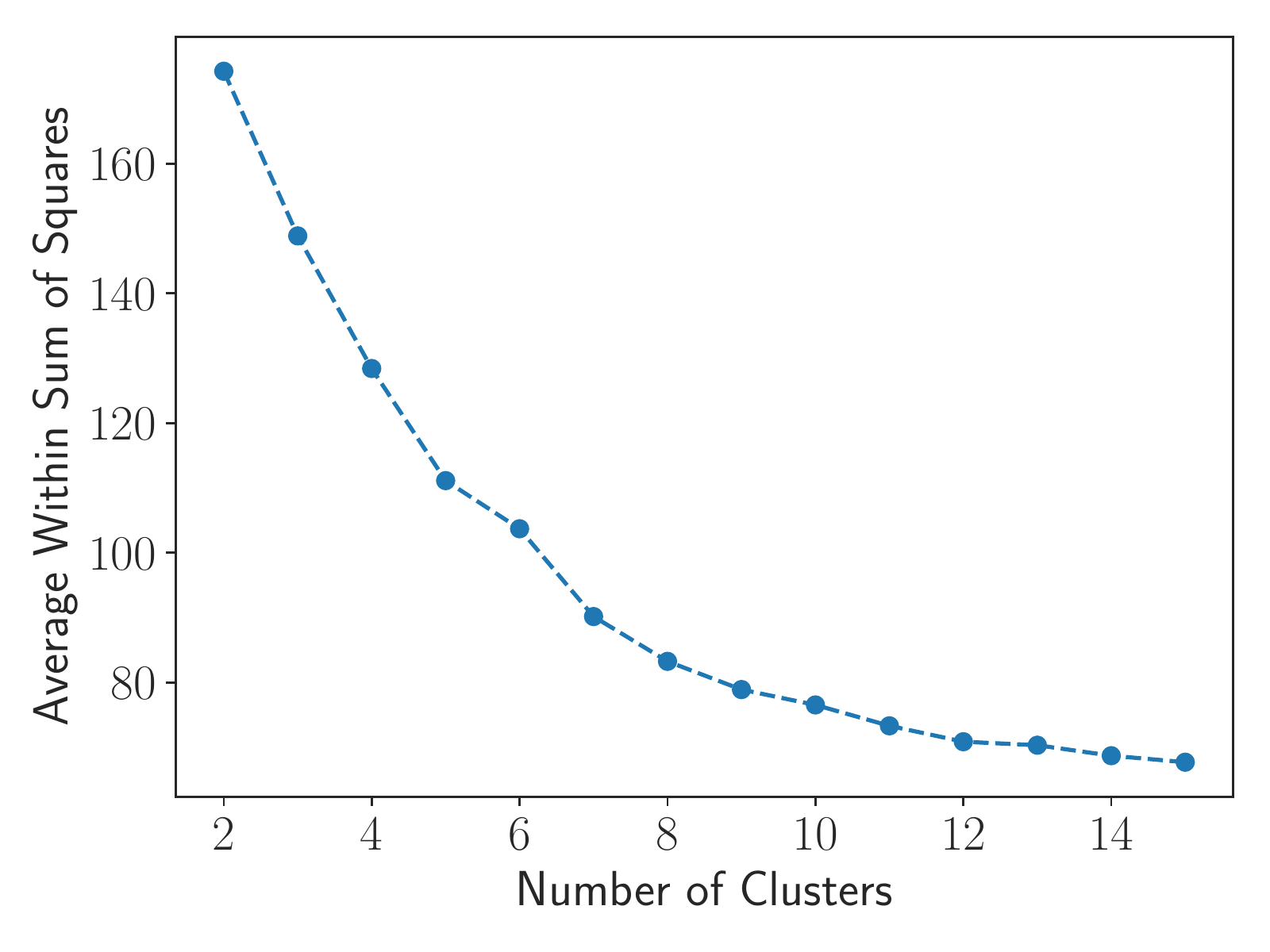}
		\centering \textbf{B}
	\end{minipage}
	\caption{Plots comparing the within cluster sum of squared distances when cluster on fixed and time-series features. Distance measures used are very different, so the numerical values on the y-axis should not be compared. \textbf{(A)}: Time invariant features, \textbf{(B)}: Time-series features }\label{fig:ClusteringWSS}
\end{figure}
From \textbf{Figure 11} we see diminishing reduction in the within cluster sum of squared distances after around 6 clusters in each case, suggesting that one should not segment the data further than this.

\section{Models Summary}\label{sec:AppendixModelSummary}

In \textbf{Table 7}, we summarize the models considered and note some key points one should consider when comparing them.	
They have been chosen to reflect existing work in the transportation literature and recent advancements in other disciplines.
\begin{table}[ht!]
  \centering
  \caption{Summary of models considered with key points and origins highlighted.}\label{table:ModelsSummary}
  \resizebox{\textwidth}{!}{%
  \renewcommand{\arraystretch}{1.25}\begin{tabular}{|c|c|c|l|}
  \hline
  Model     & \makecell{Distributional \\ Assumptions} & Complexity & \multicolumn{1}{c|}{Notes} \\
  \hline
  AFT (LN)    & Log-normal            & Linear     &  \makecell[l]{Accelerated Failure Time model assuming a log-normal \\ duration distribution, used in previous traffic studies, for \\ example \cite{analytical_method_to_estimate_accident_duration_using_archived_speed_profile_and_its_statistical_analysis}.} \\
  \hline
  AFT (W)     & Weibull               & Linear     & \makecell[l]{Accelerated Failure Time model assuming a Weibull duration \\ distribution, used in previous traffic studies, for example \\ \cite{an_exploratory_hazard_based_analysis_of_highway_incident_duration} and \cite{response_time_of_highway_traffic_accidents_in_abu_Dhabi_investigation_with_hazard_based_duration_models}. } \\
  \hline
  Cox       & Non-Parametric        & Linear     & \makecell[l]{Cox regression model, baseline hazard is non-parametric, \\ covariate effects act as an exponential of a linear regression. \\ Used in previous traffic studies such as \\ \cite{determination_of_the_risk_factors_that_influence_occurence_time_of_traffic_accidents_with_survival_analysis} and \cite{the_effect_of_earlier_or_automatic_collision_notification_on_traffic_mortality_by_survival_analysis}. } \\
  \hline
  RSF       & Non-Parametric        & Non-Linear & \makecell[l]{Random survival forest model, using an ensemble of 1000 trees. \\ Used in various other applications such as \cite{random_survival_forest_in_practice} \\ and \cite{is_random_survival_forest_an_alternative_to_cox_proportional_model}.} \\
  \hline
  NN (NP)     & Non-Parametric        & Non-Linear & \makecell[l]{Neural network model with a non-parametric distribution \\ modelled by a softmax output layer. Used in healthcare \\ applications in \cite{deephit_a_deep_learning_approach_to_survival_analysis_with_competing_risks}.} \\
  \hline
  NN (LN)     & \makecell{Mixture of \\ log-normal} & Non-Linear & \makecell[l]{Neural network model parametrizing a mixture distribution, \\ with log-normal components for each mixture component. \\ Mixture used due to work in \cite{application_of_finite_mixture_models_for_analysing_freeway_incident_clearance_time} suggesting \\ these may offer predictive power for traffic incident datasets.} \\
  \hline
  NN (Kernel) & \makecell{Non-Parametric \\ with smoothing} & Non-Linear  & \makecell[l]{Neural network model with a kernel smoothed output \\ distribution, compromising between the strong assumption of \\ NN (LN) and the complete freedom of NN (NP). } \\
  \hline
  SW (NP)     & Non-Parametric        & Non-Linear & \makecell[l]{Neural network model with a sliding window CNN architecture \\ to capture time-series features. Output is non-parametric as in \\ NN (NP). Used in healthcare applications in \cite{dynamic_prediction_in_clinical_survival_analysis_using_temporal_convolutional_networks}.} \\
  \hline
  SW (LN)     & \makecell{Mixture of \\ log-normal} & Non-Linear & \makecell[l]{Neural network model as in SW (NP), with an output layer \\ specifying a mixture of log-normals as in NN (LN). Combines \\ ideas as in \cite{dynamic_prediction_in_clinical_survival_analysis_using_temporal_convolutional_networks} with mixture observations in \\ \cite{application_of_finite_mixture_models_for_analysing_freeway_incident_clearance_time}. } \\
  \hline
  SW (Kernel) & \makecell{Non-Parametric \\ with smoothing} & Non-Linear  & \makecell[l]{Neural network model as in SW (NP), with an output layer \\ specifying a  kernel smoothed distribution. This compromises \\ between the strong assumption of SW (LN) and the complete \\ freedom of SW (NP). } \\
  \hline
  \end{tabular}%
  }
\end{table}

\section{Hyper Parameter Grids}\label{sec:AppendixHyperParameterSearch}

For each of our deep learning models, we consider a range of hyper-parameters and apply 100 iterations of random search to determine optimal choices.
In \textbf{Table 8} we detail the possible choices for various hyper-parameters.
For all fitting, we apply early stopping based on the loss on a holdout set, and we train using a batch size of 128.
\begin{table}[ht]
  \centering
  \caption{Hyper parameters considered when fitting neural network models.}\label{table:HyperParamSearch}
  \begin{tabular}{|c|c|}
  \hline
  Hyper Parameter               & Values Considered          \\
  \hline
  Dropout                       & 0.5                        \\
  \hline
  Kernel size                   & \{5, 10\}                  \\
  \hline
  $L_1, L_2$ regularization     & $10^{-4}$                  \\
  \hline
  Learning rate                 & \{ $10^{-4}$, $10^{-2}$ \} \\
  \hline
  Number of dense layers        & \{1, 2, 3\}                \\
  \hline
  Number of neurons per dense layer   & \{32, 64, 128, 256\} \\
  \hline
  Number of mixtures            & \{1, 2, 3\}                \\
  \hline
  Number of CNN layers          & \{1, 2, 3\}                \\
  \hline
  Number of filters per layer   & \{4, 8, 16\}               \\
  \hline
  Smoothing bandwidth           & 3 (minutes)                \\
  \hline
  Time-Series window size       & \{30, 60\} (minutes)       \\
  \hline
  $\eta_\sigma$                 & \{0.1, 1.0\}               \\
  \hline
  \end{tabular}
\end{table}
Learning rate for all models was decayed when learning stalled, and the Adam optimizer was used. 
Models are implemented in Keras.

\section{Joint Models}\label{sec:AppendixJointModels}

Here we discuss the main alternative for dynamic predictions in survival analysis one can find in the literature: joint models.
A more detailed discussion for interested readers can be found in \cite{joint_models_for_longitudinal_and_time_to_event_data_with_applications_in_R}.
The primary idea behind joint modelling is to couple two models, a linear mixed effects model for the time-series available, and a survival model for the incident times.
A general linear mixed effects model is written as: 
\begin{equation}
\begin{split}
y_i        &= \bm{\tilde{x}}_i'\bm{\tilde{\beta}} + \bm{z}_i'\bm{b}_i + \epsilon_i \\
b_i        &\sim \mathcal{N}\left( 0, \bm{D} \right) \\
\epsilon_i &\sim \mathcal{N}\left( 0, \sigma^2 \bm{I}_{n_i} \right). \\
\end{split}
\end{equation}
Here, $\bm{\tilde{\beta}}$ describes the average longitudinal evolution through time of the population (fixed effects) and $\bm{b}_i$ represents the deviation for a particular individual $i$ from the population average (random effects).
The error terms are assumed to be Gaussian with covariance matrix $\bm{D}$, and the random effect regression coefficients are assumed to be Gaussian distributed with some fixed variance $\sigma^2$ across the population.
Finally $\bm{\tilde{x}}_i$ and $\bm{z}_i$ are covariates of interest, repeatedly measured through time.

The survival model used in joint models is commonly a Cox model, meaning the full specification of the `basic' joint model can be written as:
\begin{equation}\label{equ:JointModel}
\begin{split}
h_i(t | \mathcal{M}_i(t), w_i ) &= h_0(t)e^{\bm{x}_i'\bm{\beta} + \alpha m(t) } \\
y_i(t)                          &= m_i(t) + \epsilon_i(t) \\
m_i(t)                          &= \bm{\tilde{x}}_i'(t)\bm{\tilde{\beta}} + \bm{z}_i'(t)\bm{b}_i  \\
\bm{b_i}   &\sim \mathcal{N}\left( 0, \bm{D} \right) \\
\epsilon_i &\sim \mathcal{N}\left( 0, \sigma^2 \right).
\end{split}
\end{equation}
where $\mathcal{M}_i(t)$ represents the history of the longitudinal process up to time $t$, and the covariates impact in the cox model are augmented by the term $\alpha m_i(t)$, representing the influence of the longitudinal data.
Given Eq.~(\ref{equ:JointModel}), we can incorporate the impact of time-varying covariates and uncertainty into the survival model, however we are really interested in dynamic predictions: determining the probability of surviving to a particular survival time $u > t$, conditioned on surviving up to $t$.
It is detailed in \cite{joint_models_for_longitudinal_and_time_to_event_data_with_applications_in_R} how to do this, requiring Monte Carlo sampling to integrate over all possible random effect values.

We do not consider such a model for comparison in this work for a number of reasons.
Firstly, particularly in extreme accident scenarios, the behaviour of traffic metrics is likely highly non-linear as a function of the known covariates.
Secondly, we would be required to specify the model for the multivariate time-series, the survival outcome and how to link to the two models when applying this, leaving significant room for model misspecification to accumulate without exhaustive searches.
Doing so without significant prior work to base our assumptions on in not desirable.
Finally, the computational complexity is prohibitive for a dataset of the size we are considering. 
Further discussions on the challenges of joint models are given in \cite{joint_modelling_of_time_to_event_and_multivariate_longitudinal_outcomes_recent_developments_and_issues}.

\section{Comparison of Temporal Convolutions to Manually Engineered Time-Series Values}\label{sec:AppendixRawVsSW}

In \textbf{Table 9} we show the C-index and Brier score comparing a sliding window model that engineers time-series features through temporal convolutions to a model that simply takes the most recent information (the level and gradient).
The later model is denoted `Raw (NP)' here as it does not derive features itself from the series, and has a non-parametric output. 
The same two models are compared in \textbf{Table 10}, showing instead the MAPE at various percentiles into an incident.
\begin{table}[ht!]
\centering
  \caption{ Results comparing a sliding window model and a model where we input the most recent time-series values and the gradient computed as previously discussed. The sliding window model achieves better C-Index and Brier values across all considered prediction time and horizon pairs. }\label{table:CindexBrierRawVsSW}
  \resizebox{\textwidth}{!}{\begin{tabular}{|c|c|c|c|c|c|c|c|c|c|c|c|}
    \hline
    Prediction Time            & \multirow{2}{*}{Metric} & \multirow{2}{*}{Model} & \multicolumn{8}{c|}{Prediction Horizon (minutes)} & Mean Over \\
    \cline{4-11}
    (minutes)				   & &            & 5 & 15 & 30 & 45 & 60 & 120 & 180 & 240 & Horizons \\
    \hline
    \multirow{4}{*}{$t = 0$}   & \multirow{2}{*}{C-Index} & SW (NP)  & - & \textbf{0.798} & \textbf{0.743} & \textbf{0.705} & \textbf{0.682} & \textbf{0.642} & \textbf{0.637} & \textbf{0.634} & \textbf{0.692} \\ 
							   &                          & Raw (NP) & - & 0.755 & 0.701 & 0.669 & 0.646 & 0.599 & 0.592 & 0.588 & 0.650 \\
 
    \cline{2-12}
                               & \multirow{2}{*}{Brier} & SW (NP)  & \textbf{0.007} & \textbf{0.082} & \textbf{0.160} & \textbf{0.213} & \textbf{0.262} & \textbf{0.290} & \textbf{0.195} & \textbf{0.113} & \textbf{0.165} \\ 
							   &                        & Raw (NP) & 0.015 & 0.122 & 0.208 & 0.260 & 0.306 & 0.310 & 0.203 & 0.116 & 0.193 \\
    \hline
    \multirow{4}{*}{$t = 15$}  & \multirow{2}{*}{C-Index} & SW (NP)  & \textbf{0.947} & \textbf{0.884} & \textbf{0.811} & \textbf{0.772} & \textbf{0.731} & \textbf{0.678} & \textbf{0.669} & \textbf{0.666} & \textbf{0.770} \\ 
							   &                          & Raw (NP) & 0.946 & 0.849 & 0.773 & 0.729 & 0.689 & 0.637 & 0.624 & 0.620 & 0.733 \\
    \cline{2-12}
                               & \multirow{2}{*}{Brier}  & SW (NP) & \textbf{0.019} & \textbf{0.069} & \textbf{0.134} & \textbf{0.180} & \textbf{0.230} & \textbf{0.259} & \textbf{0.170} & \textbf{0.096} & \textbf{0.145} \\ 
							   &                          & Raw (NP) & 0.022 & 0.091 & 0.163 & 0.210 & 0.257 & 0.272 & 0.175 & 0.098 & 0.161 \\
    \hline
    \multirow{4}{*}{$t = 32$}  & \multirow{2}{*}{C-Index} & SW (NP) & \textbf{0.960} & \textbf{0.905} & \textbf{0.803} & \textbf{0.761} & \textbf{0.733} & \textbf{0.689} & \textbf{0.681} & \textbf{0.677} & \textbf{0.776} \\ 
							   &                          & Raw (NP) & 0.948 & 0.870 & 0.773 & 0.730 & 0.702 & 0.655 & 0.645 & 0.641 & 0.746 \\
    \cline{2-12}
                               & \multirow{2}{*}{Brier} & SW (NP)  & \textbf{0.018} & \textbf{0.067} & \textbf{0.138} & \textbf{0.183} & \textbf{0.223} & \textbf{0.236} & \textbf{0.155} & \textbf{0.088} & \textbf{0.139} \\ 
							   &                        & Raw (NP) & 0.021 & 0.083 & 0.158 & 0.203 & 0.241 & 0.248 & 0.160 & 0.090 & 0.151 \\
    \hline
    \multirow{4}{*}{$t = 45$}  & \multirow{2}{*}{C-Index} & SW (NP)  & \textbf{0.967} & \textbf{0.880} & \textbf{0.813} & \textbf{0.784} & \textbf{0.749} & \textbf{0.703} & \textbf{0.692} & \textbf{0.688} & \textbf{0.785} \\ 
							   &                          & Raw (NP) & 0.953 & 0.851 & 0.777 & 0.748 & 0.715 & 0.667 & 0.654 & 0.651 & 0.752 \\
    \cline{2-12}
                               & \multirow{2}{*}{Brier} & SW (NP)  & \textbf{0.019} & \textbf{0.074} & \textbf{0.138} & \textbf{0.176} & \textbf{0.217} & \textbf{0.221} & \textbf{0.148} & \textbf{0.089} & \textbf{0.135} \\ 
							   &                        & Raw (NP) & 0.021 & 0.087 & 0.158 & 0.196 & 0.234 & 0.230 & 0.152 & 0.090 & 0.146 \\
    \hline
    \multirow{4}{*}{$t = 60$}  & \multirow{2}{*}{C-Index} & SW (NP) & \textbf{0.953} & \textbf{0.903} & \textbf{0.830} & \textbf{0.787} & \textbf{0.755} & \textbf{0.711} & \textbf{0.702} & \textbf{0.699} & \textbf{0.793} \\ 
							   &                          & Raw (NP) & 0.938 & 0.870 & 0.799 & 0.763 & 0.729 & 0.679 & 0.667 & 0.665 & 0.764 \\
    \cline{2-12}
                               & \multirow{2}{*}{Brier} & SW (NP)  & \textbf{0.023} & \textbf{0.073} & \textbf{0.136} & \textbf{0.176} & \textbf{0.212} & \textbf{0.210} & \textbf{0.139} & \textbf{0.089} & \textbf{0.132} \\ 
							   &                        & Raw (NP) & 0.025 & 0.087 & 0.153 & 0.192 & 0.228 & 0.221 & 0.144 & 0.091 & 0.143 \\
    \hline
    \multirow{4}{*}{$t = 120$} & \multirow{2}{*}{C-Index} & SW (NP)  & \textbf{0.968} & \textbf{0.896} & \textbf{0.852} & \textbf{0.824} & \textbf{0.791} & \textbf{0.744} & \textbf{0.739} & \textbf{0.735} & \textbf{0.819} \\ 
							   &                          & Raw (NP) & 0.958 & 0.847 & 0.795 & 0.773 & 0.745 & 0.697 & 0.692 & 0.689 & 0.775 \\
    \cline{2-12}
                               & \multirow{2}{*}{Brier} & SW (NP)  & \textbf{0.023} & \textbf{0.082} & \textbf{0.128} & \textbf{0.157} & \textbf{0.189} & \textbf{0.183} & \textbf{0.130} & \textbf{0.089} & \textbf{0.123} \\ 
							   &                        & Raw (NP) & 0.026 & 0.101 & 0.159 & 0.180 & 0.207 & 0.194 & 0.135 & 0.091 & 0.137 \\
	\hline
  \end{tabular}}
  \vspace{4mm}
  \caption{Results comparing a sliding window model and a model where we input the most recent time-series values and the gradient computed as previously discussed. The Sliding window model achieves better MAPE at the 30th and 50th percentiles, but marginally worse at the 70th and 90th percentiles.}\label{table:DynamicErrorAtPercentilesRawVsSW}
  \begin{tabular}{|c|c|c|c|c|}
    \hline
    \multirow{3}{*}{Model} & \multicolumn{4}{c|}{MAPE Dynamic Model} \\
    \cline{2-5}
                & \multicolumn{4}{c|}{Percentile Into Incident Prediction Made at}   \\
    \cline{2-5}
                & 30th   & 50th   & 70th   & 90th \\
    \hline
    SW (NP)     & \textbf{31.660} & \textbf{21.998} & 17.069 & 10.399 \\ 
    \hline
    Raw (NP)    & 33.922 & 23.140 & \textbf{16.762} & \textbf{10.134} \\
	\hline
  \end{tabular}
\end{table}

\section{SHAP Discussion}

Shapley values arise from a theory from coalition games, discussed at length in \cite{multiagent_systems_algorithmic_game_theoretic_and_logical_foundations}, stating that there must be a unique way to divide the pay-off between a set of players whilst satisfying reasonable constraints.
In the machine learning context, the 'game' becomes the model output, and the 'players' are the features.
To explain a complex model, in our case a neural network, one writes a simpler 'explainer model' of the form:
\begin{equation}\label{equ:SHAPExplainerModelAppendix}
g(z') = \phi_0 + \sum_{i=1}^M\phi_iz_i'.
\end{equation}
Here, $\phi_0$ represents the baseline model output, and then each feature $i$ shifts this output up or down, until a final value is reached, and we have $M$ features in total. 
This output can be measured for each neuron in the output layer.
Recall from the main text that une computes the SHAP value for feature $i$ as: 
\begin{equation}\label{equ:SuppTexShapValue}
\phi_i =\sum_{S \subseteq \mathcal{M} \text{\textbackslash} {i} }  \frac{|S|!(|M| - |S| - 1)!}{M!} \left[ F(S \cup \{i\}) - F(S) \right]
\end{equation} 
where $\mathcal{M}$ is the set of all features.
We can intuitively consider how Eq.~(\ref{equ:SuppTexShapValue}) arises as follows.
Imagine at first we have an empty feature vector.
Then, at random, we start adding features to it.
Whenever we happen to add feature $i$, the change in output will be:
\begin{equation}\label{appendix:equ:ShapExplain1}
F(S \cup \{i\}) - F(S)
\end{equation} 
Now consider, at the point we add $i$ into $S$, how many ways could $S$ have formed into its current state? 
There are $|S|!$ ways this could have happened.
Further, there are $(|N| - |S| - 1)!$ ways that the remaining features could be added after $i$ has been.
This suggests for a single instance we have:
\begin{equation}\label{appendix:equ:ShapExplain2}
|S|!(|N| - |S| - 1)! \left[ v(S \cup \{i\}) - v(S) \right]
\end{equation} 
However, this is just for a single $S$, so we then sum over all possible sets $S$ and then average the value by dividing by the the total number of possible orderings of players.
Combining all of this, we attain Eq.~(\ref{equ:SuppTexShapValue}).

However, there is a clear problem in that neural networks cannot take arbitrary missing features as an input, so using Eq.~(\ref{equ:SuppTexShapValue}) alone one cannot compute the SHAP values for such models.
To address this, one considers some `reference' or `background' dataset, from which values are taken as replacements when considering alterations of feature vectors.
Such an approach is discussed in \cite{a_unified_approach_to_interpreting_model_predictions}.
The idea is that, since we cannot omit a feature fully, we instead set its value to a reasonable reference value.
The appropriate choice of background is an open problem in applying these methods.
For image tasks, it might be clear than one can use a blank image as a reference, and then we are evaluating a pixels importance relative to it being blank.
However, in many domains some reasonable and intuitive reference value does not exist.
Instead, it is common to provide a dataset with many records in as the background, and average over it, which is the approach we take in our work.
Due to computational constraints, we set the background dataset to 10000 random instances from the training dataset, and compute the SHAP values for 1000 randomly selected points in the unseen data at specific prediction times.

A second complication of note is that there is often structure to feature vectors in problems, and elements are not simply a random collection of possible values for each entry.
One can consider this structure most clearly when considering a one-hot encoded example. 
Suppose our feature vector contained three binary values, which indicate morning, afternoon or evening, which of course correspond to a time of day being discretised and then one-hot encoded.  
In the raw methodology for SHAP in \cite{a_unified_approach_to_interpreting_model_predictions}, one would go through each feature, and consider setting it to a reference value and use the change in model output as a measure of importance. 
However, suppose we had our time of day encoding, and an entry occurred during the morning, meaning the subset of our feature vector was $[1, 0, 0]$.
We may then consider altering the entry corresponding to the binary label for afternoon, which might lead us to considering a feature vector of $[1, 1, 0]$.
Of-course, such an input is not practically possible, as a data-point can only come from one time of day, and so the model output for it is irrelevant to our problem.
To avoid this, recent work in \cite{shap_github} incorporates the structure in the data before performing any perturbations.
A partition of the data into highly correlated components is performed, and then when considering feature importance, instead of altering a single element, a set are altered if they are grouped together.
This is a natural requirement for our problem where locations, time of day and incident types have been one-hot encoded.
We do however note that we compute the SHAP values both incorporating this structure and neglecting it, and see similar results in both cases.  

\section{SHAP Values for Categorical Features}\label{sec:AppendixSHAPCategorical}

In the main text, we considered SHAP values as a measure for variable importance and gave example interpretations of why the neural network model was outputting particular predictions.
Here we plot further examples of what the categorical values contribute overall to the model.
In \textbf{Figure 12}, we show the results for the time of day variable.
\begin{figure}[ht!]
	\centering
	\begin{minipage}[t]{.48\textwidth}
		\includegraphics[width=\textwidth]{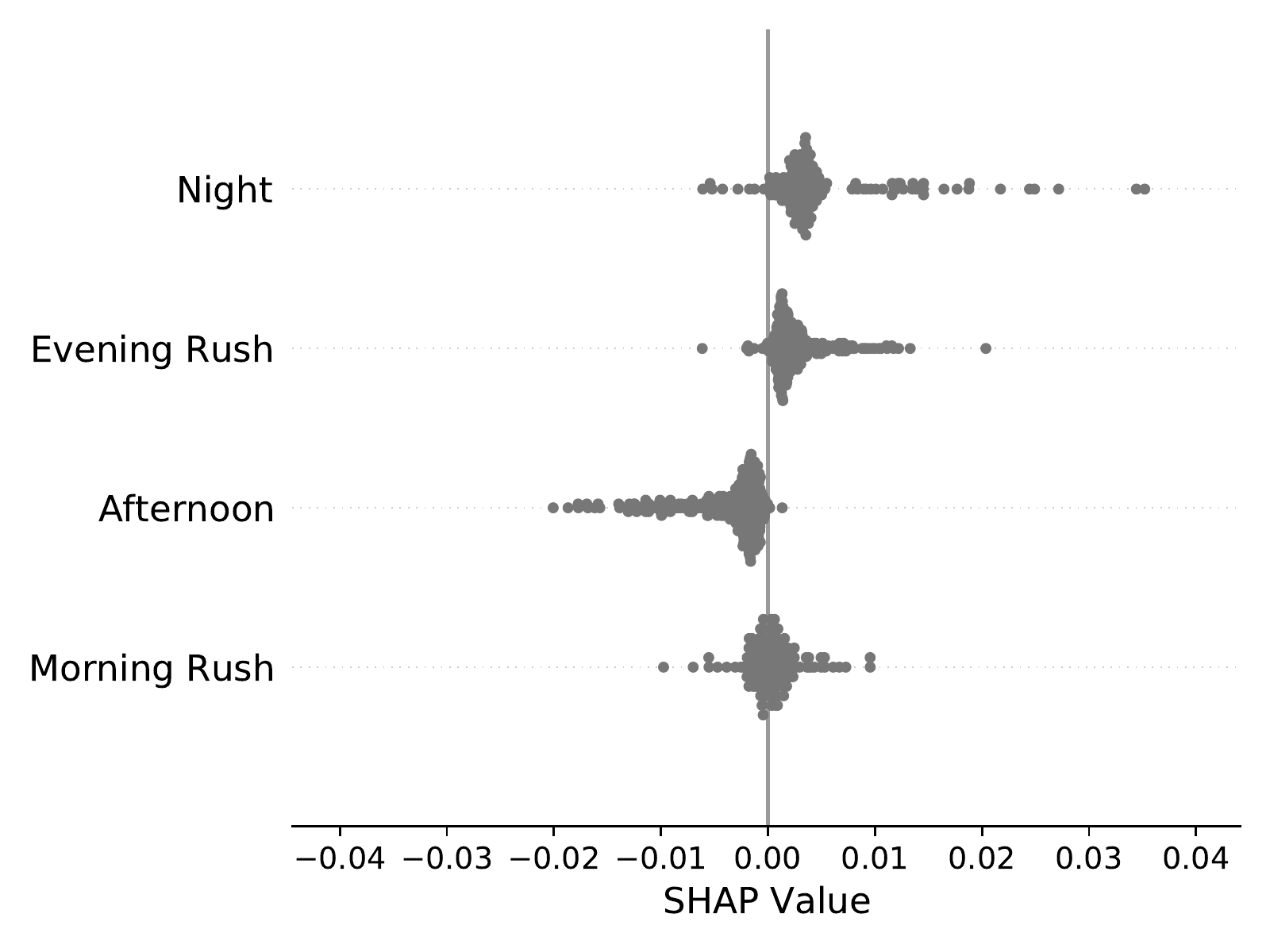}
		\centering \textbf{A}
	\end{minipage}
	~
	\begin{minipage}[t]{.48\textwidth}
		\includegraphics[width=\textwidth]{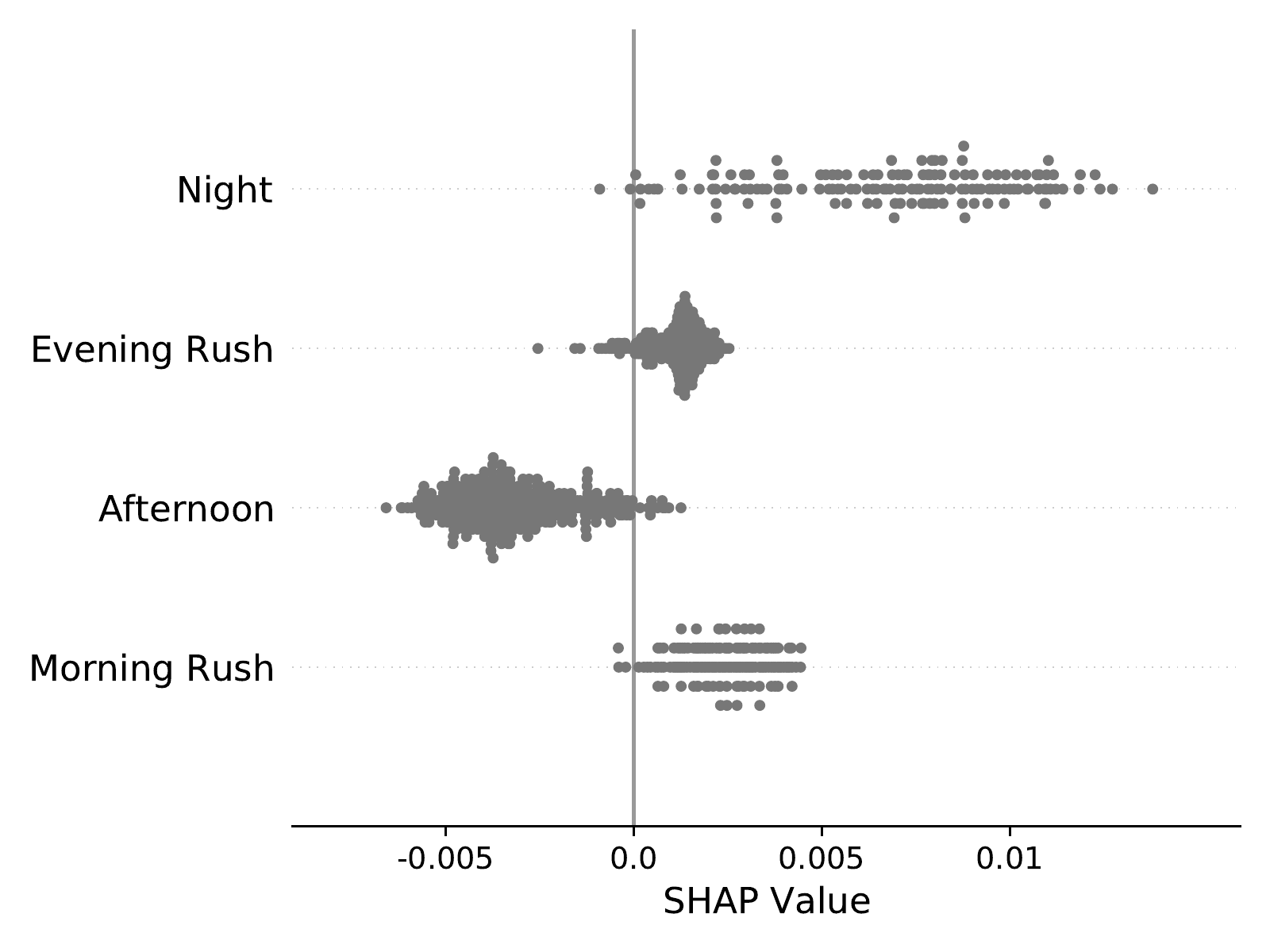}
		\centering \textbf{B}
	\end{minipage}

	\begin{minipage}[t]{.48\textwidth}
		\includegraphics[width=\textwidth]{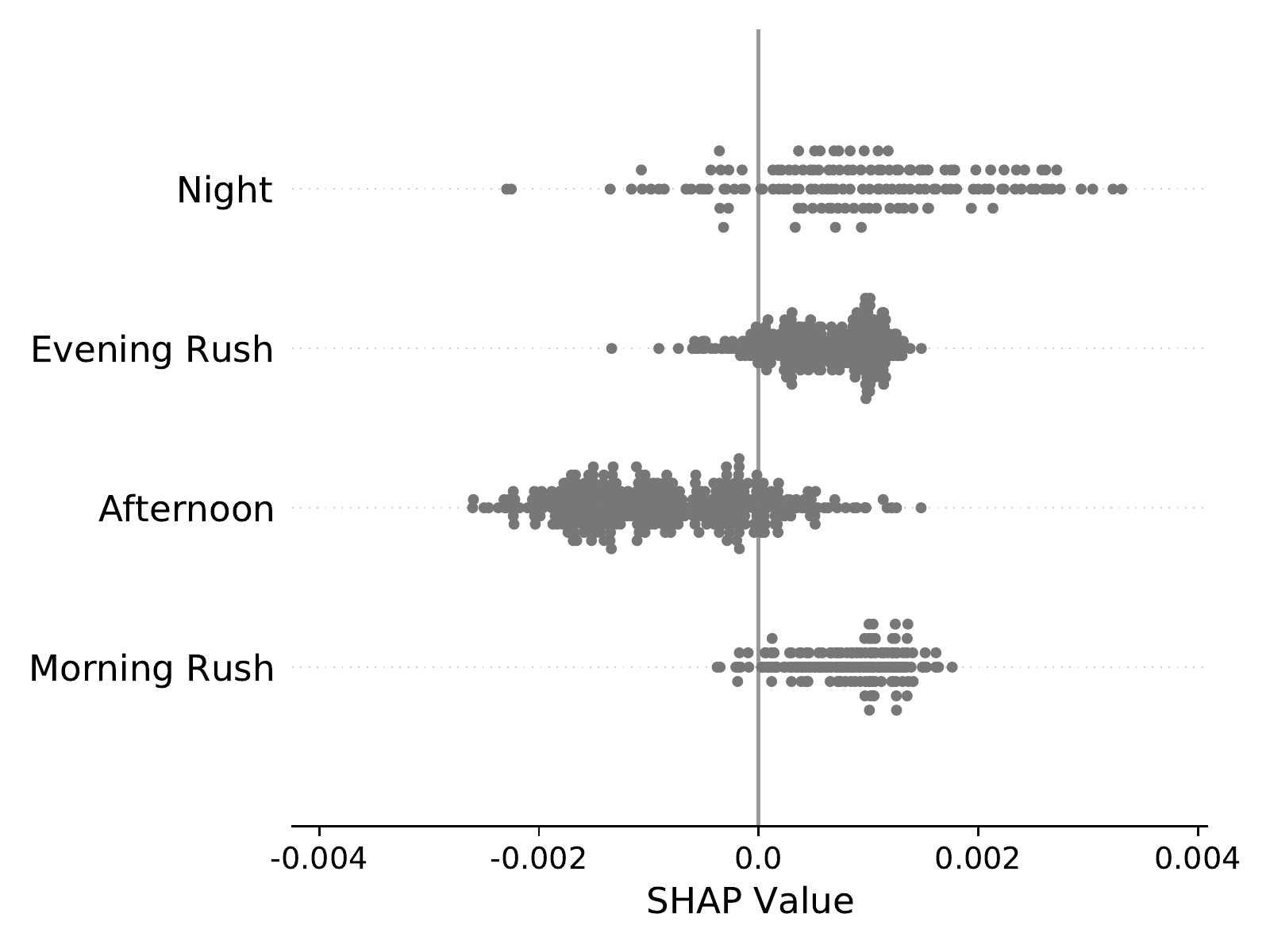}
		\centering \textbf{C}
	\end{minipage}
	~
	\begin{minipage}[t]{.48\textwidth}
		\includegraphics[width=\textwidth]{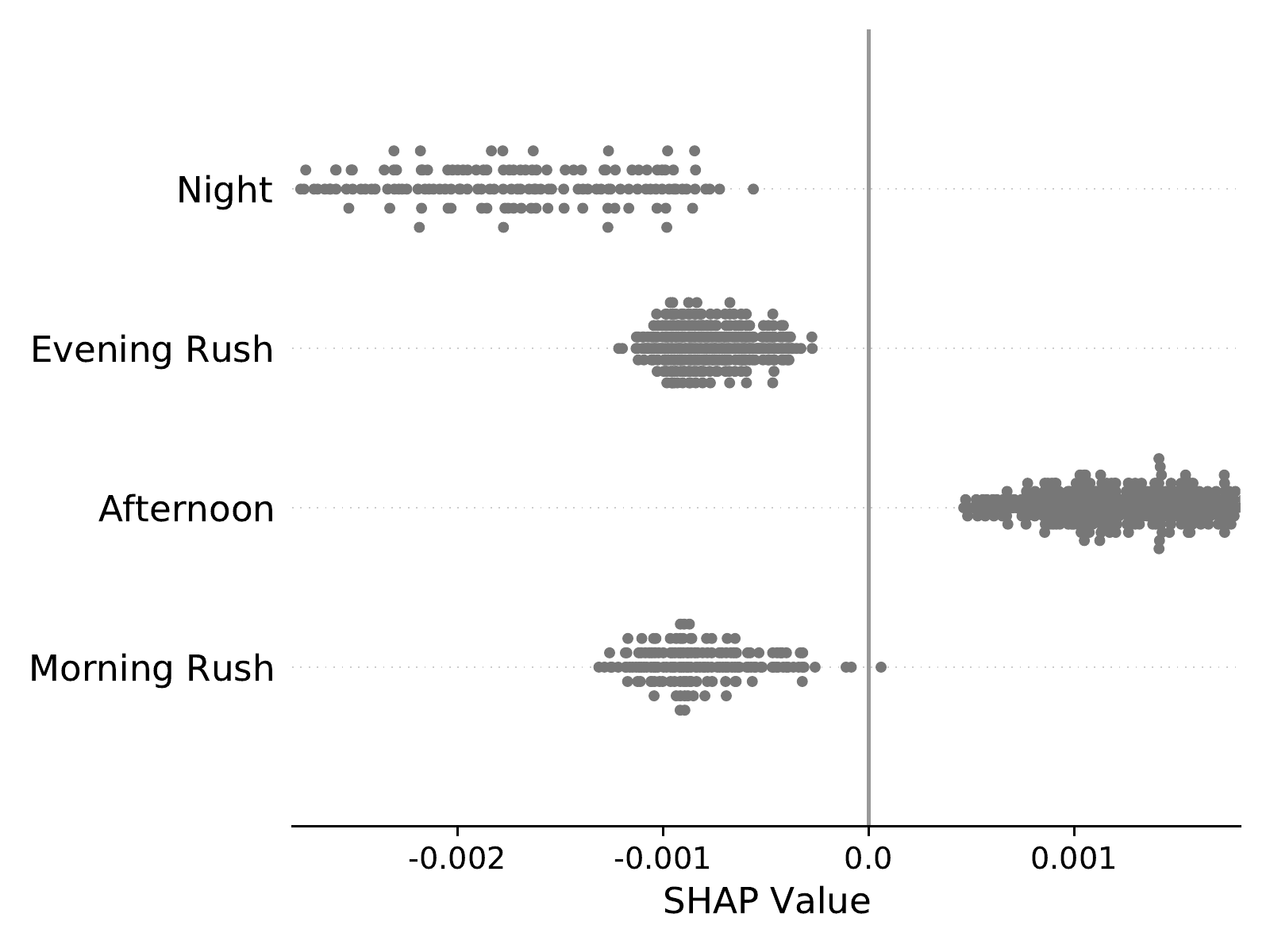}
		\centering \textbf{D}
	\end{minipage}

	\caption{SHAP values for the time of day feature. In each plot, we have summed the SHAP values for each one-hot encoded value, and then only plotted the resulting value in the row corresponding to each data-points true feature value. This shows the overall impact of the location feature, and allows one to view this impact separately for each value it attains. \textbf{(A)}: $h=5$, \textbf{(B)} $h=30$, \textbf{(B)} $h=60$, \textbf{(B)} $h=180$ }\label{fig:SHAPCategoricalTimeOfDay}
\end{figure}
Generally, we see that the impact of the time of day variable attaining the value `afternoon' pushes down predictions at short horizons (5, 30 minutes in \textbf{Figures 12A, 12B}).
At a horizon of 60 minutes (\textbf{Figure 12C}), we see that sometimes the model result is deceased and sometimes it is increased, and at a horizon of 180 minutes (\textbf{Figure 12D}) every instance had the model output increased.
From this we might be lead to believe that incidents occurring in the afternoon can spill over into the evening rush hour and last a significant amount of time.
The same analysis can be done for alternative times of day, for example we see that incidents in the night have the output increased at horizons of 5, 30 and 60 minutes (\textbf{Figures 12A, 12B, 12C}) and decreased at horizons of 180 minutes (\textbf{Figure 12D}).

If we consider instead the incident type, we see results as shown in \textbf{Figure 13}.
\begin{figure}[ht!]
	\centering
	\begin{minipage}[t]{.48\textwidth}
		\includegraphics[width=\textwidth]{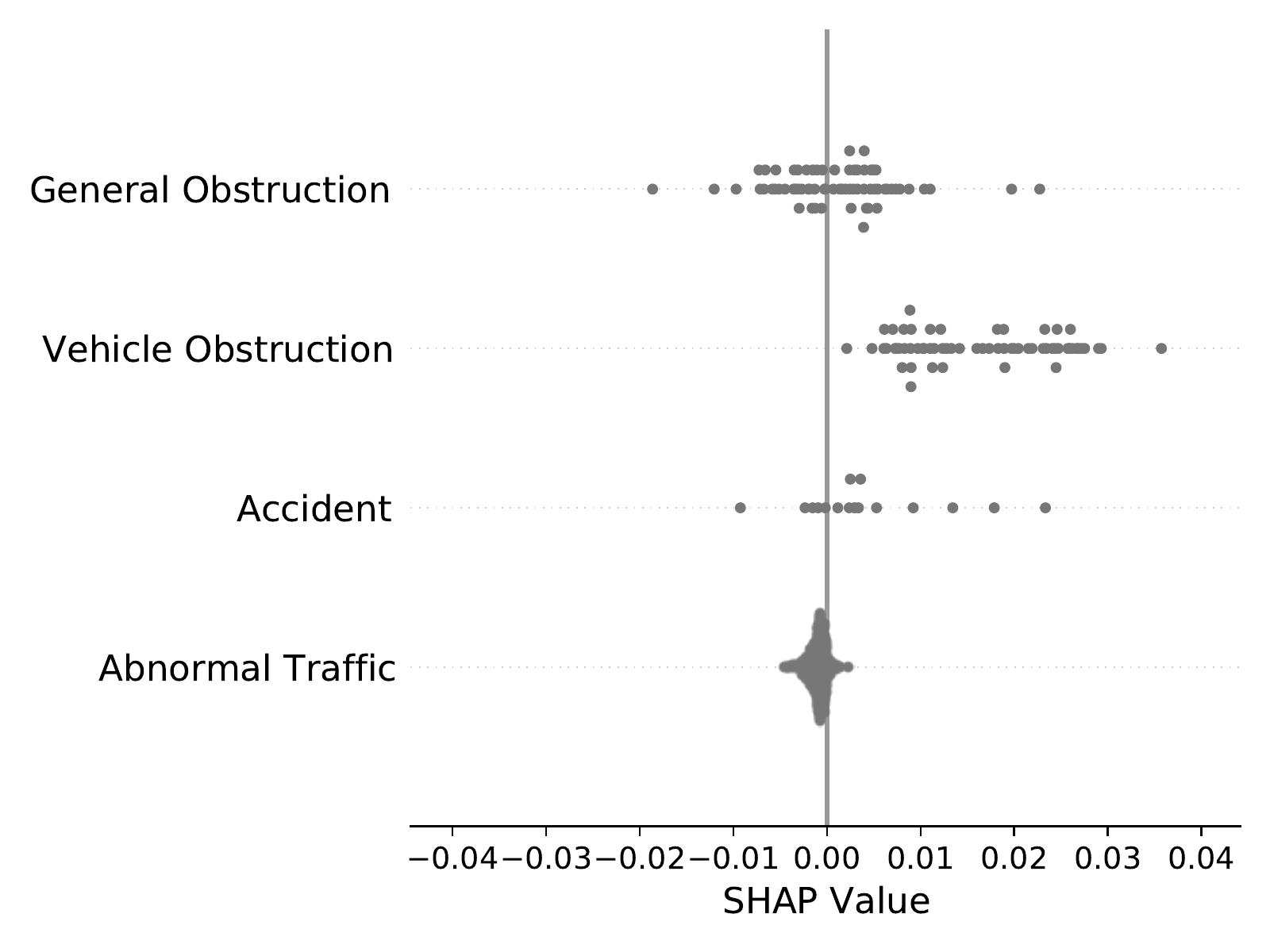}
		\centering \textbf{A}
	\end{minipage}
	~
	\begin{minipage}[t]{.48\textwidth}
		\includegraphics[width=\textwidth]{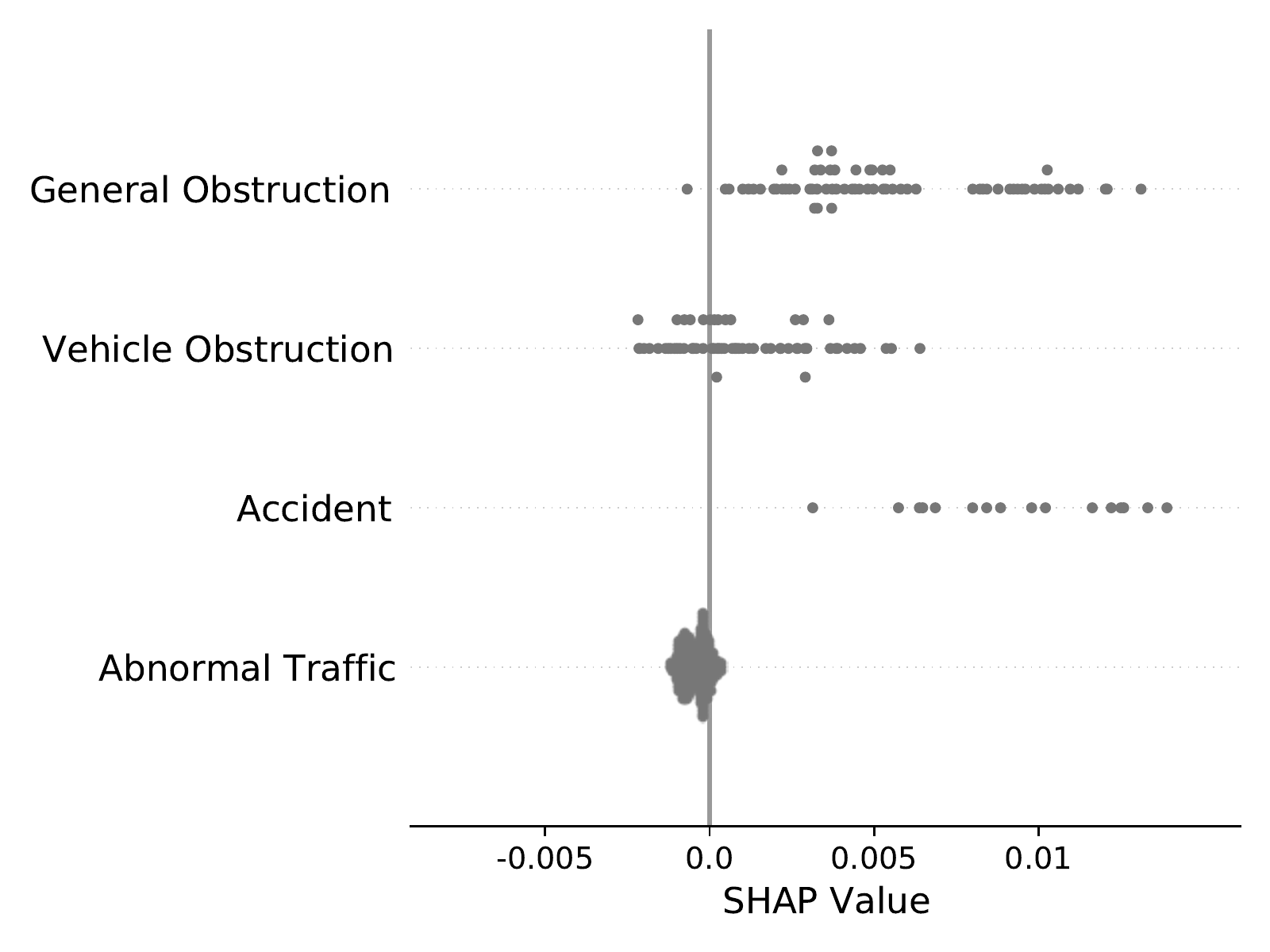}
		\centering \textbf{B}
	\end{minipage}

	\begin{minipage}[t]{.48\textwidth}
		\includegraphics[width=\textwidth]{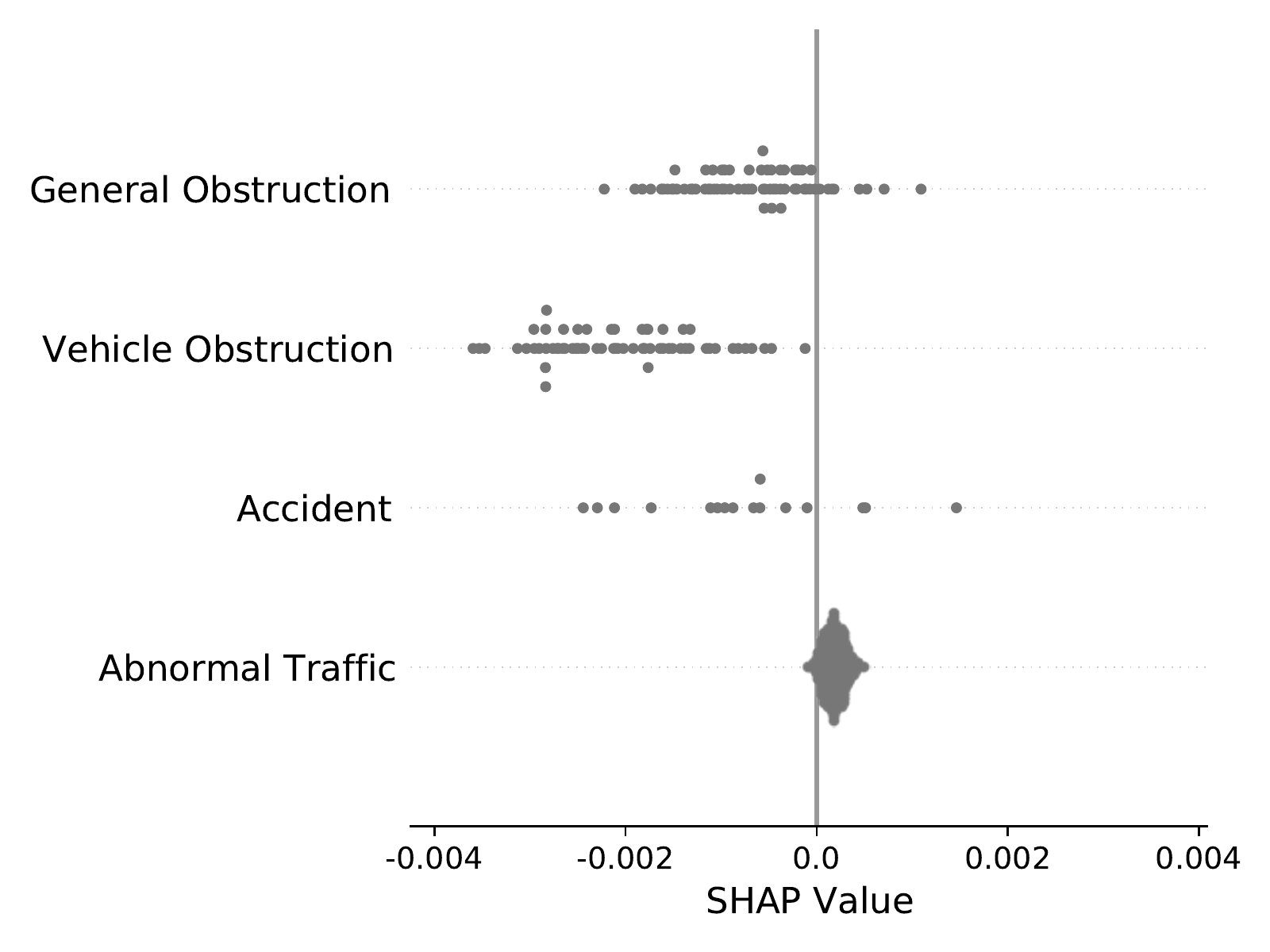}
		\centering \textbf{C}
	\end{minipage}
	~
	\begin{minipage}[t]{.48\textwidth}
		\includegraphics[width=\textwidth]{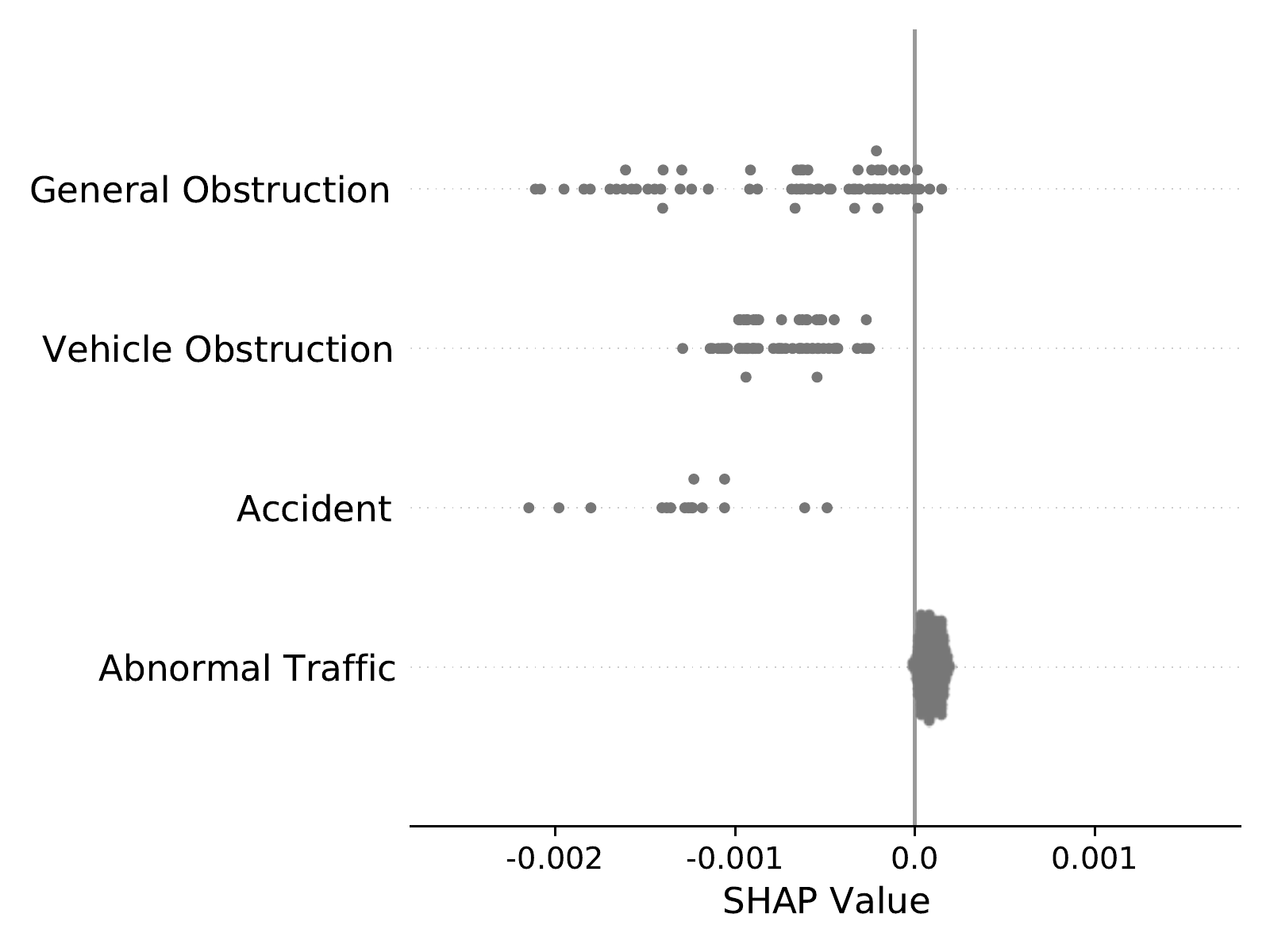}
		\centering \textbf{D}
	\end{minipage}

	\caption{SHAP values for the incident type feature. In each plot, we have summed the SHAP values for each one-hot encoded value, and then only plotted the resulting value in the row corresponding to each data-points true feature value. This shows the overall impact of the location feature, and allows one to view this impact separately for each value it attains. \textbf{(A)}: $h=5$, \textbf{(B)} $h=30$, \textbf{(B)} $h=60$, \textbf{(B)} $h=180$ }\label{fig:SHAPEventTypeSplit}
\end{figure}
We first note that the NTIS incidents data is dominated by abnormal traffic incidents, which is clearly seen by the random sample of 1000 incidents for explanation containing a large number of abnormal traffic cases.
When we inspect the data, the abnormal traffic cases are quite heavy tailed, suggesting either they are not always turned off by the operator in a timely manner, or there are some significant drops in link performance that are not directly attributed to a physical incident on the link.
We generally see that general obstructions increases the model output at horizons of 30 minutes (\textbf{Figure 13B}) but decreases it after this.
Abnormal traffic incidents generally increase the model output at higher horizons and decrease it at lower ones, however the time-series and other information could then account for some of the short-lived abnormal traffic incidents we observe, or indeed some of the heavy tailed aspects we see.

Finally, we can also inspect the seasons impact on the model, with results shown in \textbf{Figure 14}.
\begin{figure}[ht!]
	\centering
	\begin{minipage}[t]{.48\textwidth}
		\includegraphics[width=\textwidth]{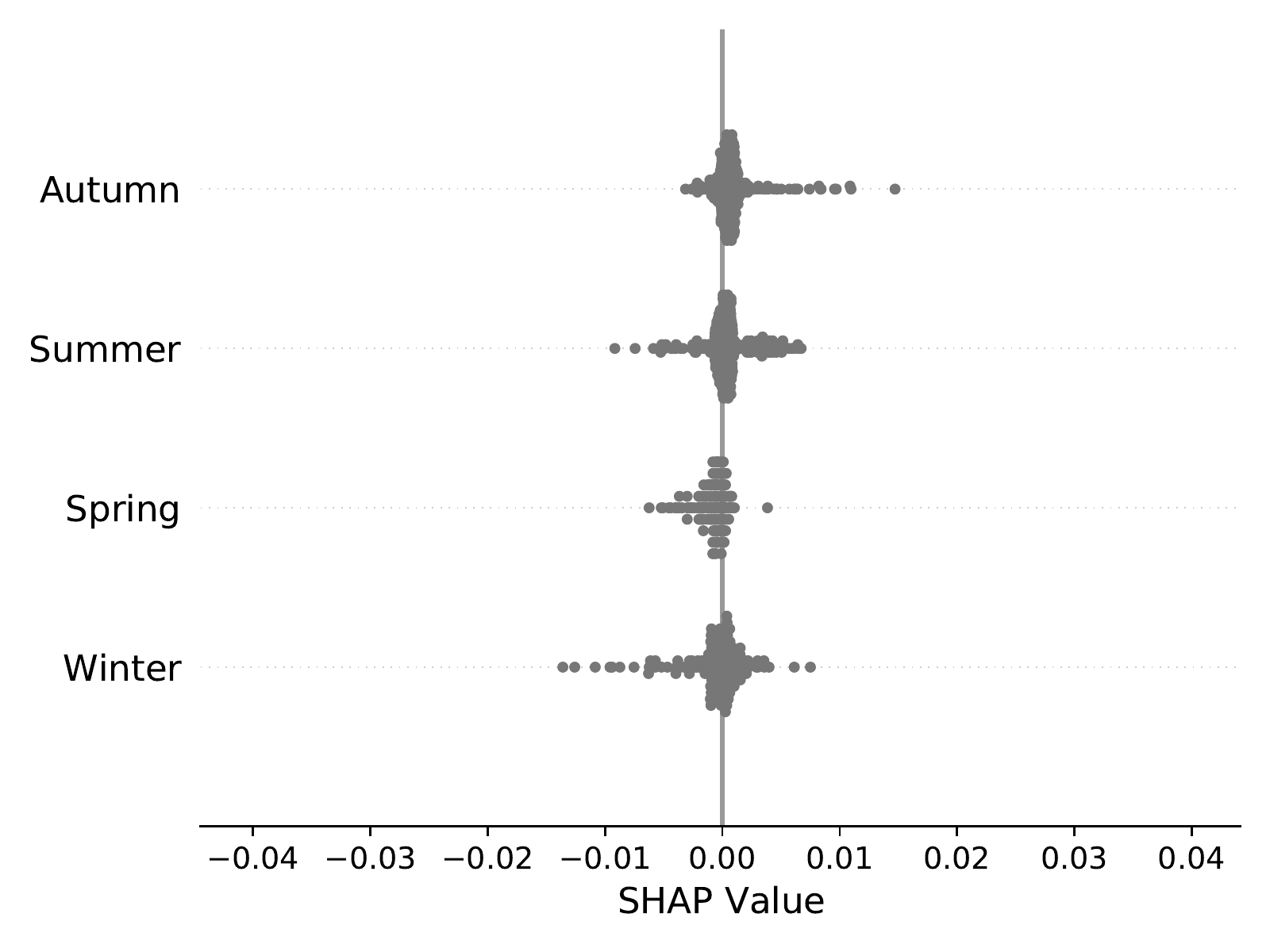}
		\centering \textbf{A}
	\end{minipage}
	~
	\begin{minipage}[t]{.48\textwidth}
		\includegraphics[width=\textwidth]{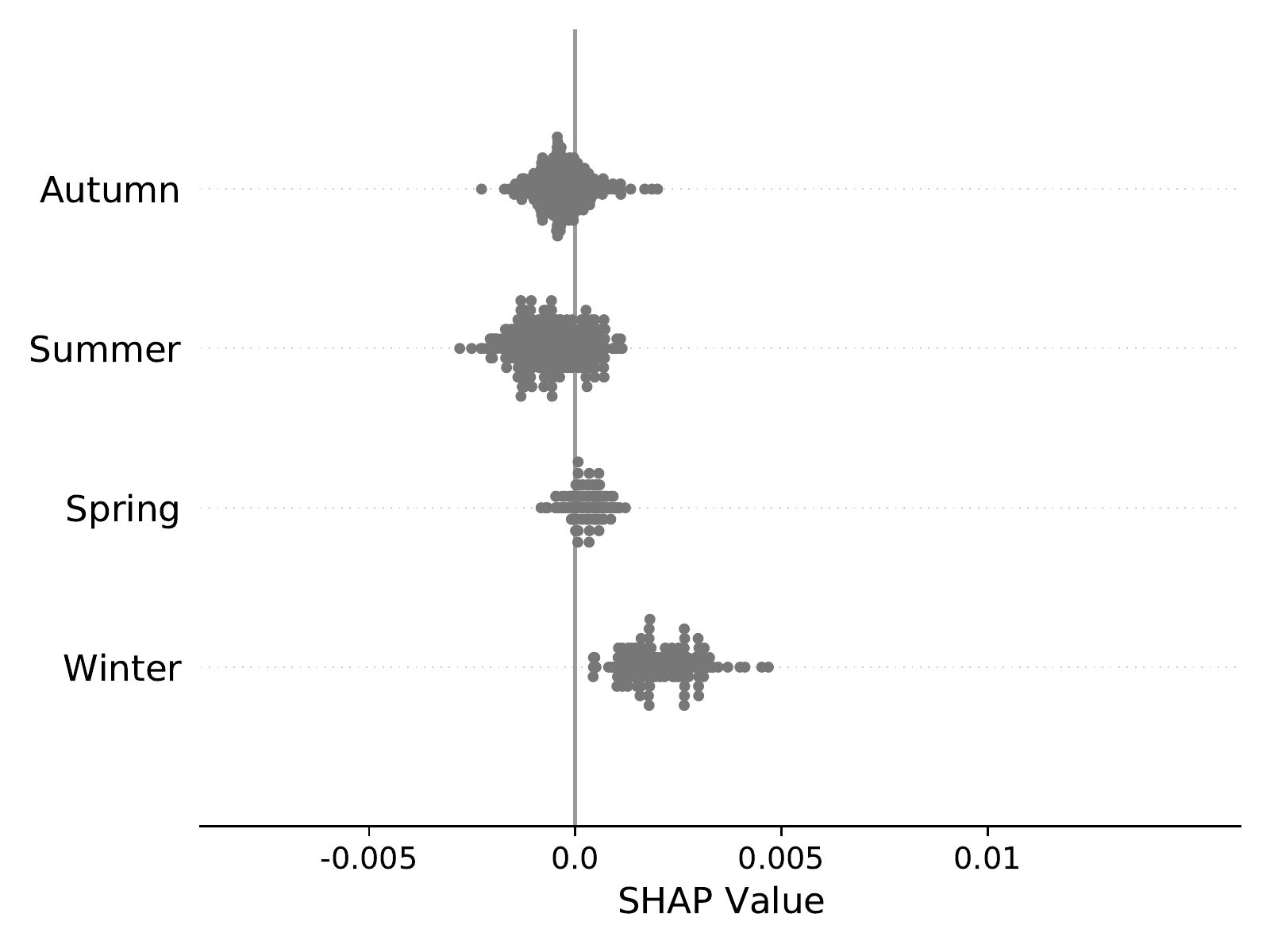}
		\centering \textbf{B}
	\end{minipage}

	\begin{minipage}[t]{.48\textwidth}
		\includegraphics[width=\textwidth]{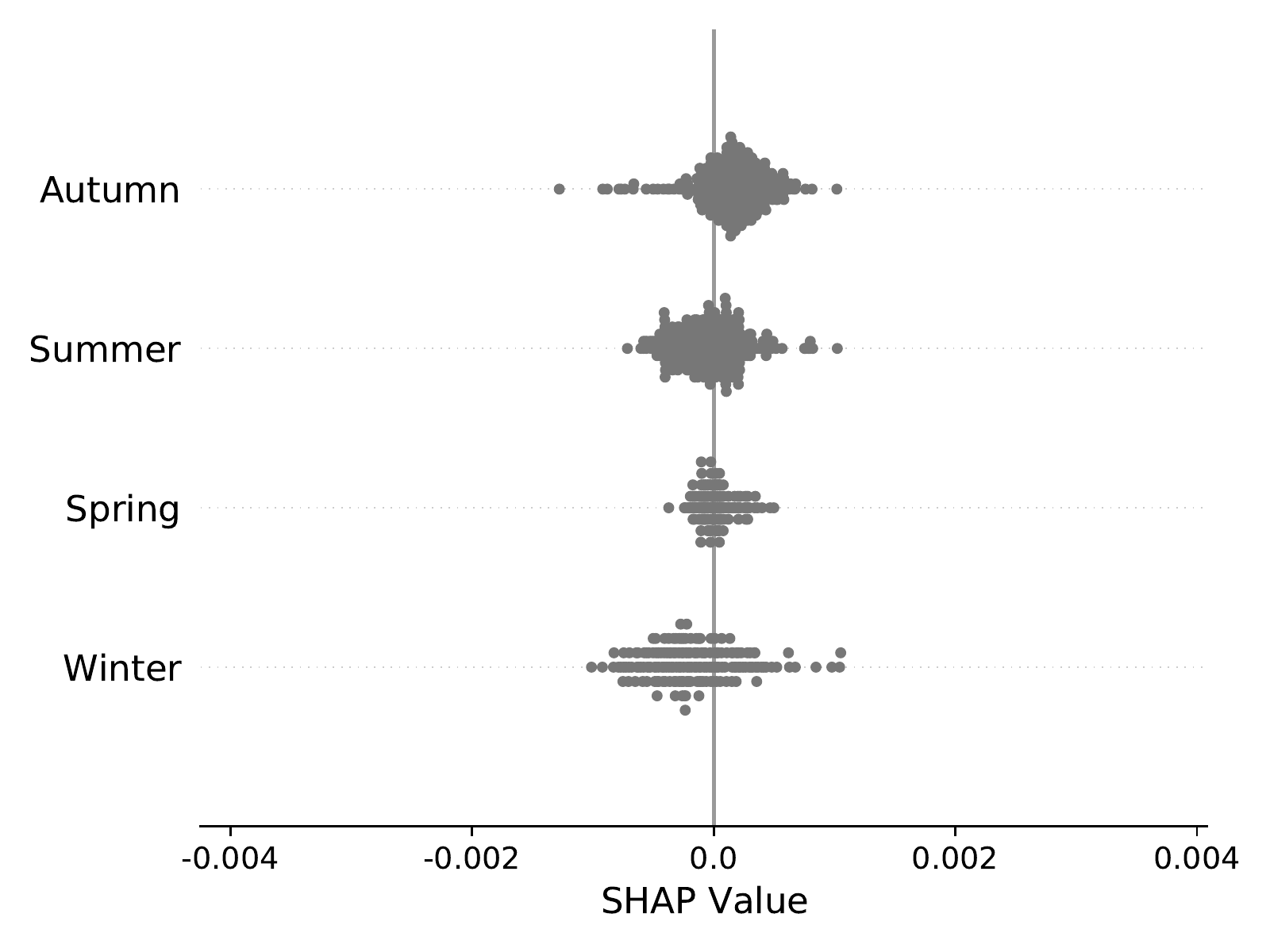}
		\centering \textbf{C}
	\end{minipage}
	~
	\begin{minipage}[t]{.48\textwidth}
		\includegraphics[width=\textwidth]{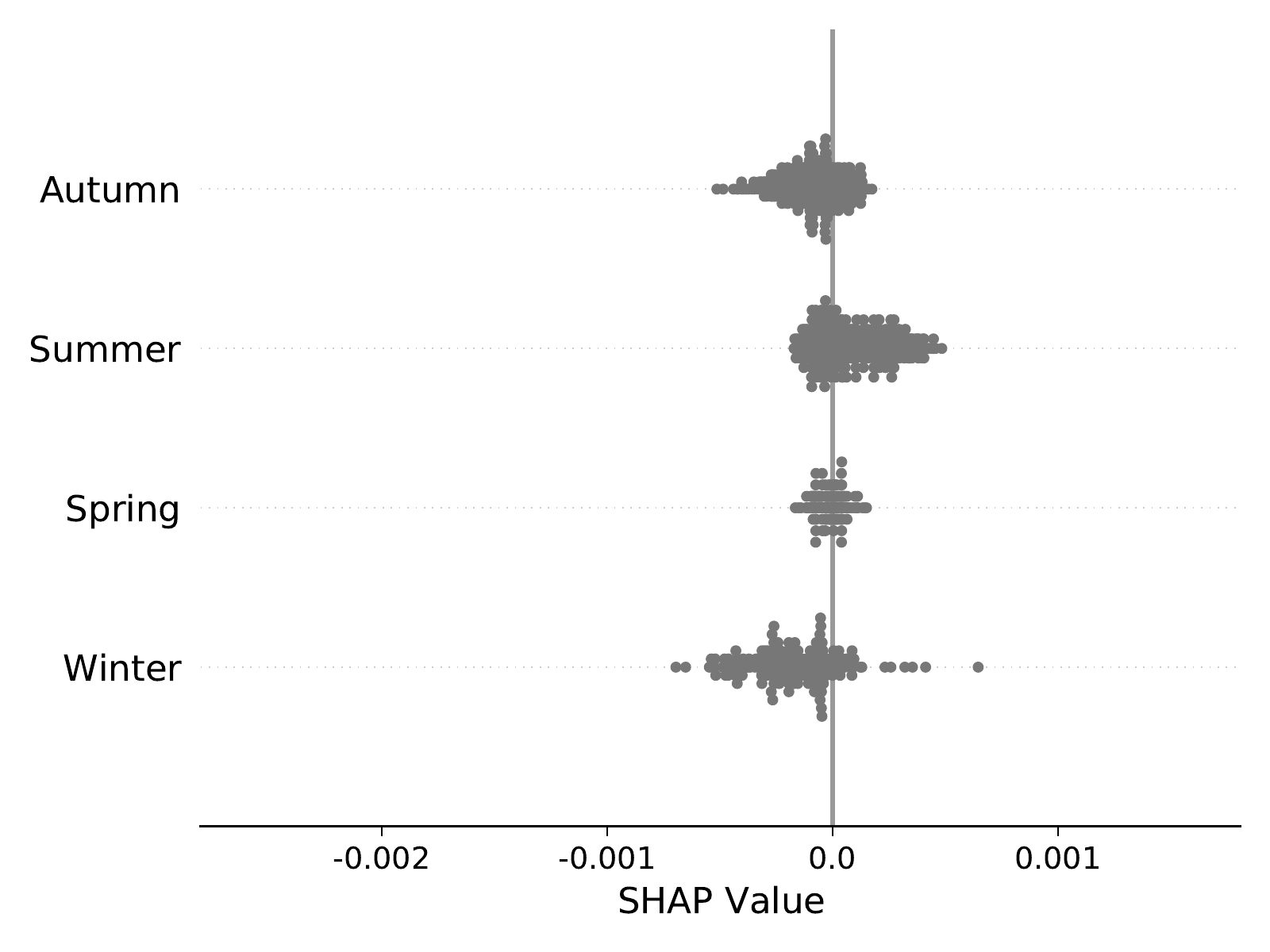}
		\centering \textbf{D}
	\end{minipage}

	\caption{SHAP values for the season. In each plot, we have summed the SHAP values for each one-hot encoded value, and then only plotted the resulting value in the row corresponding to each data-points true feature value. This shows the overall impact of the location feature, and allows one to view this impact separately for each value it attains. \textbf{(A)}: $h=5$, \textbf{(B)} $h=30$, \textbf{(B)} $h=60$, \textbf{(B)} $h=180$ }\label{fig:SHAPSeasonSplit}
\end{figure}
It is first important to note that we only have one year of data, so the season variable here could actually capture some long term systematic change in response time or incident severity.
However, we are not told to expect this from industry experts.
When we consider incidents in the winter, we see the output is mixed at a 5 minute horizon (\textbf{Figure 14A}), increased at a horizon of 30 minutes (\textbf{Figure 14B}) where as it is decreased at the later horizons.
The effect of spring appears to be systematically smaller in size than the other seasons.

\section{AUROC Validation}

Whilst we have looked at C-index and Brier score, for different prediction time, prediction horizon pairs as was done in \cite{dynamic_deephit_a_deep_learning_approach_for_dynamic_survival_analysis_with_competing_risks_based_on_longitudinal_data}, we can equally look at the area under the receiver operator curve as a performance metric, as was done in \cite{dynamic_prediction_in_clinical_survival_analysis_using_temporal_convolutional_networks}. 
We show results for this, at particular horizons, averaged over all input windows, in \textbf{Table 11}, computed using the methods discussed in \cite{scikit_survival} and \cite{summary_measure_of_discriminiation_in_survival_models_based_on_cumulative_dynamic_time_dependent_roc_curves}.

\begin{table}
\centering
  \caption{AUROC values for the dynamic models. Larger values indicate a better model. We see similar conclusions are present here that were made when inspecting the time-varying C-index values.}\label{table:AUROCTable}
  \renewcommand{\arraystretch}{1.25}\begin{tabular}{|c|c|c|c|c|c|c|c|c|}
    \hline
    \multirow{2}{*}{Model} & \multicolumn{7}{c|}{Prediction Horizon (minutes)} & Mean Over \\
    \cline{2-8} 
                & 15 & 30 & 45 & 60 & 120 & 180 & 240 & Horizons \\
    \hline
    Cox         & 0.693 & 0.716 & 0.655 & 0.615 & 0.558 & 0.557 & 0.540 & 0.619 \\
    RSF         & 0.805 & 0.745 & 0.692 & 0.653 & 0.582 & 0.585 & \textbf{0.572} & 0.662 \\ 
    SW (LN)     & 0.857 & 0.798 & 0.759 & 0.714 & 0.622 & 0.580 & 0.552 & 0.697 \\ 
	SW (NP)     & 0.862 & 0.808 & 0.771 & 0.731 & \textbf{0.642} & \textbf{0.602} & 0.568 & 0.712 \\ 
	SW (Kernel) & \textbf{0.878} & \textbf{0.818} & \textbf{0.780} & \textbf{0.735} & 0.640 & 0.594 & 0.558 & \textbf{0.715} \\ 
	SW (Raw)    & 0.877 & 0.809 & 0.764 & 0.712 & 0.613 & 0.571 & 0.544 & 0.698 \\ 
	\hline
  \end{tabular}
\end{table}

\section{MAPE as a Function of Time}\label{appendix:ErrorPerMinute}

We have looked at the error at different percentiles into an incident because of the national criteria set by Highways England when considering incident duration modelling.
However, we note that some other papers consider error at different minutes into an incident.
These are clearly two different methods of aggregation.
The first averages across different events, taking predictions from potentially very different time points, but that represent some fraction of the entire incident.
The second averages across the same prediction time for every incident, but disregards the fact that this might be a very short time relative to the total duration of some events, and a very long time relative to others.
We present results for the second averaging here to give clear comparison and show our results align generally with the expected behaviour of other dynamic prediction works.
Absolute percentage error (APE), computed as a function of minutes into an incident for some of the models considered is given in \textbf{Figure 15} for events longer than 60 minutes.
This behaviour is generally consistent with that seen in Figure 2 in \cite{competing_risk_mixture_model_and_text_analysis_for_sequential_incident_duration_prediction}.
\begin{figure}[ht!]
	\centering
	\includegraphics[width=0.46\textwidth]{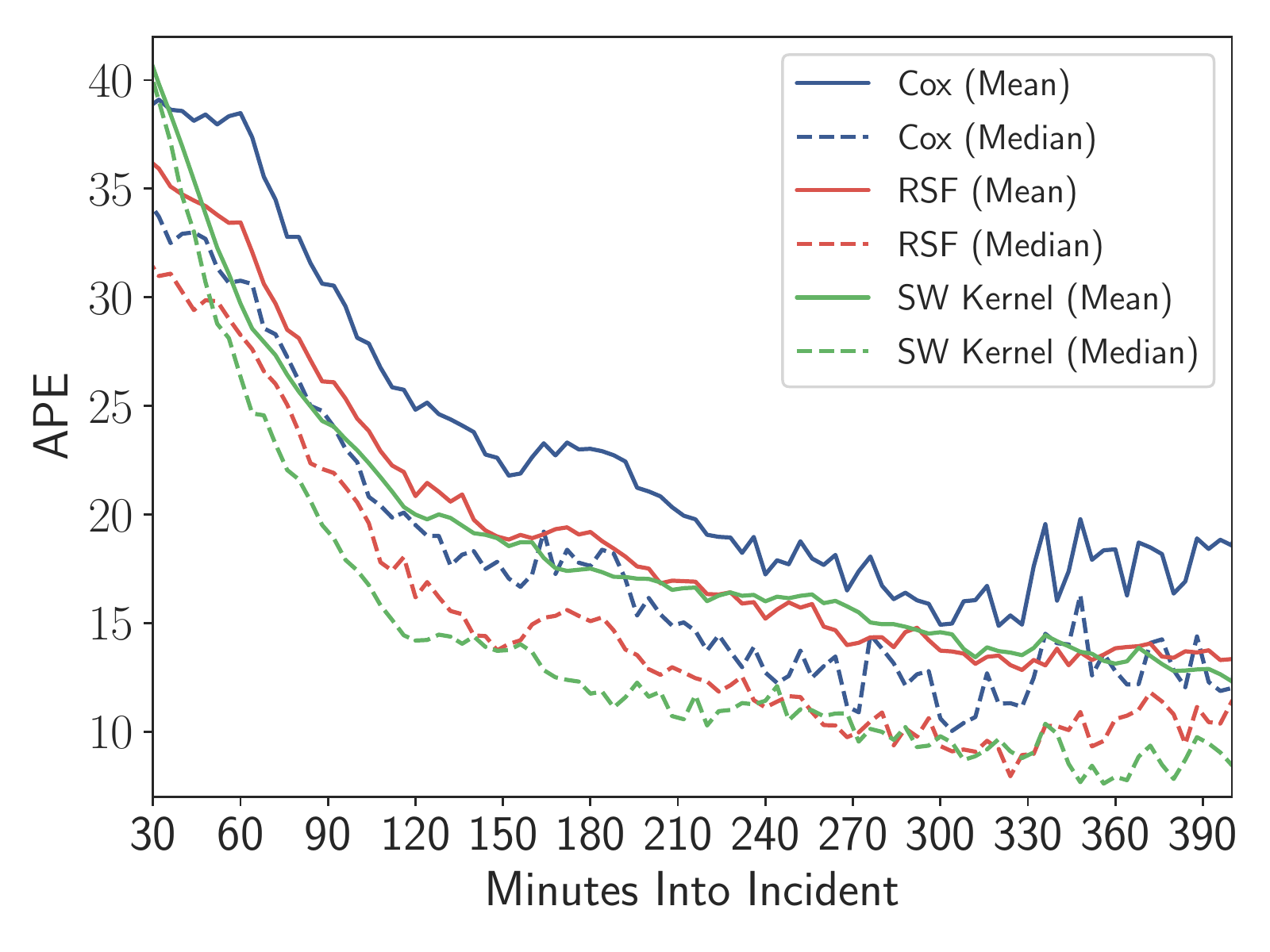}
	\caption{APE (both mean and median shown) for some dynamic models as a function of the minutes into an incident. We show 30 minutes onwards, which is the point where only incident related information is being fed into the sliding window model, and the landmarking models are no longer fit with minor events, as these have already ended.}\label{fig:MAPEVsMintues}
\end{figure}

\cleardoublepage

\bibliographystyle{unsrtnat}
\bibliography{references}

\end{document}